\documentclass[journal=aesccq,manuscript=article]{achemso}
\usepackage{array}
\usepackage{ulem}
\usepackage{titlesec}

\title{A Systematic Study on the Absorption Features of Interstellar Ices in Presence of Impurities}

\author{P. Gorai}
\affiliation{Indian Centre for Space Physics, 43 Chalantika, Garia Station Road, Kolkata 700084, India}
\email{prasanta.astro@gmail.com}
\author{M. Sil}
\affiliation{Indian Centre for Space Physics, 43 Chalantika, Garia Station Road, Kolkata 700084, India}
\email{milansil93@gmail.com}
\author{A. Das}
\affiliation{Indian Centre for Space Physics, 43 Chalantika, Garia Station Road, Kolkata 700084, India}
\email{ankan.das@gmail.com}

\author{B. Sivaraman}
\affiliation{Atomic Molecular and Optical Physics Division, Physical Research Laboratory, Ahmedabad, India}

\author{S. K. Chakrabarti}
\affiliation{Indian Centre for Space Physics, 43 Chalantika, Garia Station Road, Kolkata 700084, India}

\author{S. Ioppolo}
\affiliation{School of Electronic Engineering and Computer Science, Queen Mary University of London, Mile End Road, London E1 4NS, UK}
\alsoaffiliation{Department of Physical Sciences, The Open University, Milton Keynes, UK}
\email{s.ioppolo@qmul.ac.uk}

\author{C. Puzzarini}
\affiliation{Dipartimento di Chimica “Giacomo Ciamician”, Via F. Selmi 2, 40126 Bologna, Italy}
\email{cristina.puzzarini@unibo.it}

\author{Z. Kanuchova}
\affiliation{Astronomical Institute of Slovak Academy of Sciences, SK-059 60 Tatranska Lomnica, Slovakia}

\author{A. Dawes}
\affiliation{Department of Physical Sciences, The Open University, Milton Keynes, UK}

\author{M. Mendolicchio}
\author{G. Mancini}
\author{V. Barone}
\affiliation{Scuola Normale Superiore, piazza dei Cavalieri 7, 56125 Pisa, Italy}
\email{vincenzo.barone@sns.it}

\author{N. Nakatani}
\affiliation{Institute for Catalysis, Hokkaido University, N21W10 Kita-ku, Sapporo, Hokkaido 001-0021, Japan}
\alsoaffiliation{Department of Chemistry, Graduate School of Science and Engineering, Tokyo Metropolitan University, 1-1 Minami-Osawa, Hachioji, Tokyo 192-0397, Japan}

\author{T. Shimonishi}
\affiliation{Frontier Research Institute for Interdisciplinary Sciences, Tohoku University, Aramakiazaaoba 6-3, Aoba-ku, Sendai, Miyagi, 980-8578, Japan}
\alsoaffiliation{Astronomical Institute, Tohoku University, Aramakiazaaoba 6-3, Aoba-ku, Sendai, Miyagi, 980-8578, Japan}

\author{N. Mason}
\affiliation{School of Physical Sciences, University of Kent, Canterbury, Kent, CB2 7NH, United Kingdom}
\alsoaffiliation{Department of Physical Sciences, The Open University, Milton Keynes, UK}

\begin{document}

\begin{abstract}
Spectroscopic studies play a key role in the identification and analysis of interstellar ices and their structure. Some molecules have been identified within the interstellar ices either as pure, mixed, or even as layered structures. Absorption band features of water ice can significantly change with the presence of different types of impurities (CO, $\rm {CO_2}$, $\rm{CH_3OH}$, $\rm{H_2CO}$, etc.). In this work, we carried out a theoretical investigation to understand the behavior of water band frequency, and strength in the presence of impurities. The computational study has been supported and complemented by some infrared spectroscopy experiments aimed at verifying the effect of HCOOH, $\rm{NH_3}$, and $\rm{CH_3OH}$ on the band profiles of pure $\rm{H_2O}$ ice. Specifically, we explored the effect on the band strength of libration, bending, bulk stretching, and free-OH stretching modes. Computed band strength profiles have been compared with our new and existing experimental results, thus pointing out that vibrational modes of $\rm{H_2O}$ and their intensities can change considerably in the presence of impurities at different concentrations. In most cases, the bulk stretching mode is the most affected vibration, while the bending is the least affected mode. HCOOH was found to have a strong influence on the libration, bending, and bulk stretching band profiles. In the case of NH$_3$, the free-OH stretching band disappears when the impurity concentration becomes 50\%. This work will ultimately aid a correct interpretation of future detailed spaceborne observations of interstellar ices by means of the upcoming JWST mission.
\end{abstract}


{
\section{Introduction}
\label{sec:intro}
Interstellar grains mainly consist of nonvolatile silicate or carbonaceous compounds covered by icy mantle layers. Interstellar ices play a crucial role in the chemical enrichment of the interstellar medium (ISM). While the existence of interstellar ice was first proposed by \citet{eddi37} in 1937, a turning point was marked, more than 40 years later, when \citet{tiel82} introduced a combined gas-grain chemistry for the chemical evolution of the ISM. More recently, it has been demonstrated that even pre-biotic molecules can be produced in UV-irradiated astrophysical relevant ices \cite{woon02}. For instance, \citet{nuev14} experimentally showed that nucleobases can be formed by UV irradiation of pyrimidine in H$_2$O-rich ice mixtures containing NH$_3$, CH$_3$OH, and CH$_4$.

The composition of interstellar ices can be determined through their absorption spectra in the infrared (IR) region. Since the composition of ISM grain mantles strongly depends on physical conditions
\cite{das08,das10,das11,das16}, the observed spectra can be very different in different astrophysical regions. $\rm{H_2O}$ is the most dominant ice component in dense molecular clouds \cite{gibb04}, accounting for $60-70\%$ of the icy mantels \cite{whit03}. Water ice was firstly detected through the comparison of ground-based observations of its O-H stretching band at $3278.69$ cm$^{-1}$ ($3.05$ $\mu$m) toward Orion-KL \cite{gill73} and laboratory work by \citet{irvi68}. Since then, several ground-based observations were carried out to identify the signatures of water ice in different astrophysical environments, with further laboratory studies supporting such observations \cite{merr76,lege79,hage79}. More recently, water was detected by the space-borne Infrared Space Observatory (ISO) mission through its Short-Wavelength Spectrometer (SWS) and Long-Wavelength Spectrometer (LWS) in the mid- and far-infrared spectral region. In the mid-IR, along with its strong $\rm{O-H}$ stretching mode ($3.05$ $\mu$m), water shows weaker bending and combination bands at $1666.67$ cm$^{-1}$ (6.00 $\mu$m) and $2222.22$ cm$^{-1}$ (4.50 $\mu$m), respectively, and the libration mode at $769.23$ cm$^{-1}$ (13.00 $\mu$m), which is usually blended with
the grain silicate spectroscopic features along the line of sight to star forming regions in the ISM \cite{gibb04}.

After H$_2$, water is the second most abundant molecular species in the Universe and its gas-phase abundance in the ISM is even comparable to that of CO. Due to the high abundance of water in interstellar ices \cite{dart05}, the amount of the other species is very often expressed in terms of the relative abundance with respect to $\rm{H_2O}$, and thus considered as impurities. Among other solid species, CO, CO$_2$, CH$_3$OH, H$_2$CO, HCOOH, NH$_3$, CH$_4$, and OCS have been unambiguously identified \cite{gibb04}, while theoretical studies suggest that N$_2$ and O$_2$ might be trapped in the ice matrix as well \cite{vand93}. It should be noted that although homonuclear molecules are IR inactive, they can become IR active when embedded in ice matrices. Interstellar ice matrices are usually classified as (i) polar ices, if dominated by polar molecules like H$_2$O, CH$_3$OH, NH$_3$, OCS, H$_2$CO, HCOOH, and (ii) apolar ices, if they are dominated by molecules like CO, CO$_2$, CH$_4$, N$_2$, and O$_2$. Interstellar ices are believed to be a combination of both with a first polar (water-rich) layer and an apolar CO-dominated layer deposited on top of it during the catastrophic freeze-out of CO molecules in the cold core of molecular clouds \cite{boog15}.

Infrared spectroscopy is a suitable technique for identifying interstellar species, particularly, in condensed phases. However, it requires that vibrations are IR active, condition which is fulfilled when the dipole moment changes during vibration. The IR spectrum of a water cluster is one of the primary tools to analyze the features of the aggregation processes in a water matrix \citep{ohno05,bouw07,ober07}. Moreover, four vibrational modes of water, namely libration, bending, bulk stretching, and free-OH stretching, are essential to obtain relevant information about the water cluster itself in various astrophysical environments \cite{gera95,ohno05,bouw07,ober07}.

However, there are some difficulties for the observation of interstellar ices in the mid-IR, such as the need for a background illuminating source being required for absorption, e.g. a protostar or a field star. Furthermore, peak positions, line widths, and intensities of molecular ice features need to be known and compared to laboratory spectra, which further depend on ice temperature, crystal structure of the ice, and mixing or layering with other species \cite{ehre97,schu99,cook16}. As a result, only a very limited number of species have been unambiguously detected in interstellar ices. CO is routinely observed from various
ground-based facilities. In the solid phase, its abundance may vary from $3\%$ to $20\%$ of the water-ice. CO absorbance shows both polar and apolar band profiles. \citet{soif79} reported the detection of the fundamental vibrational band of CO at 4.61 $\mu$m (2169.20 cm$^{-1}$) in absorption toward W33A, based on the laboratory work of \citet{mant75}. The corresponding band profile consists of a broad (polar) component peaking at $2136.75$ cm$^{-1}$ ($4.68$ $\mu$m) and a narrow (non-polar) component peaking at $2141.33$ cm$^{-1}$ ($4.67$ $\mu$m) \cite{chia95,chia98}. $\rm{CO_2}$ was detected in absorption at $657.89$ cm$^{-1}$ (15.20 $\mu$m) toward several IRAS  sources by \citet{dhen89}, based on their laboratory work. The presence of $\rm{CO_2}$ in ice mantles was found on very few astrophysical objects before the launch of ISO  \cite{dart05}, which allowed to firmly establish the ubiquitous nature of CO$_2$ \cite{degr96,guer96,gera99}. In the ice phase, CH$_3$OH abundance varies between 5\% and 30\% with
respect to $\rm{H_2O}$. Its abundance can be even lower in some sources, such as Sgr A and Elias 16 \cite{gibb00}. From ground-based observations, the stretching band of methanol at $2832.80$ cm$^{-1}$ ($3.53$ $\mu$m) was detected in massive protostars \cite{baas88,grim91,alla92}. The first attempt of H$_2$CO observation was made by \citet{schu96}, based on the absorption feature at $2881.84$ cm$^{-1}$ ($3.47$ $\mu$m), towards the protostellar source GL 2136 using the United Kingdom Infrared Telescope (UKIRT). They estimated the abundance of $\rm{H_2CO}$ to be $\sim 7\%$ with respect to $\rm{H_2O}$; however, only a small fraction of $\rm{H_2O}$ is mixed with $\rm{H_2CO}$. Space-based ISO observations \cite{kess96,kess03} estimated the formaldehyde abundance ranging between 1\% and 3\%  in five high mass protostellar envelopes \cite{kean01}. HCOOH was detected both in the solid and gas phase \cite{schu99,vand95,iked01}. CH$_4$ was simultaneously detected in both gas and ice phases toward NGC 7538 IRS 9 \cite{lacy91}. Infrared spectra from the Spitzer Space Telescope show a feature corresponding to the bending mode of solid $\rm{CH_4}$ at $1298.70$ cm$^{-1}$ ($7.7$ $\mu$m) \cite{ober08}; in that work they derived its abundance to range from 2\% to 8\%, with the exception of some sources where abundances were found to be as high as $11-13\%$. \citet{knac82} claimed the first identification of $\rm{NH_3}$ in interstellar grains from an IR absorption feature at $3367.00$ cm$^{-1}$ (2.97 $\mu$m) (NH stretching mode), a detection later proved to be wrong \cite{knac87}. Eventually, the detection of $\rm{NH_3}$ was reported by \citet{lacy98}, who assigned an absorption feature at $1109.88$ cm$^{-1}$ ($9.01$ $\mu$m) toward NGC 7538 IRS 9. \citet{palu95} and \citet{palu97}, based on their laboratory work, identified toward a number of sources an absorption feature at $2040.82$ cm$^{-1}$ ($4.90$ $\mu$m) that can be assigned to OCS when mixed with $\rm{CH_3OH}$.

Recently, the ROSINA mass spectrometer onboard the ESA's Rosetta spacecraft has discovered an abundant amount of molecular oxygen, $\rm{O_2}$, in the coma of the 67P/Churyumov-Gerasimenko comet, thus deriving the ratio $\rm{O_2/H_2O}$ = $3.80\pm0.85\%$ \cite{biel15}. Neutral mass spectrometer data obtained during the ESA's {\it Giotto} flyby are consistent with abundant amounts of $\rm{O_2}$ in the coma of comet 1P/Halley, the $\rm{O_2/H_2O}$ ratio being evaluated to be $3.70\pm1.7\%$ \cite{rubi15}. This makes $\rm{O_2}$ the third most abundant species. In the ISM, O$_2$ and N$_2$ are nearly absent in the gas phase because they are depleted on grains in the form of solid \cite{vand93}. Since N$_2$ and O$_2$ do not possess dipole moment, they cannot be detected using radio observations. However, $\rm{N_2}$ and  $\rm{O_2}$ might be detected by their weak IR active fundamental transition in solid phase, which lies around $4.3 \ \mu$m for N$_2$ and around $6.4 \ \mu$m for O$_2$ \cite{sand01,ehre97}.

To date, infrared observations suggest that the ice mantles in molecular clouds are unambiguously composed of the few aforementioned molecules \cite{herb09}. However, more complex species, such as complex organic molecules (COMs), are also expected to be frozen on ice grains in dense cores. The low sensitivity or low resolution of available observations combined with spectral confusion in the infrared region can cause the weak features due to solid COMs to be hidden by those due to more abundant ice species. The upcoming NASA's James Webb Space Telescope (JWST; \url{https://jwst.stsci.edu}) space mission set to explore the molecular nature of the Universe and the habitability of planetary systems promises to be a giant leap forward in our quest to understand the origin of molecules in space. The high-resolution of the spectrometers onboard the JWST will enable the search of new COMs in interstellar ices and will shed lights on different ice morphologies, thermal histories, and mixing environments. JWST will be able to map the sky and see right through and deep into massive clouds of gas and dust that are opaque in the visible. However, the large amount of spectral data provided by JWST could be analyzed only if extensive spectral laboratory and modeling datasets are available to interpret such data. The work presented here aims at gaining information on the effect of intermolecular interactions in interstellar relevant ices, thus providing some valuable
new laboratory and computed absorption spectra of water-rich ices. These will be useful for the interpretation of future observations in the mid-infrared spectral region.

In this paper, a detailed systematic study of the four fundamental vibrational modes of water in presence of various molecular species with different concentration ratios has been carried out. Since water is the major component of interstellar ice matrix, the latter is considered as composed of water molecules with the other compounds being impurities or pollutants. Since the hydrogen bonding network of pure water clusters in the solid state is strongly affected by increasing concentration of impurities, the spectra of the water pure ice is remarkably different from that of ice containing other species. Indeed, ice bands are very sensitive to intermolecular interactions \cite{sand98}, with both strength and band profiles being affected \cite{knez05, bouw07, ober07}. The changes in the spectral behavior and band strengths are primarily due to molecular size, proton affinity, and polarity of the pollutants. {  To the best of our knowledge, there are not similar studies on liquid water systems as some of the impurity species chosen here are water-soluble. A large number of studies have been devoted to vibrational spectra of diluted aqueous solutions of species corresponding to the polar impurities we considered in the present study. However, the attention is usually focused on the variation of the vibrational properties of the solute and not of the solvent \citep{choi11,capp11,blas13}. In a broader context, some infrared studies are available for the whole solubility range of some species \citep{max03,max04}.}

This paper is organized as follows. In Section \ref{sec:method}, we describe the methodology. In Section \ref{sec:experiment}, we briefly discuss the experimental details. Results and discussions are presented in Section \ref{sec:results_discussions}, and finally, in Section \ref{sec:conclusions}, the concluding remarks are reported.}

\section{Methodology}
\label{sec:method}

There are three established structures of water ice (with a local density of $0.94$ g cm$^{-3}$) formed by vapor deposition at low-pressure. Two of them are crystalline (hexagonal and cubic) and one is a low-density amorphous form \cite{blak94}. High-density amorphous water ice (with a local density $1.07$ g cm$^{-3}$) also exists and can be formed by the vapor deposition at low temperatures \cite{blak94}. \citet{prad12} experimentally analyzed the number of water molecules needed to generate the smallest ice crystal. According to that study, the appearance of crystallization is first observed for $275 \pm 25$ and $475 \pm 25$ water molecules, these aggregates showing the well-known band of crystalline ice around $3200$ cm$^{-1}$ (in the OH-stretching region). \citet{blak94} found experimentally that the onset of crystallization occurs at $148$ K. Since we aim to validate our calculations for the low-temperature and low-pressure regime, we focus on the amorphous ices showing a peak in the IR spectrum at around $3400$ cm$^{-1}$. However, since it is not clear how many water molecules are necessary to mimic the amorphous nature of water ice, we considered pure water clusters of different sizes and studied their absorption spectra. For this purpose, we have optimized water clusters of increasing size at different levels of theory. The water clusters considered are: 2H$_2$O (dimer), 4H$_2$O (tetramer), 6H$_2$O (hexamer), 8H$_2$O (octamer), and 20H$_2$O, with their structures being optimized with three different methods (B3LYP, B2PLYP, and QM/MM) as explained in Section \ref{sec:comp_details}. The specific choice of these cluster models is based on experimental outcomes. Experimentally, it has been demonstrated that the water dimer has a nearly linear hydrogen bonded structure \cite{odut80}. The water clusters with 4 H$_2$O molecules are cyclic in the gas phase \cite{vian97}. Moving to a 6 H$_2$O cluster, different structures are available: a three-dimensional cage in the gas phase \cite{liu97} and cyclic (chair) in the liquid helium droplet \cite{naut00}. For 8H$_2$O, an octamer cube has been found in gaseous states \cite{ohno05}. Finally, the 20H$_2$O cluster has been considered to check the direct effect of the environment, with more details being
provided in the computational detail section.

Using the optimized structures of the series of clusters above, harmonic frequencies have been computed and the band strengths of the four fundamental modes have been calculated by assuming the integration bounds as shown in Table \ref{tab:integration_bounds}. Similar integration bounds (except for free-OH stretching mode) were considered in \citet{bouw07} and \citet{ober07}. Similar absorption profiles of four fundamental modes of pure water have been obtained from our calculations, with their intensity, band positions and
strengths varying for the different cluster sizes and levels of theory used. It is thus essential to find the best compromise between accuracy and computational cost. This means to understand which is the smallest cluster and the cheapest level of theory able to provide a reliable description of water ice. To this aim, we have compared the band positions and the corresponding band strengths of the four vibrational fundamental modes of water obtained with different cluster sizes and different methodologies, to experimental work.

\begin{table}
\centering
\scriptsize{
\caption{Integration bounds for the four fundamental modes of vibration.}
\label{tab:integration_bounds}
\begin{tabular}{cccc}
 \hline
 \hline
&&\multicolumn{2}{c}{\bf Integration bounds}\\
\cline{3-4}
Species & Assignment & Lower (cm$^{-1}$) & Upper (cm$^{-1}$)\\
 \hline
 \hline
 & $\mathrm{\nu_{libration}}$&500 &1100\\
 & $\mathrm{\nu_{bending}}$&1100  &1900\\
H$_2$O& $\mathrm{\nu_{bulk-stretching}}$&3000&3600  \\
 &$\mathrm{\nu_{free-OH-stretching}}$ &3600&4000\\
 \hline
 \hline
\end{tabular}}
\end{table}

While the outcome of this comparison will be discussed later in the text, here we anticipate that the 4H$_2$O cluster in the c-tetramer configuration will be chosen as water ice unit. To investigate the effect of impurities, a number of impurity molecules have been added in order to obtain the desired ratio, as shown in Table \ref{tab:ice_mixture_composition}. For example, in order to get 2:1 ratio of water:impurity(x), we considered $4$ water molecules hooked up with $2$ `$\rm{x}$' molecules. However, for some systems, to have more realistic features of the water cluster, we needed to consider more water molecules. Since it is known that the water ice clusters containing six $\rm{H_2O}$ molecules are the form of all natural snow and ice on Earth \citep{abas05}, we also present a case with six H$_2$O molecules together as a 
unit (see Figure \ref{fig:optimized_structure_6H2O} in the Supporting Information, SI). The cyclic hexamer (chair) configuration of the water cluster containing six $\rm{H_2O}$ has been found the most stable
\cite{ohno05}, and considered in our calculations.

\begin{table}
\centering
\scriptsize{
\caption{Ice mixture composition details.}
\label{tab:ice_mixture_composition}
\begin{tabular}{cccc}
 \hline
 \hline
{\bf H$_2$O:X} & {\bf Total no.} & {\bf No. of} & {\bf No. of}\\
& {\bf of molecules} & {\bf water molecules} & {\bf pollutant molecules} \\
\hline
\hline
1:0.25 & 5 & 4 (80.0\%) & 1 (20.0\%) \\
1:0.50 & 6 & 4 (66.7\%) & 2 (33.3\%) \\
1:0.75 & 7 & 4 (57.1\%) & 3 (42.9\%) \\
1:1.00 & 8 & 4 (50.0\%) & 4 (50.0\%) \\
\hline
\hline
\end{tabular}}\\
\vskip 0.2 cm
{\bf Notes.} Contributions in percentage are provided in the parentheses.
\end{table}

\begin{figure}
\centering
\includegraphics[width=\textwidth]{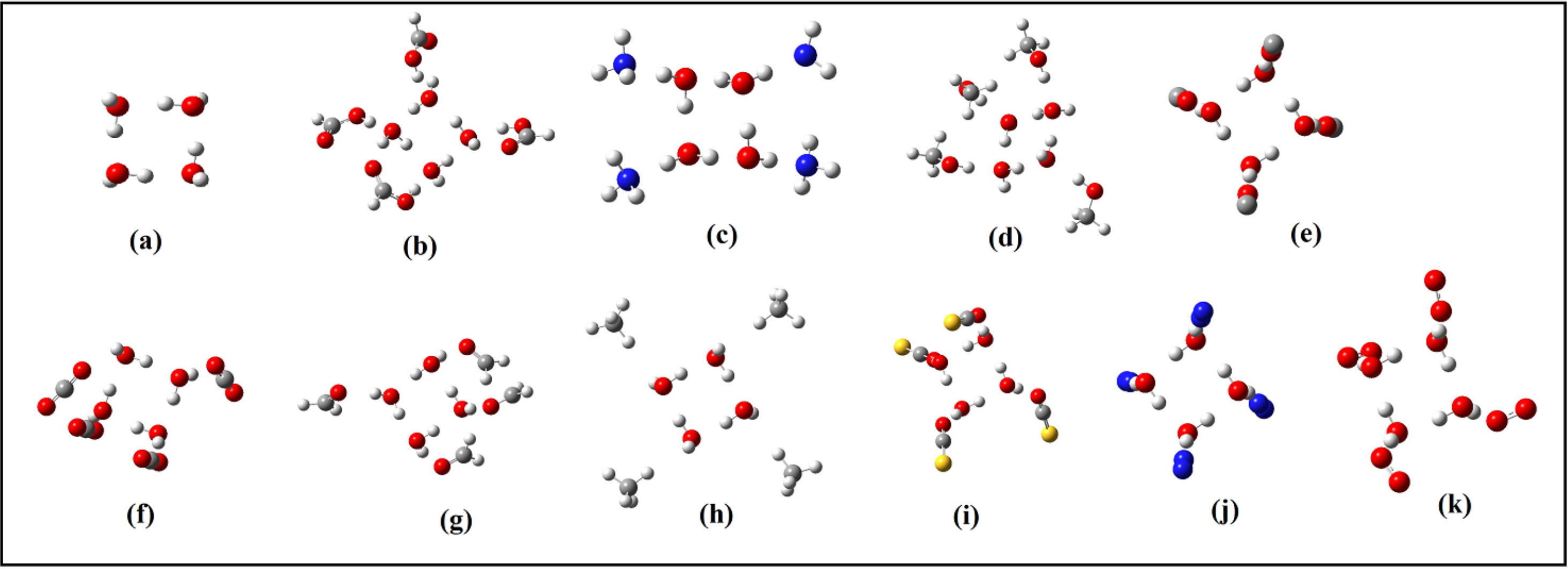}
\caption{{Optimized structures for (a) pure water and for the $4:4$ concentration ratio: (b) $\rm{H_2O-HCOOH}$, (c)
$\rm{H_2O-NH_3}$, (d) $\rm{H_2O-CH_3OH}$,  (e) $\rm{H_2O-CO}$, (f) $\rm{H_2O-CO_2}$, (g) $\rm{H_2O-H_2CO}$, (h) $\rm{H_2O-CH_4}$,
(i) $\rm{H_2O-OCS}$, (j) $\rm{H_2O-N_2}$, (k) $\rm{H_2O-O_2}$ clusters.}}
\label{fig:optimized_structure}
\end{figure}

In Figure \ref{fig:optimized_structure}a, we present the optimized water clusters for the c-tetramer configuration. The same structure was considered by \citet{ohno05} and others \citep{sil17,das18,nguy19}. Since four H atoms are available for interacting with the impurities by means of hydrogen bond, in our
calculations we can reach up to a $1:1$ ratio between the water and the impurity (i.e., we can reach up to 50\% concentration of the impurity in the ice mixture).

In order to understand the effect of impurities on the band strengths of the four fundamental bands considered, we have calculated the area under the curve for each band for different mixtures of pure water and pollutants. The band strength has then been derived using the following relation (introduced in \citet{bouw07} and \citet{ober07}):

\begin{equation}
A_{H_2O:x=1:y}^{band} = \int_{band} I_{H_2O:x=1:y} \times \frac{A_{band}^{H_2O}}{ \int_{band} I_{H_2O}},
\end{equation}
where $A_{H_2O:x}^{band}$ is the calculated band strength of the vibrational water mode in the $1:y$ mixture, $I_{H_2O:x=1:y}$ is its integrated area, $A_{band}^{H_2O}$ is the band strength of the water modes available from the literature, and $\int_{band} I_{H_2O}$ is the integrated area under the vibrational mode for pure water ice. The experimental absorption band strengths of the three modes of pure water ice are taken from \citet{gera95}, who carried out measurements with amorphous water at $14$ K. The adopted values are $2\times10^{-16}$, $1.2\times10^{-17}$, $3.1\times10^{-17}$ $\mathrm{cm \ molecule^{-1}}$, for the bulk stretching ($3280$ cm$^{-1}$), bending ($1660$ cm$^{-1}$), and libration mode ($760$ cm$^{-1}$), respectively. Our {\it ab-initio} calculations refer to the temperature at $0$ K. For the calculation of the band strengths, we are considering the strongest feature of that band. Since for the free-OH stretching
mode no experimental values exist, we consider the result $A_{free-OH}^{H_2O}= 2.09 \times 10^{-17}$ and $2.52 \times 10^{-17}$ $\mathrm{cm \ molecule^{-1}}$ for the c-tetramer and hexamer water clusters, respectively.

{

\subsection{Computational details}
\label{sec:comp_details}

As already mentioned, quantum-chemical calculations have been performed to evaluate the changes of the absorption features of four different fundamental modes, namely, (i) libration, (ii) bending, (iii) bulk stretching, and (iv) free-OH stretching of water in the presence of impurities (CO, $\rm{CO_2}$, $\rm{CH_3OH}$, $\rm{H_2CO}$, HCOOH, $\rm{CH_4}$, NH$_3$, OCS, $\rm{N_2}$, and $\rm{O_2}$). High-level quantum chemical calculations (such as CCSD(T) method and hybrid force field method) are proven to be the best suited for reproducing the experimental data \cite{puzz14,baro15a}. However, due to the dimension of our targeted species, these levels of theory are hardly applicable.

As already anticipated, different DFT functionals have been tested. Most computations have been carried out using the B3LYP hybrid functional \cite{beck88,lee88} in conjunction with the 6-31G(d) basis set (Gaussian 09 package \cite{fris13}). Some test computations have also been performed by using the B2PLYP double-hybrid functional \cite{grim06} in conjunction with the the m-aug-cc-pVTZ basis set \cite{papa09}, in which the $d$ functions have been removed on hydrogen atoms (maug-cc-pVTZ-$d$H). In this case, harmonic force fields have been obtained employing analytic first and second derivatives \cite{bicz10} available in the Gaussian 16 suite of programs \cite{fris16}. The reliability and effectiveness of this computational model in the evaluation of vibrational frequencies and intensities have been documented in several studies (see, for example, ref. \citenum{baro15b}). We have also performed anharmonic calculations (at the B3LYP/6-31G(d) level) for the H$_2$O-CO and H$_2$O-NH$_3$ systems in order to check the effect of anharmonicity on the band strength profiles of the four water fundamental modes.}

\begin{figure}
\centering
\begin{minipage}{0.23\textwidth}
\includegraphics[width=\textwidth]{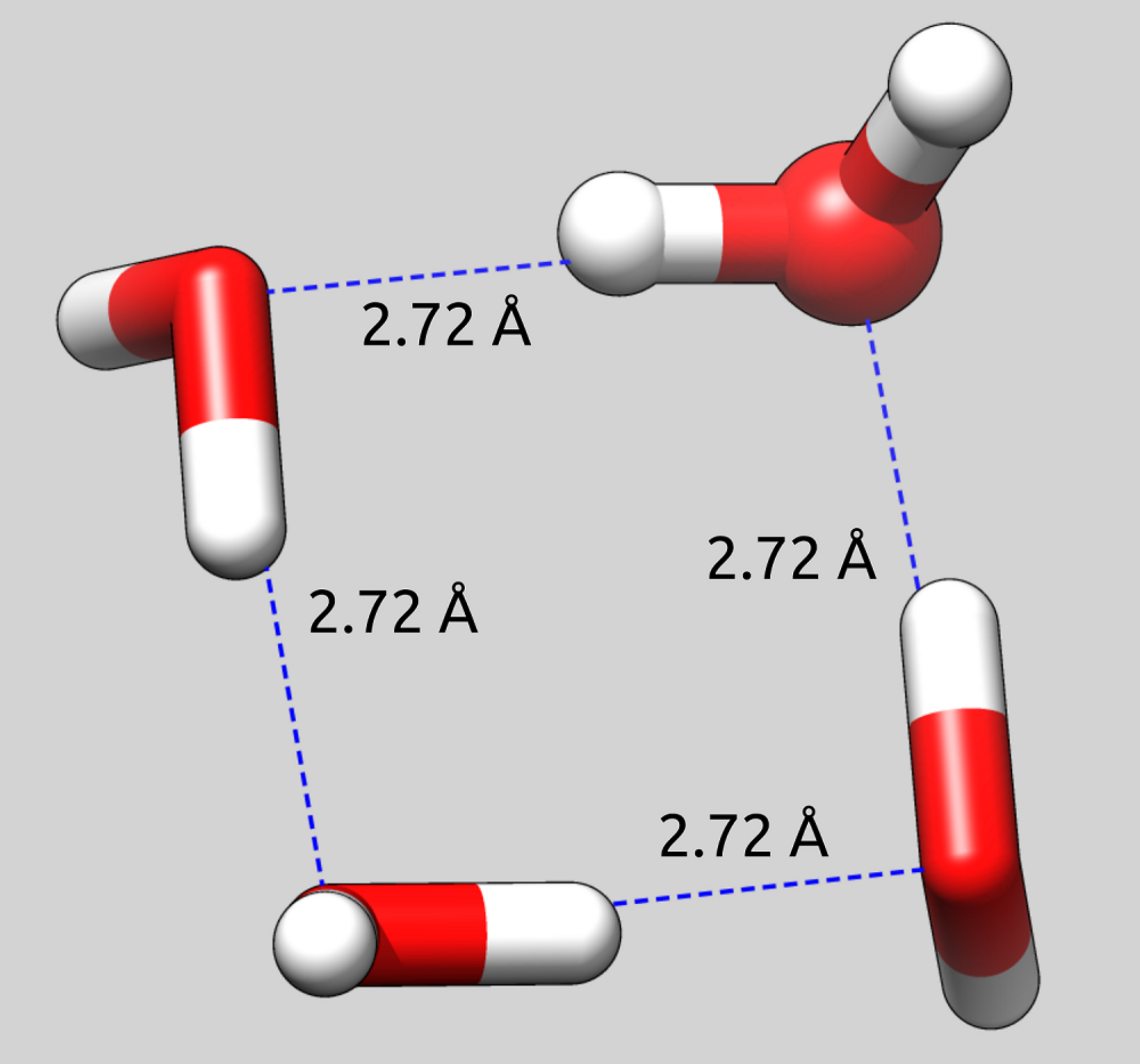}
\end{minipage}
\hskip 2cm
\begin{minipage}{0.50\textwidth}
\includegraphics[width=\textwidth]{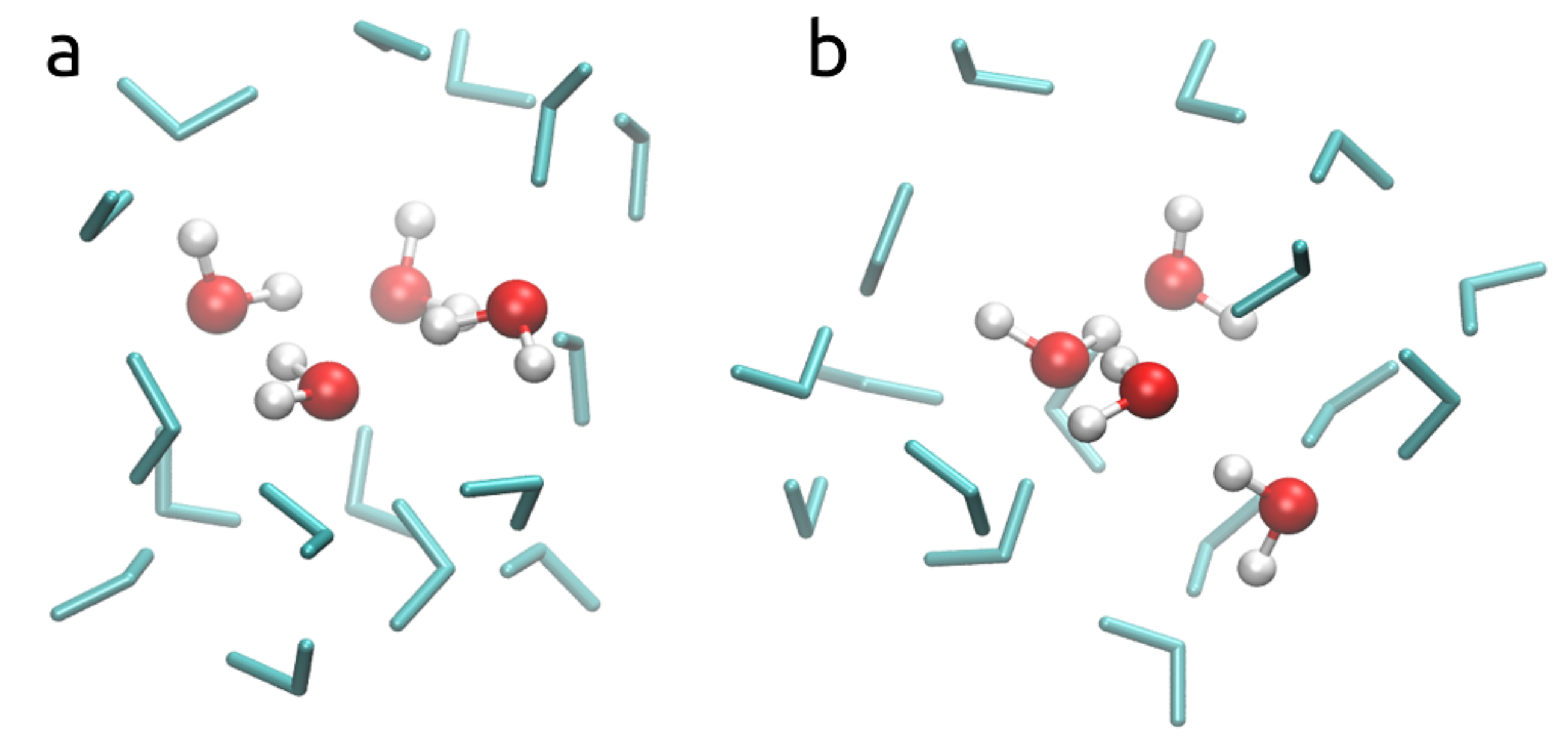}
\end{minipage}
\caption{ {\bf Left:} 4 water system; the single QM water molecule is depicted in ball and stick representation
    and the 3 MM molecules in licorice representation; O-H distances are indicated too. It is to be noted that
    the four H$_2$O can be considered equivalent, {\bf Right:} Innermost water molecules described at QM level
    (ball and stick) and surrounding molecules described at MM level (lines) for the 20H$_2$O system. (a) Configuration 1.
    (b) Configuration 2.}
\label{fig:qm-mm}
\end{figure}

The spectral features of the astrophysical ices can be altered in both active (direct) and passive (bulk) ways. Following a consolidated practice \cite{sanf20}, to include the passive contribution of the bulk ice on the spectral properties of the ice mixtures considered, we embedded our explicit cluster in a continuum solvation field to represent local effects on the ice mixture. To this end, we resorted to the integral equation formalism (IEF) variant of the Polarizable Continuum Model (PCM) \cite{toma05}. The solute cavity has been built by using a set of interlocking spheres centered on the atoms with the following radii (in \AA): $1.443$ for hydrogen, $1.925$ for carbon, $1.830$ for nitrogen, and $1.750$ for oxygen, each of them scaled by a factor of $1.1$, which is the default value in Gaussian. For the ice dielectric constant, that of bulk water ($\varepsilon=78.355$) has been used, although any dielectric constant larger than about $30$ would lead to very similar results. In addition, we have also performed QM/MM geometry optimizations of a pure water cluster containing 4 H$_2$O molecules, in which all but one molecule at the square vertexes were put in the MM layer (see Figure \ref{fig:qm-mm}, left panel). A pure water cluster system containing
20 H$_2$O molecules has also been considered. For this, we started from the coordinates of the full QM optimization and selected two alternative sets of four innermost molecules at the center of the cluster with a complete hydrogen bond network (determined with a geometric criterion \cite{pagl17}) with first neighbor water molecules; the remaining 16 molecules were described at the MM level (see Figure \ref{fig:qm-mm}, right panel). All QM/MM \cite{chun15} calculations were carried out with the Gaussian 16 \cite{fris16} code (rev. C01) using the hybrid B3LYP functional in conjunction with the 6-31 G(d) basis set. Atom types and force field parameters for water molecules in the MM layer were assigned according to the SPC-Fw flexible water model \cite{wu06}; the choice was driven by (i) the necessity for a flexible, 3-body classical water model and (ii) the accuracy with which the selected model reproduced ice I$_h$ properties. Solvent effects were mimicked by using PCM \cite{canc97}. 
The vibrational analysis results from QM/MM calculations are provided in Tables \ref{tab:4H2O_MM-QM}, \ref{tab:20H2O_MM-QM_Conf_1}, and 
\ref{tab:20H2O_MM-QM_Conf_2} in the SI.

\section{Experimental Methods}
\label{sec:experiment}
Literature laboratory data are here used whenever possible to constrain simulations \cite{bouw07,ober07}. In the cases of formic acid, ammonia, and methanol in water ice, new experiments have been performed using the high vacuum (HV) Portable Astrochemistry Chamber (PAC) at the Open University (OU) in the United Kingdom. A detailed description of the system is reported elsewhere \cite{dawe16}. Briefly, the main chamber is a commercial conflat flange cube (Kimball Physics Inc.) connected to a turbo molecular pump (300 l/s), a custom made stainless steel dosing line through an all metal leak valve, a cold finger of a closed-cycle He cryostat (Sumitomo Cryogenics) and two ZnSe windows suitable for IR spectroscopy. During operation,  the base pressure in the chamber is in the 10$^{-9}$ mbar range, and the base temperature of the cold finger is 20 K. In thermal contact with the cryostat, the substrate is a ZnSe window (20 mm x 2 mm). A DT-670 silicon diode temperature sensor (LakeShore Cryotronics) is connected to the substrate to measure its temperature, while a Kapton flexible heater (Omegalux) is used to change its temperature. Diode and heater are both connected to an external temperature controller (Oxford Instruments).

Gaseous samples were prepared and mixed in a pre-chamber (dosing line) before being dosed into the main chamber through an all metal leak valve. A mass-independent pressure transducer was used to control the amount of gas components mixed in the pre-chamber. Chemicals were purchased at Sigma-Aldrich with the highest purity available [HCOOH ($>$95\%), NH$_3$ (99.95\%), and CH$_3$OH (99.8\%)]. Ices were grown $in situ$ by direct vapour deposition onto the substrate at normal incidence via a 3 mm nozzle that is 20 mm away from the sample. Infrared spectroscopy was performed in transmission using a Fourier Transform infrared (FTIR; Nicolet Nexus 670) spectrometer with an external Mercury cadmium telluride (MCT) detector. A background spectrum comprising 512 co-added scans was acquired before deposition at 20 K and used as
reference spectrum for all the spectra collected after deposition to remove all the infrared signatures along the beam pathway that were not originated by the ice sample. Each IR spectrum is a collection of 256 co-added scans. The IR path was purged with dry compressed air to remove water vapour.

\begin{figure}
\includegraphics[width=8cm, height=6cm]{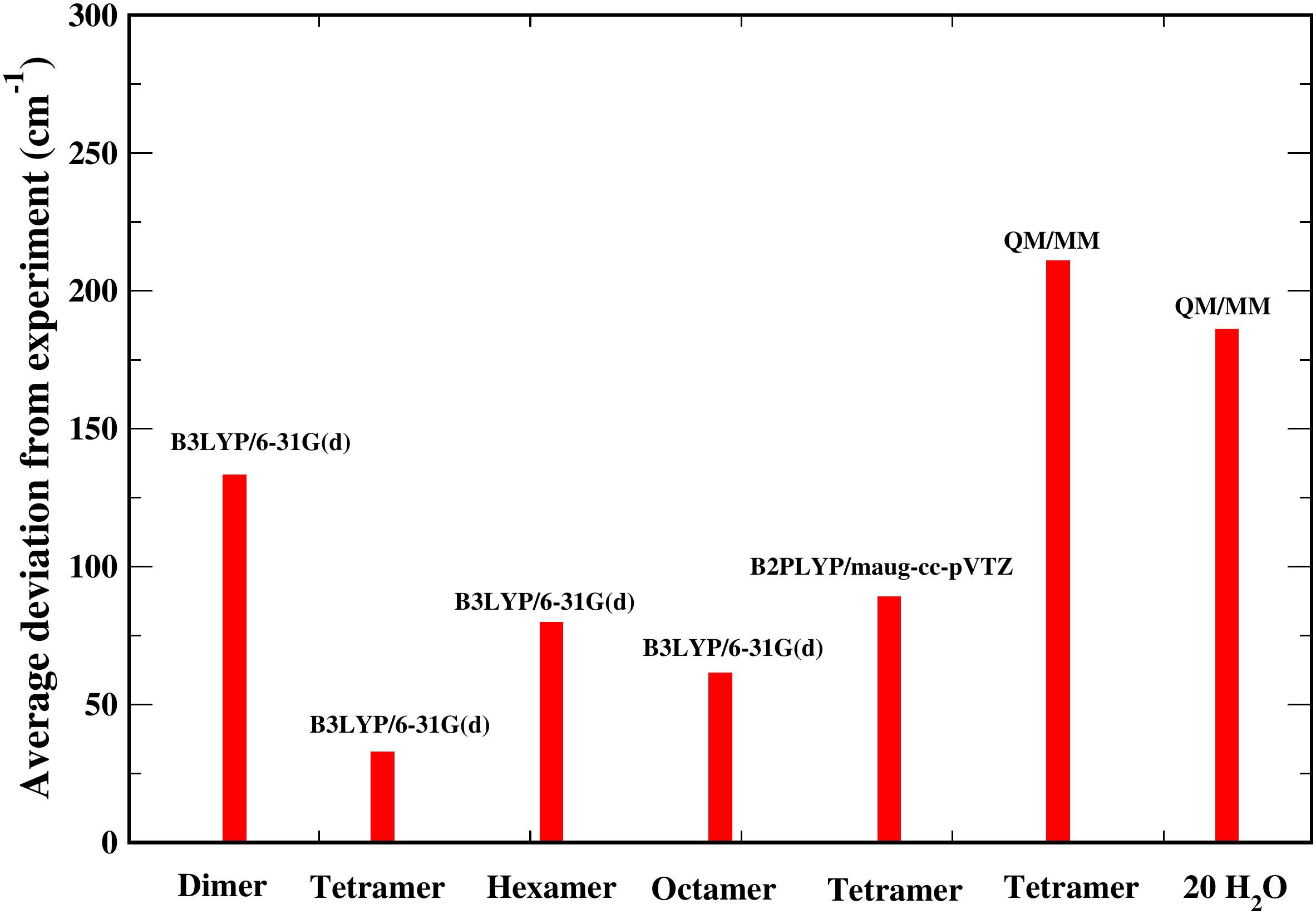}
\includegraphics[width=8cm, height=6cm]{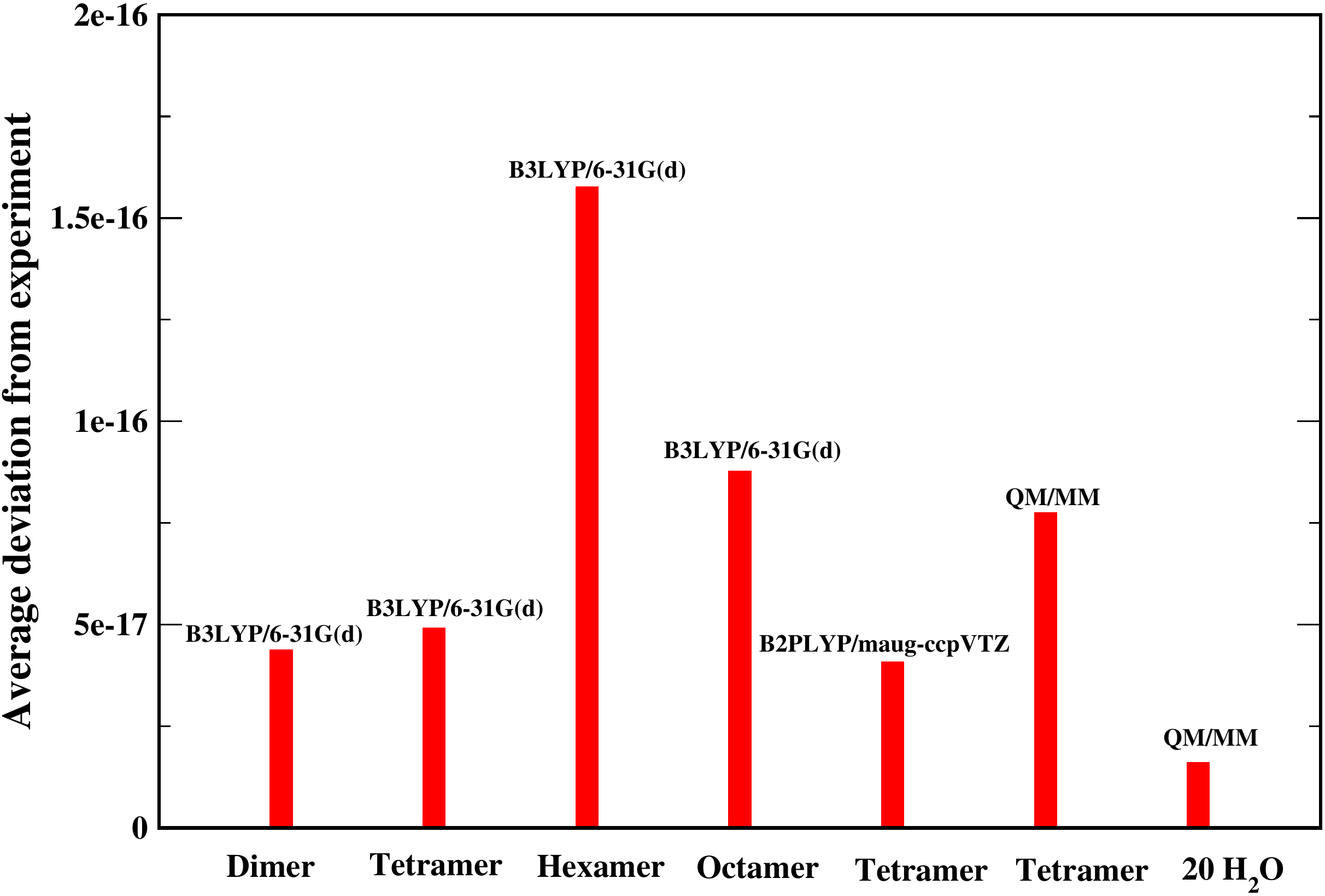}
\caption{{Deviation of computed band positions (left panel) and band strengths (right panel) from experiments.}}
\label{fig:histo-h2o-cluster-compare}
\end{figure}

{
\section{Results and Discussions}
\label{sec:results_discussions}

In this section, first of all, the pure water ice will be addressed in order to establish the best compromise between accuracy and computational cost for the description of the water ice unit cell. To this aim, we will resort on the comparison with experiment. Then, we will move to the ice containing impurities. To further proceed with the validation of our protocol, water ice containing HCOOH, NH$_3$, CH$_3$OH, CO, and CO$_2$ as impurities will be investigated, thus exploiting the comparison between experiment and computations. This will also involve, as mentioned above, new measurements. Finally, in the last part, our protocol will be extend to the study of water ices with H$_2$CO, CH$_4$, N$_2$, and O$_2$ as
impurities.

\subsection{Part 1. Validation}

\subsubsection{Band strength of pure water}
\label{band_strength_pure_water}
In Table \ref{tab:band_strength_pure_water}, the water band positions obtained with different methods and different sizes of the water cluster are compared with experimental data. Since computations provide several frequencies corresponding to a single mode of vibration, for the sake of comparison, we have reported the computed frequencies of the four fundamental modes after convolving them with a Gaussian function with an adequate width \citep{lica15} (all transition frequencies are collected in the Appendix,
{  Table \ref{table:comparison-different-water-cluster}}). The comparison of Table \ref{tab:band_strength_pure_water} is graphically summarized in Figure \ref{fig:histo-h2o-cluster-compare}.
The left panel shows the average deviation of the band position of three fundamental modes of water (libration, bending, and stretching) from the experimental counterpart \citep{gera95}. It is interesting to note that the band positions obtained using the tetramer configuration and the B3LYP/6-31G(d) level of theory provides the best agreement. The right panel shows the average deviation of the band strengths
from experiments. QM/MM calculations for the 20 water-molecule cluster (as described in the computational details) show the minimum deviation from experimental data. The results obtained for the tetramer configuration, both at the B3LYP and B2PLYP level, also provide small deviations. Based on the results of the comparison carried out, the B3LYP/6-31G(d) level of theory and the tetramer configuration have been found to be a suitable combination to describe the water cluster with a limited computational cost.}

\begin{table}
{
\scriptsize{
\centering
\caption{Absorption band strengths and band positions (within parentheses; in cm$^{-1}$) of pure water ice.}
\label{tab:band_strength_pure_water}
\begin{tabular}{c|p{1.4cm}|c|c|c|c}
\hline
\hline
{Vibration}&{Experiment}& \multicolumn{4}{c}{Computed values in cm molecule$^{-1}$ and band position in cm$^{-1}$} \\
\cline{3-6}
{mode} & {\citet{gera95}} & {Dimer} & {c-Tetramer } & {c-Hexamar (chair)}& {Octamer (cube)}\\
 &&B3LYP-631G(d) &B3LYP-631G(d)&B3LYP-631G(d)& B3LYP-631G(d)\\
& &{}&{}&{}&{}\\
\hline
Libration &$3.1\times 10^{-17}$ (760)&$2.70 \times 10^{-17} (670)$&$2.40\times 10^{-17} (733)$&$1.17 \times 10^{-16}(870)$&$1.27 \times 10^{-17} (848)$\\
Bending   &$1.2\times 10^{-17}$ (1660)&$2.23 \times 10^{-17} (1710)$&$4.18\times 10^{-17} (1714)$&$4.62 \times 10^{-17} (1730)$&$6.85 \times 10^{-17} (1717)$\\
Stretching &$2.0\times 10^{-16}$ (3280)&$8.27 \times 10^{-17}(3540)$&$3.11\times 10^{-16} (3298)$&$5.53 \times 10^{-16} (3220)$&$3.89 \times 10^{-16} (3320)$\\
Free-OH & $-$                  & $1.35 \times 10^{-17} (3810)$&$2.10\times 10^{-17} (3775)$&$2.52 \times 10^{-17} (3780)$&$1.52 \times 10^{-17} (3788)$\\
\hline
\hline
{Vibration}&{Experiment}& \multicolumn{3}{c}{Computed values in cm molecule$^{-1}$ and band position in cm$^{-1}$} \\
\cline{3-6}
{mode} & {\citet{gera95}} & {c-Tetramer} & {1H$_2$O(QM)+3H$_2$O(MM)} & {4H$_2$O(QM)+16H$_2$O(MM)}& \\
& &B2PLYP/m-aug-cc-pVTZ&B3LYP-631G(d)& B3LYP-631G(d)\\
\hline
Libration &$3.1\times 10^{-17}$ (760) &$3.94 \times 10^{-17} (714)$&$7.91\times 10^{-17} (669)$&$4.9 \times 10^{-17} (720)$&\\
Bending   &$1.2\times 10^{-17}$ (1660)&$2.93 \times 10^{-17} (1635)$&$2.37\times 10^{-17} (1431)$&$1.88 \times 10^{-17} (1426)$&\\
Stretching &$2.0\times 10^{-16}$ (3280) &$2.97 \times 10^{-16}(3477)$&$2.68\times 10^{-17} (3593)$&$1.76\times 10^{-16} (3565)$&\\
Free-OH & $-$                  & $2.40 \times 10^{-17} (3865)$&$6.24\times 10^{-18} (3797)$&$2.52 \times 10^{-17} (3636)$&\\
\hline
\hline
\end{tabular}}}
\end{table}

{
In the following sections, the results for water ice with HCOOH and NH$_3$ as impurities are first reported and discussed, thereby exploiting the outcomes of new experiments.  Then, we move to the CH$_3$OH-H$_2$O ice for which new experimental results have been obtained. For the last two cases addressed, namely CO-H$_2$O and CO$_2$-H$_2$O, the experimental data for the comparison have been taken from the literature. Unless otherwise stated, we use the c-tetramer configuration for the rest of our calculations.}

\subsubsection{HCOOH ice}
\label{HCOOH_ice}
\begin{figure}
\includegraphics[width=8cm, height=6cm]{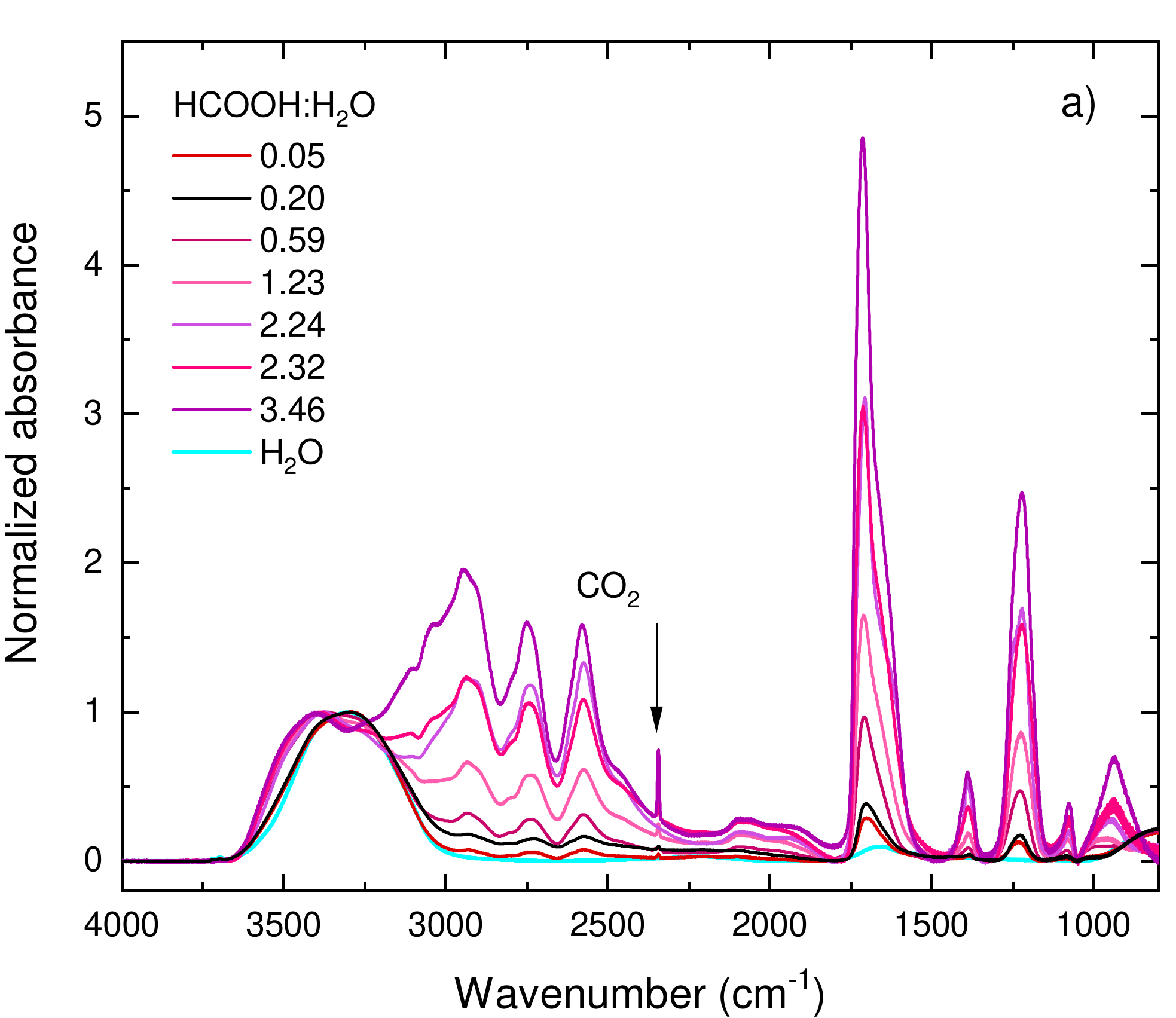}
\includegraphics[width=8cm, height=6cm]{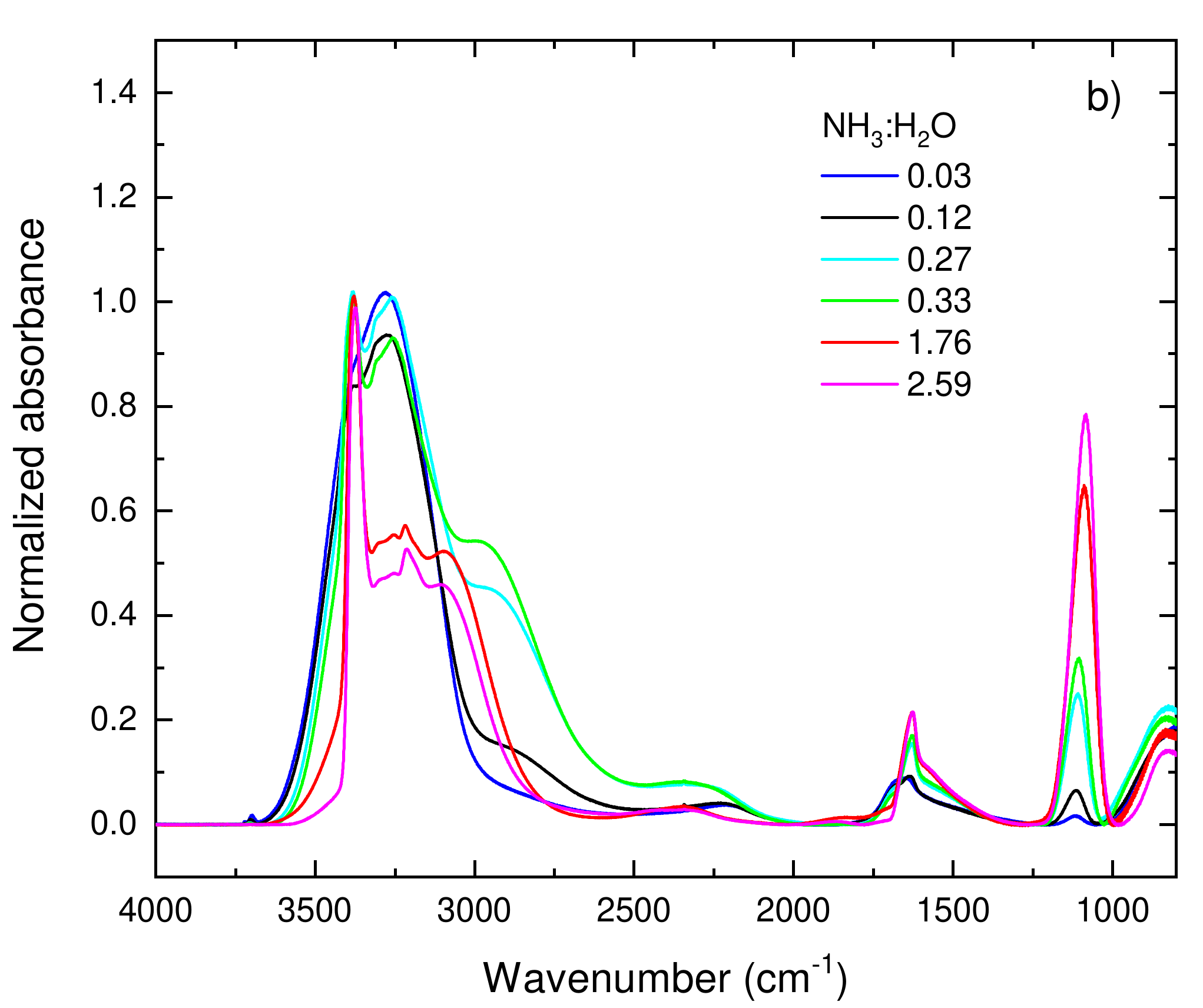}
\vskip 0.5cm
\includegraphics[width=8cm, height=6cm]{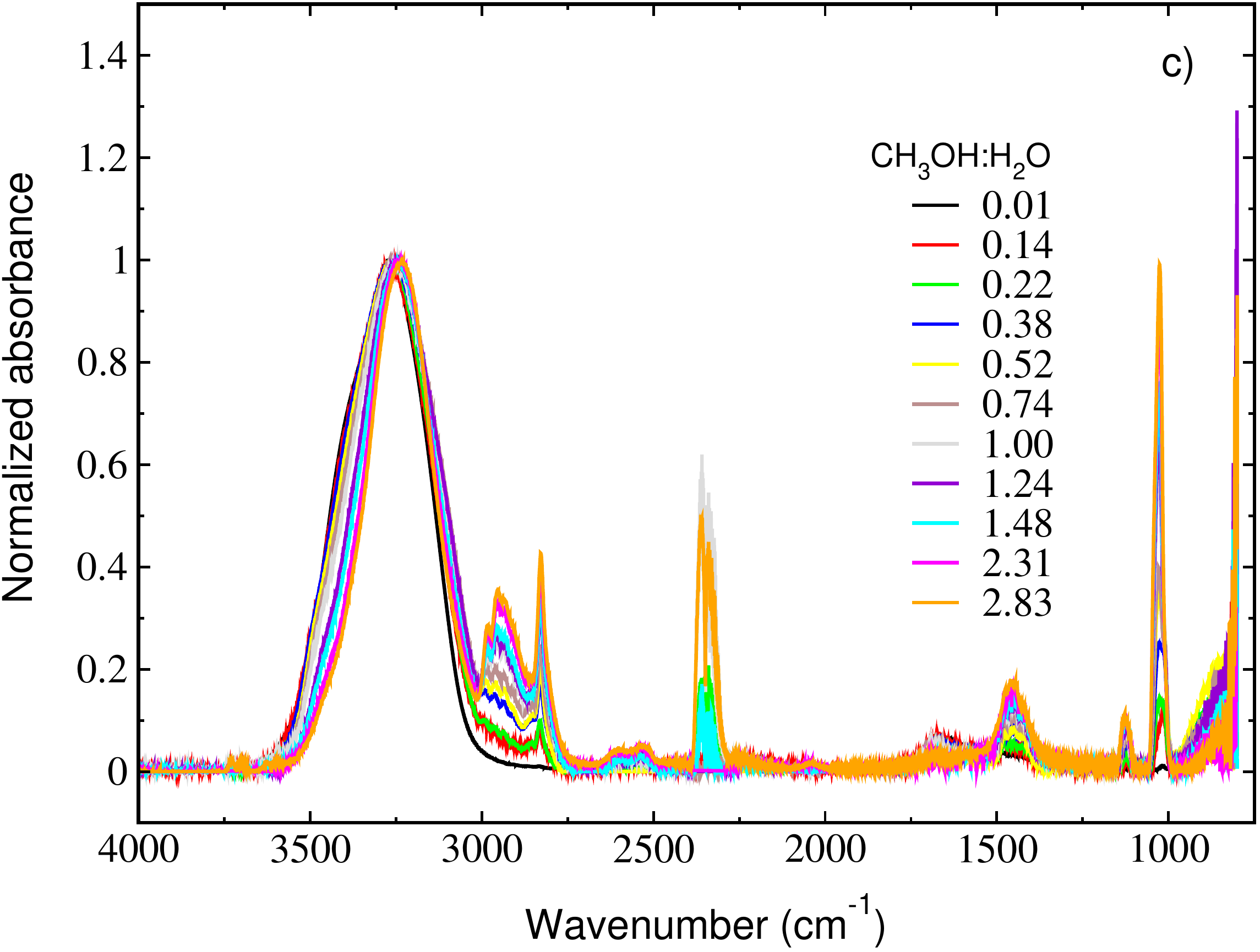}
\caption{(a) IR spectra for different HCOOH:H$_2$O ice mixtures deposited at T$=20$ K. (b) IR spectra for different NH$_3$:H$_2$O ice mixtures deposited at T$=20$ K. (c) IR spectra of different CH$_3$OH:H$_2$O ice mixtures deposited at T$=30$ K. The color legend is explained in the insets. All IR spectra are normalized with respect to the O-H stretching band.}
\label{fig:experiment}
\end{figure}

Infrared spectra were measured for various mixtures of H$_2$O and HCOOH ice deposited at $20$ K, as explained in the experimental details section (see Section \ref{sec:experiment}). These, normalized with respect to the O-H stretch, are shown in Figure \ref{fig:experiment}a. A minor contamination due to CO$_2$ was detected in some experiments. In all experiments, the amount of CO$_2$ deposited in the ice was found to be between 1000 and more than 100 times less abundant than H$_2$O and HCOOH, respectively. Therefore, we do not expect that the CO$_2$ contamination affects the recorded IR spectra profiles.

The mixture ratios were determined from the fit of the the spectrum of a selected mixture, the measurement of the area of the water band at $3333.33$ cm$^{-1}$ (3.00 $\mu$m), and the comparison with the pure water counterparts. For HCOOH, the absorption area is measured at 1700 cm$^{-1}$. In fact, HCOOH has the strongest mode at $1694.92$ cm$^{-1}$ ($5.90$ $\mu$m) which corresponds to its C=O stretching mode. But the feature overlaps with the position of the OH bending mode of solid water at $1666.67$ cm$^{-1}$ ($6.00$ $\mu$m). The contribution from the water bending mode at $\sim$1700 cm$^{-1}$ has been subtracted from the total area before the band strength mentioned above being used to calculate the amount of HCOOH in the ice mixture. The band strengths used here are $2.0\times10^{-16}$ for H$_2$O \cite{gera95} and $6.7\times10^{-17}$ for HCOOH \cite{mare87,schu99}. Another relatively weaker mode of HCOOH at $1388.89$ cm$^{-1}$ ($7.20$ $\mu$m) was also considered because the corresponding region is free from interfering transitions \cite{schu99}. As seen in Figure \ref{fig:experiment}a, the HCOOH:H$_2$O ratios cover the 0.05 to 3.46 range. In this respect, it is worthwhile noting that the abundances of solid phase HCOOH in the interstellar ices vary between 1\% to 5\% with respect to the $\rm{H_2O}$ ice \cite{biss07}.

Moving to the computational study, Figure \ref{fig:optimized_structure}b shows how the HCOOH molecules are bonded to the water molecules to form the $4:4$ H$_2$O-$\rm{HCOOH}$  mixture used in our calculations. In the Appendix, the absorption band profiles of the H$_2$O-HCOOH clusters with different impurity 
concentrations are shown (see Figure \ref{fig:H2O-HCOOH}). The transition frequencies and the
corresponding strongest intensity values, obtained at the B3LYP/6-31G(d) level, are given in the Appendix (see Table \ref{tab:H2O_X}). Calculations have also been carried out using the B2PLYP functional, 
the results being summarized in Tables \ref{tab:4H2O_B2PLYP}, \ref{tab:4H2O_1HCOOH_B2PLYP}, \ref{tab:4H2O_2HCOOH_B2PLYP}, \ref{tab:4H2O_3HCOOH_B2PLYP}, 
and \ref{tab:4H2O_4HCOOH_B2PLYP} in the SI.

To investigate how the band strength varies with impurity concentrations, the data are fitted with a linear function  A$_{eff}$ = a$\cdot$[X] + b, 
where X = HCOOH, NH$_3$, $\rm{CH_3OH}$, CO, $\rm {CO_2}$, $\rm{H_2CO}$, CH$_4$, OCS, N$_2$, and O$_2$. The coefficient `a' provides the information 
whether the band strength increases or decreases by increasing the concentration of X, [X], and the coefficient `b'  indicates the band strength of 
the vibration mode in the absence of impurities. The fitting coefficients, for all impurity considered, are provided in Table \ref{tab:linear_coeff}.
In Figure \ref{fig:band_strength}a, the band strength profile as a function of the concentration of HCOOH is shown.

\begin{figure}
\begin{minipage}{0.49\textwidth}
\includegraphics[width=\textwidth]{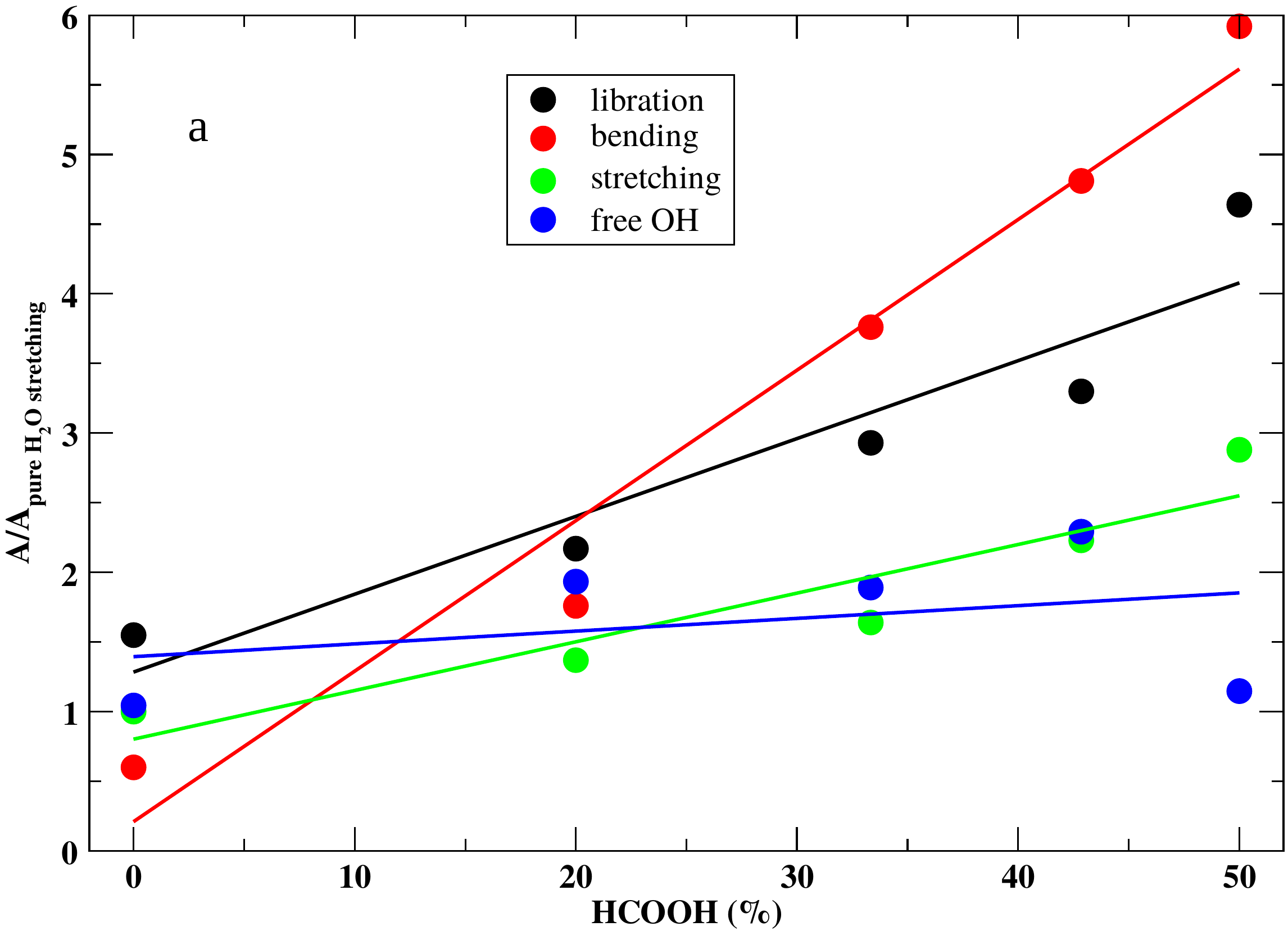}
\end{minipage}
\begin{minipage}{0.49\textwidth}
\includegraphics[width=\textwidth]{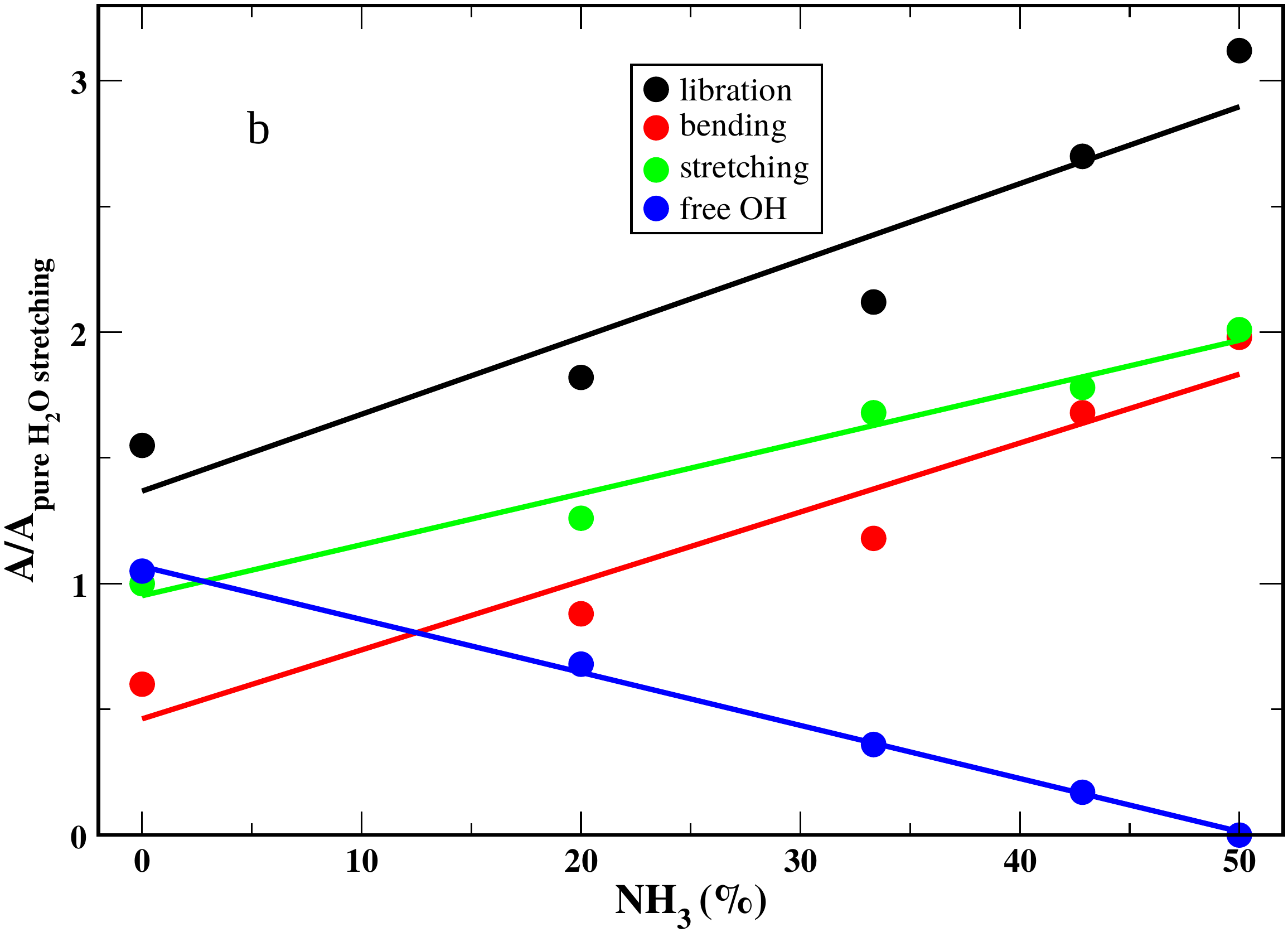}
\end{minipage}
\begin{minipage}{0.49\textwidth}
\includegraphics[width=\textwidth]{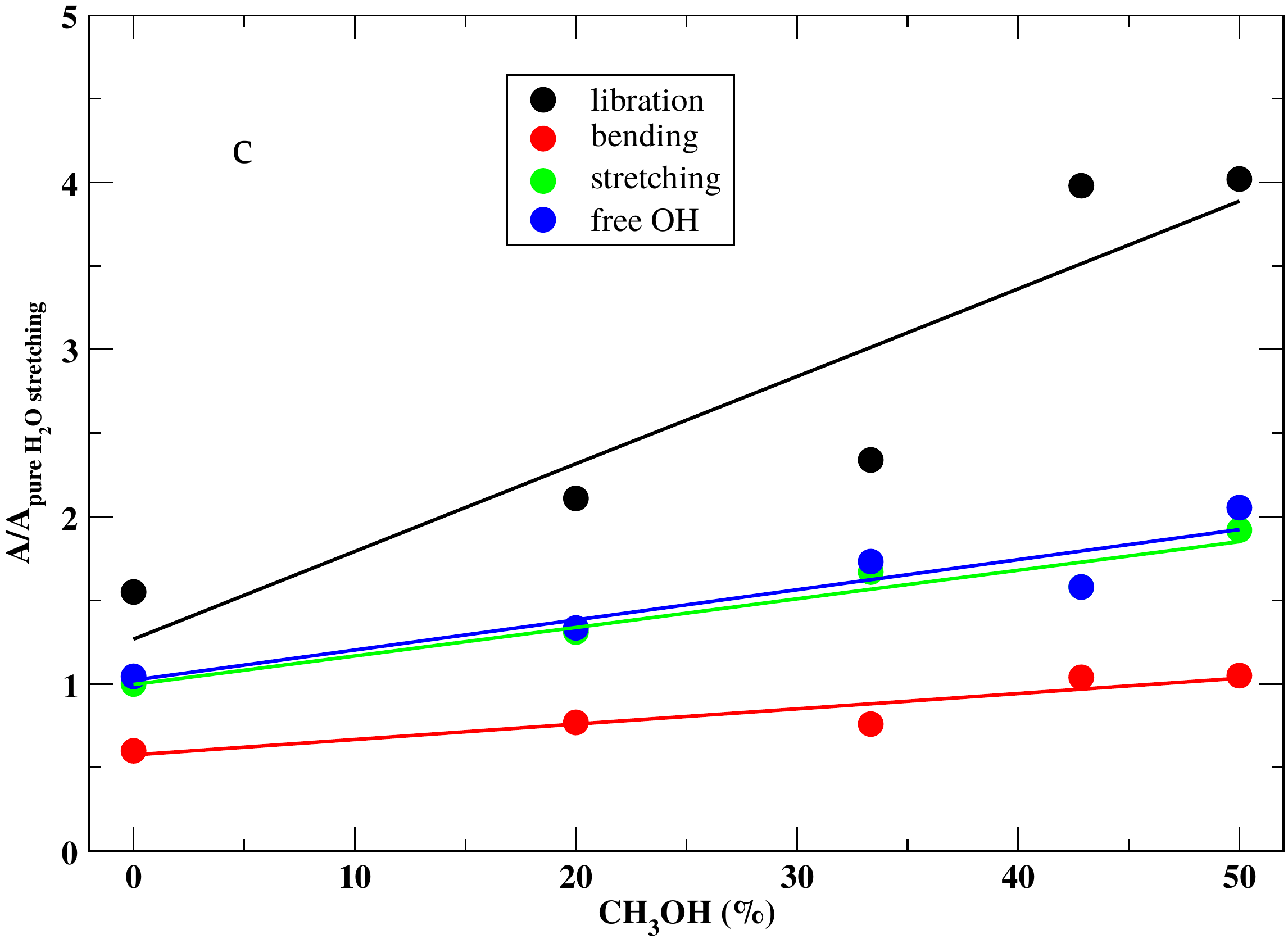}
\end{minipage}
\begin{minipage}{0.49\textwidth}
\includegraphics[width=\textwidth]{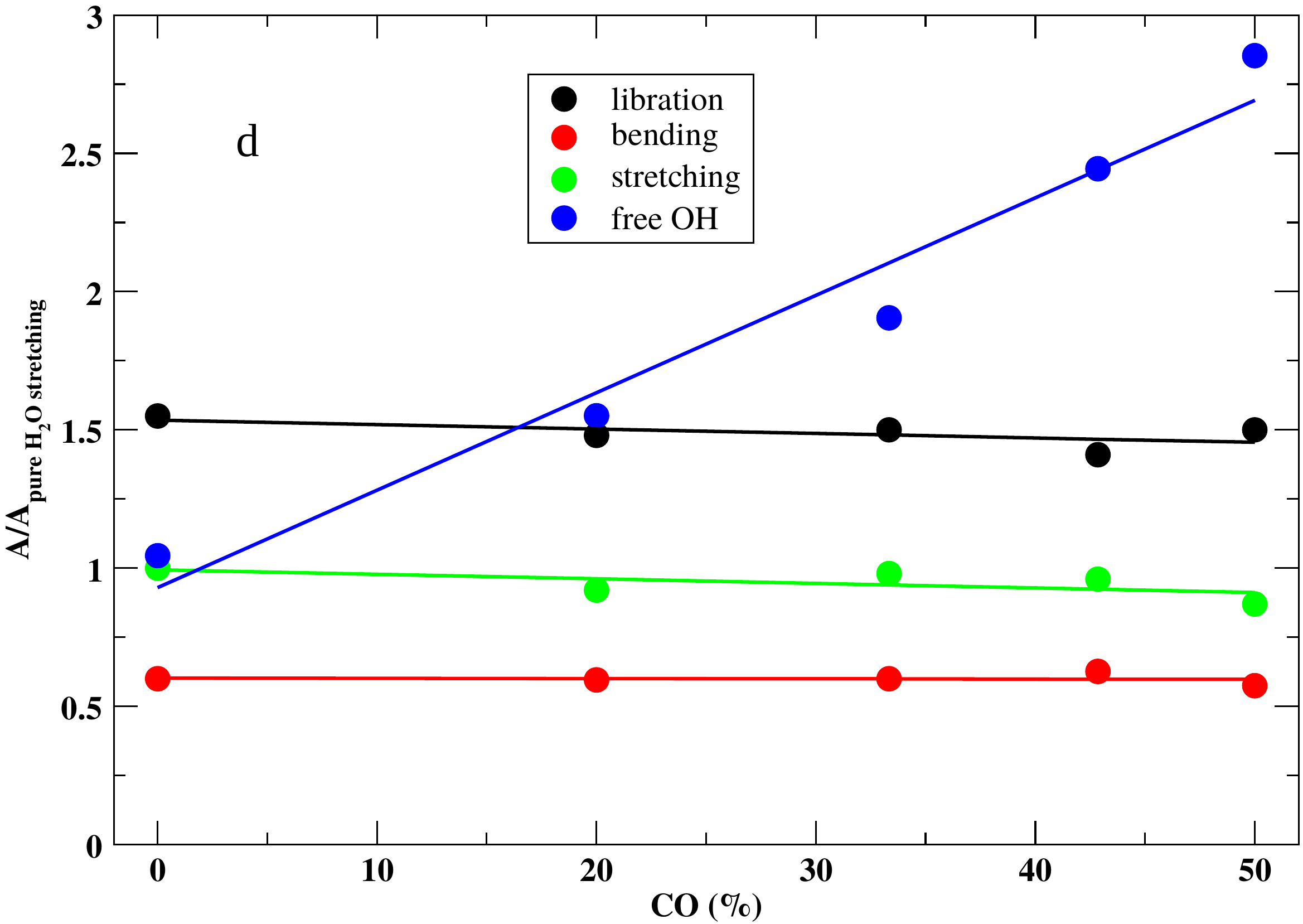}
\end{minipage}
\begin{minipage}{0.49\textwidth}
\includegraphics[width=\textwidth]{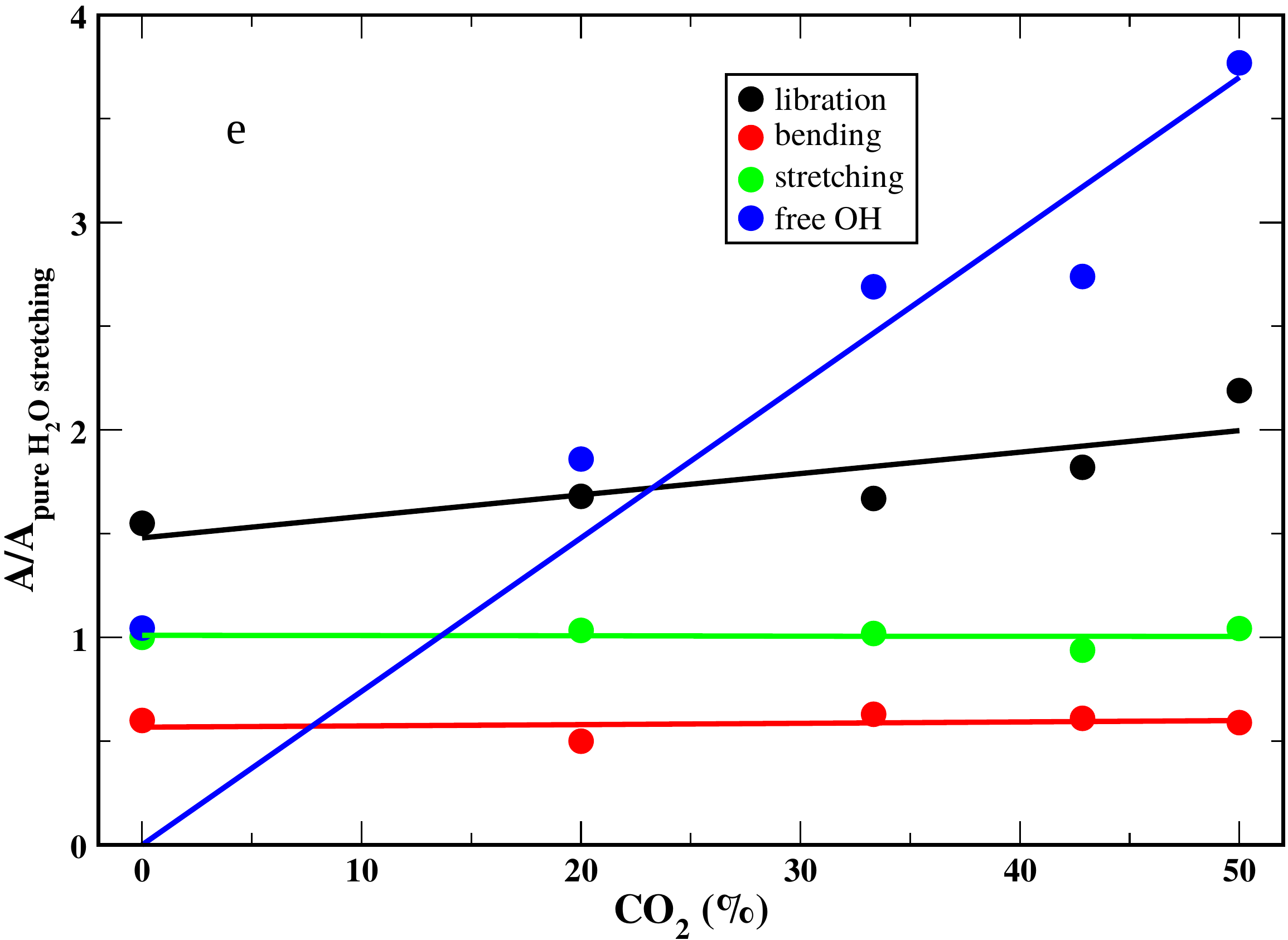}
\end{minipage}
\begin{minipage}{0.49\textwidth}
\includegraphics[width=\textwidth]{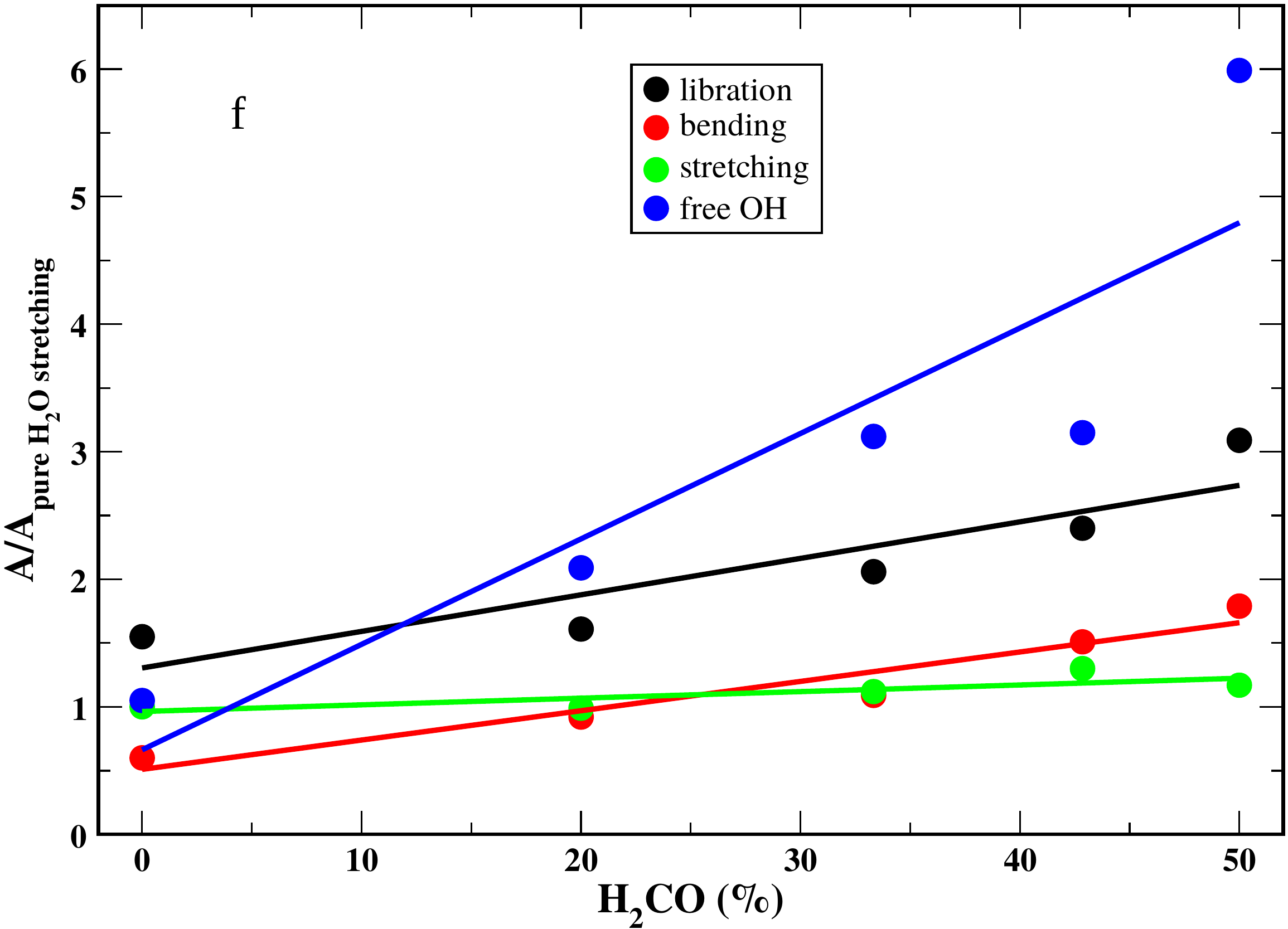}
\end{minipage}
\begin{minipage}{0.49\textwidth}
\includegraphics[width=\textwidth]{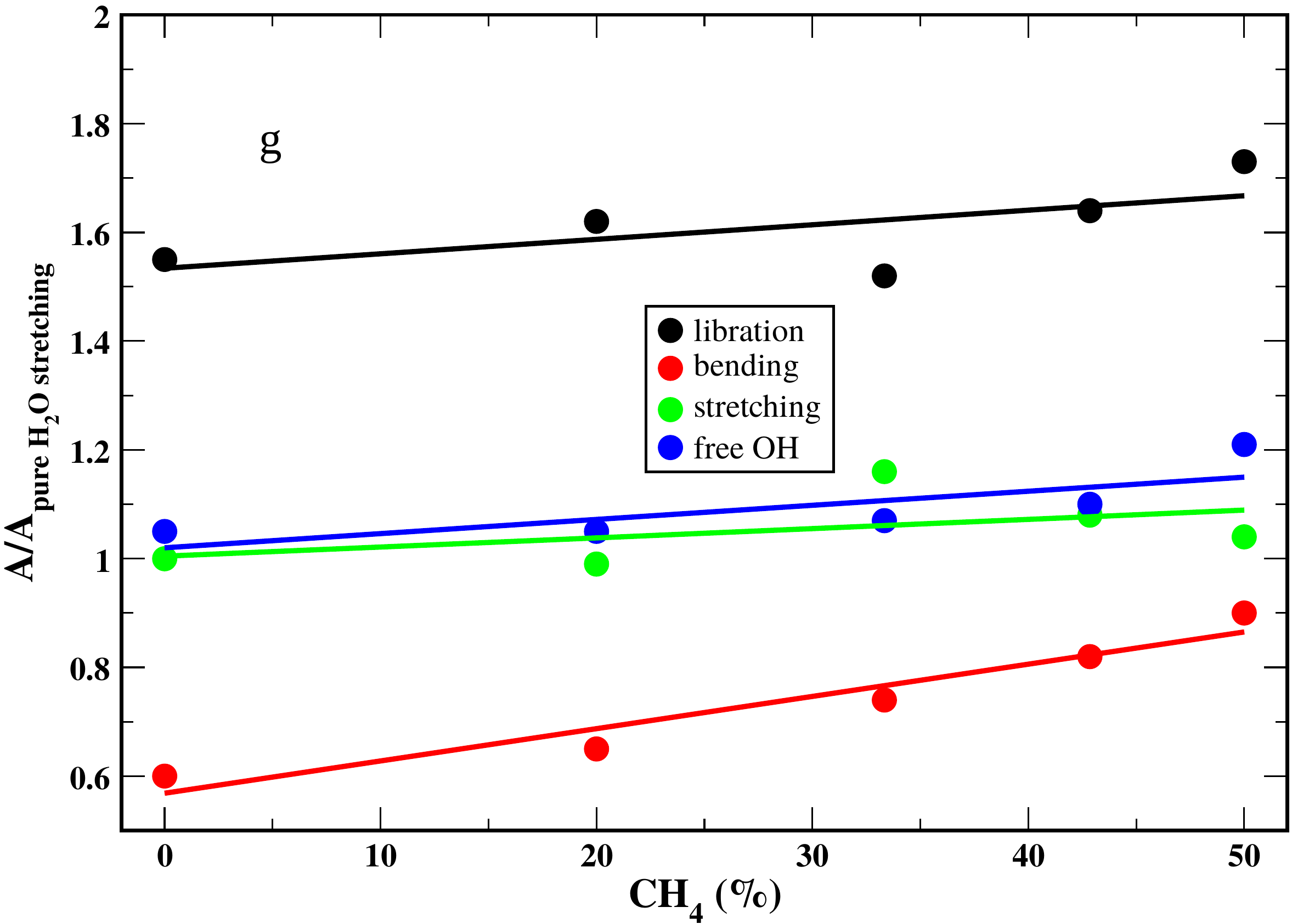}
\end{minipage}
\begin{minipage}{0.49\textwidth}
\includegraphics[width=\textwidth]{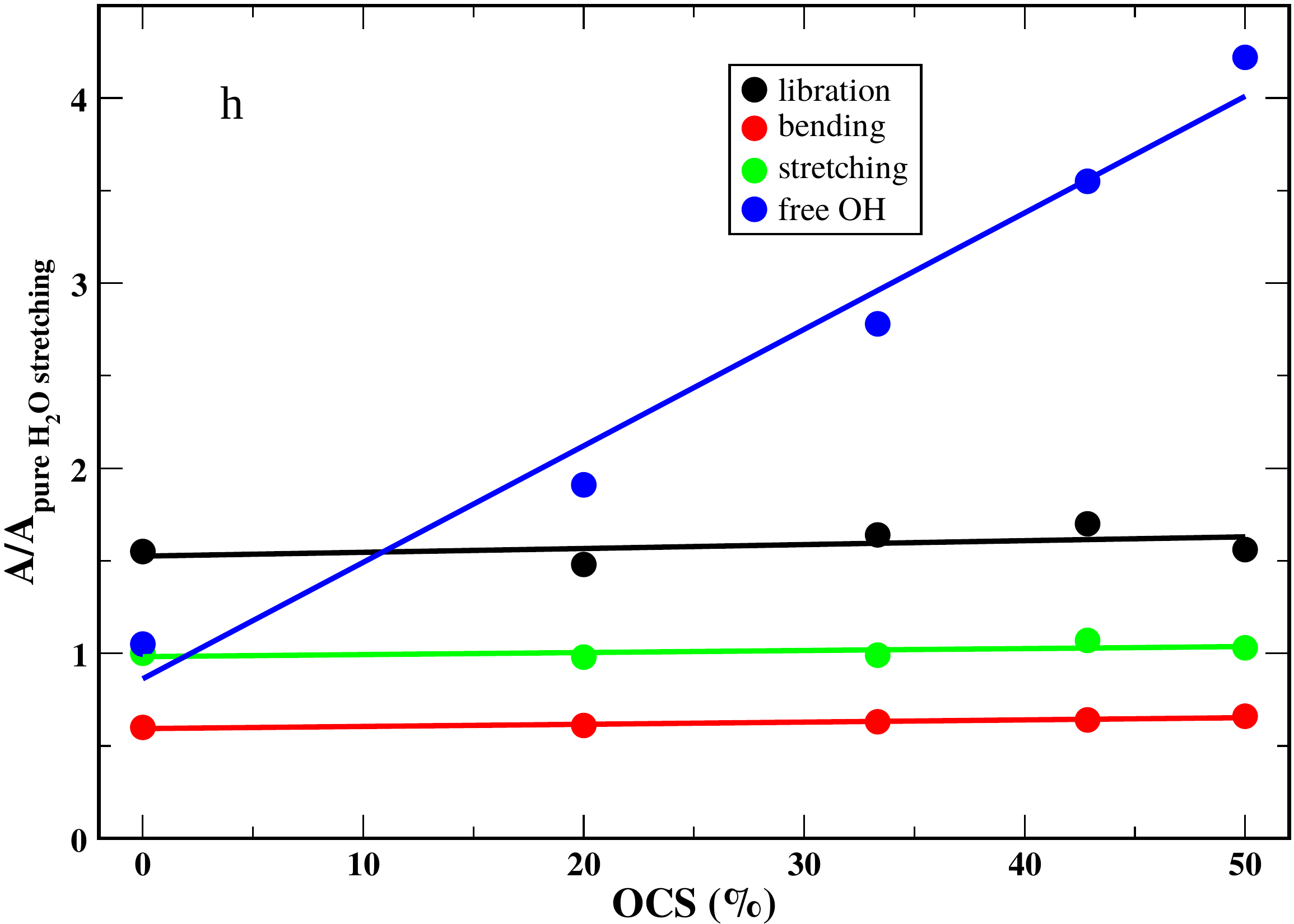}
\end{minipage}
\end{figure}

\begin{figure}
\begin{minipage}{0.49\textwidth}
\includegraphics[width=\textwidth]{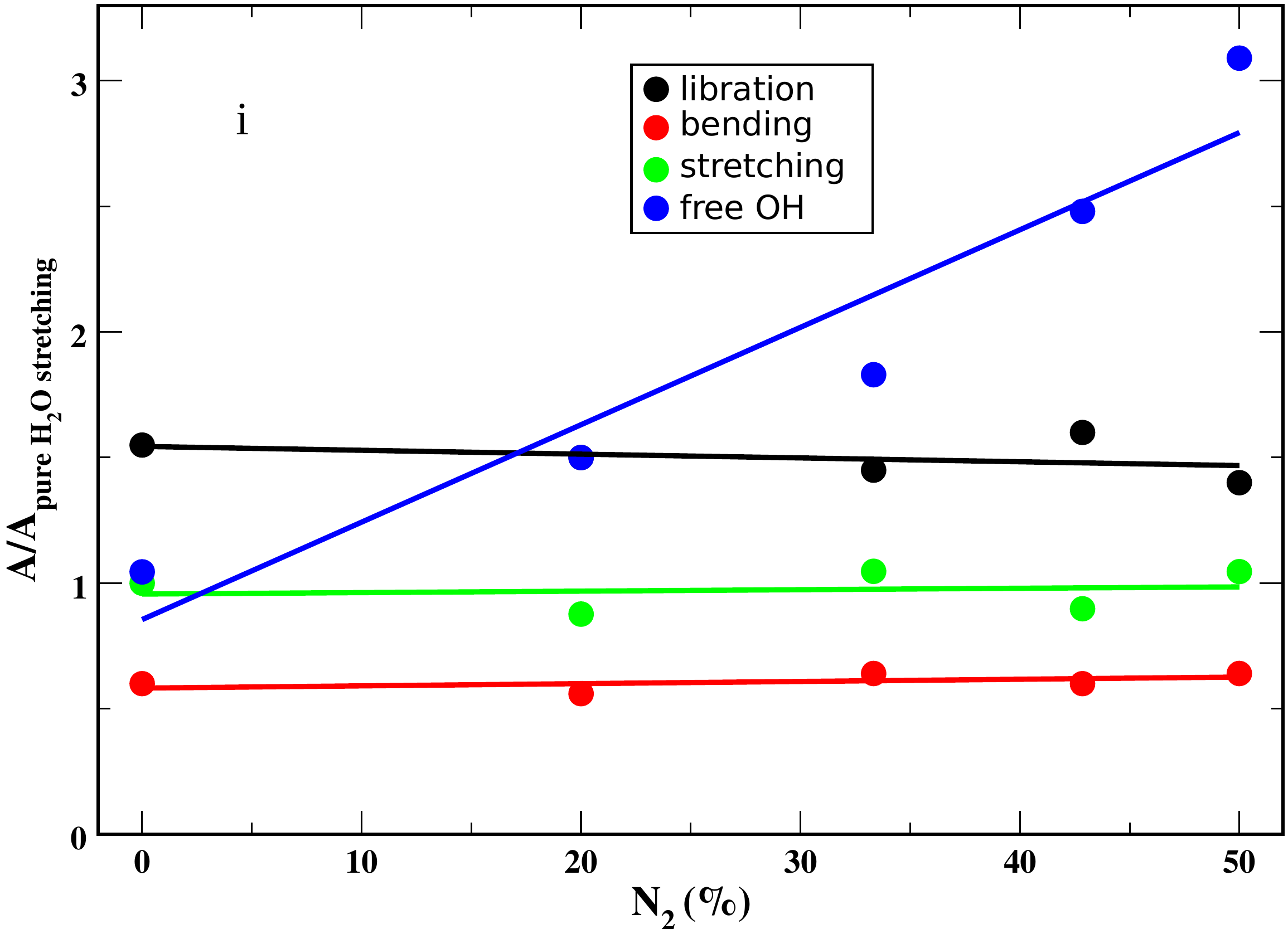}
\end{minipage}
\begin{minipage}{0.49\textwidth}
\includegraphics[width=\textwidth]{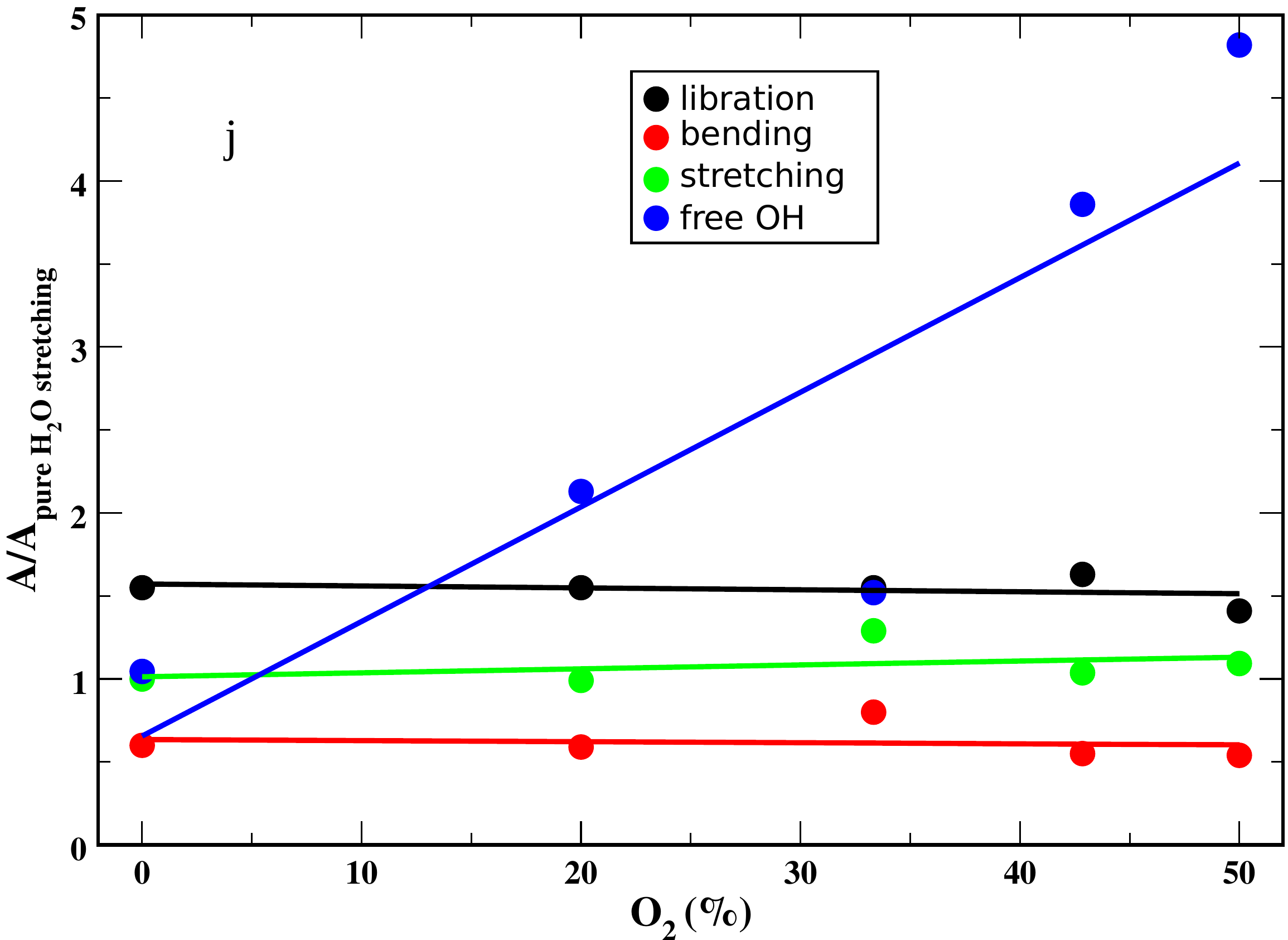}
\end{minipage}
\caption{{Band strengths of the four fundamental vibration modes of water for
(a) $\rm{H_2O-HCOOH}$, (b) $\rm{H_2O-NH_3}$, (c) $\rm{H_2O-CH_3OH}$, (d) $\rm{H_2O-CO}$, (e) $\rm{H_2O-CO_2}$, (f) $\rm{H_2O-H_2CO}$, (g) $\rm{H_2O-CH_4}$, (h) $\rm{H_2O-OCS}$, (i) $\rm{H_2O-N_2}$, and (j) $\rm{H_2O-O_2}$ clusters with various concentrations. The water c-tetramer configuration was used for pure water.}}
\label{fig:band_strength}
\end{figure}

\subsubsection{NH$_3$ ice}
\label{NH3_ice}
Most of the intense modes of ammonia overlap with the dominant features due to water and silicates. However, when ammonia is 
mixed with $\rm{H_2O}$ ice, it forms hydrates that show an intense mode at $2881.84$ cm$^{-1}$ ($3.47$ $\mu$m) \cite{dart05}, which 
lies in a relative clear region. Another characteristic feature of ammonia is the umbrella mode at $1111.11$ cm$^{-1}$ ($9.00$ $\mu$m), which is relatively intense, but it often overlaps with the $\rm{CH_3}$ rocking mode of methanol, thus leading to an 
overestimation of the abundance of ammonia.

In this work, infrared spectra were recorded for various mixing ratios of H$_2$O-NH$_3$ ice
deposited at 20 K. The IR spectra, normalized with respect to the most intense bend (i.e. the O-H stretching mode), are shown in Figure \ref{fig:experiment}b. Mixing ratios were derived by measuring the areas of the selected bands for H$_2$O band (at 2220 cm$^{-1}$) \cite{mast09} and for NH$_3$ (umbrella mode band at 1070 cm$^{-1}$) \cite{dhen86}, with a procedure analogous to that introduced in the previous section for HCOOH.

Figure \ref{fig:optimized_structure}c shows the optimized geometry of the H$_2$O-$\rm{NH_3}$ system with a $4:4$ ratio as obtained from our 
quantum-chemical calculations. In the Appendix, Figure \ref{fig:H2O-NH3} depicts the absorption band profiles of H$_2$O-$\rm{NH_3}$ mixtures 
with various concentrations. The transition frequencies and the corresponding intensity values are provided in the Appendix as well
(see Table \ref{tab:H2O_X}). The vibrational analysis has also been carried out at a  higher level of theory, thereby using the B2PLYP functional.
The results are reported in Tables \ref{tab:4H2O_1NH3_B2PLYP}, \ref{tab:4H2O_2NH3_B2PLYP}, \ref{tab:4H2O_3NH3_B2PLYP}, and 
\ref{tab:4H2O_4NH3_B2PLYP} in the SI. Figure \ref{fig:band_strength}b shows the band strengths as a function of the 
concentration of the impurity under consideration, i.e. NH$_3$.

{
\subsubsection{Comparison between experiment and simulations}
\label{comparison}
\begin{figure}
\centering
\includegraphics[width=\textwidth]{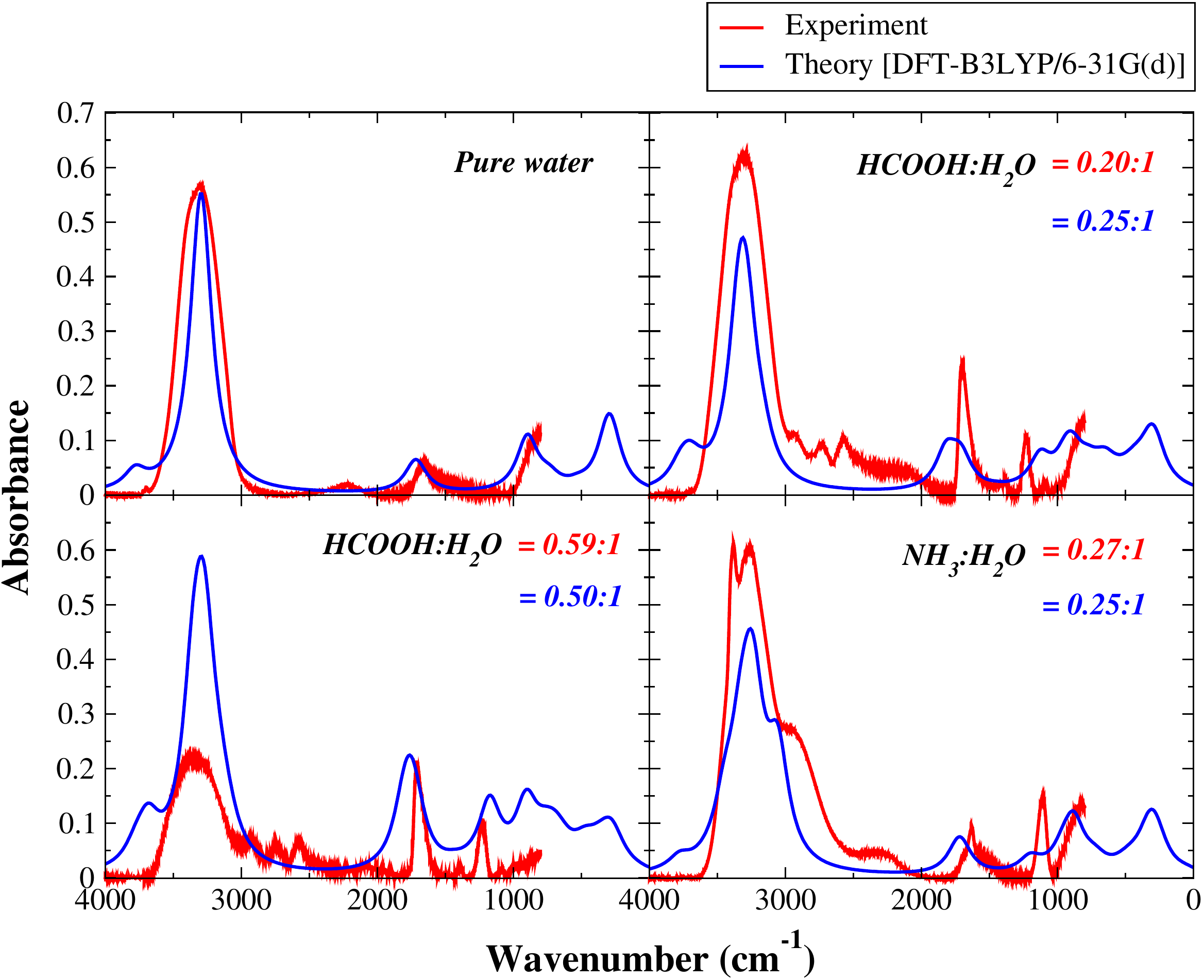}
\caption{Comparison of computed and experimental IR spectra (0 - 4000 cm$^{-1}$) for pure water as well as water with HCOOH and NH$_3$ as impurities. We have used harmonic frequencies for the computed spectra and the intensity is scaled with a factor 1000 to have best match with the experimental one.}
\label{fig:comparison_IR}
\end{figure}

In Figure \ref{fig:comparison_IR}, the comparison between experimentally obtained spectra and our computed spectra for pure water, H$_2$O-HCOOH mixture, and H$_2$CO-NH$_3$ mixture is shown. We note a good agreement between experimental and theoretical absorption spectra. Figure \ref{fig:HCOOH-NH3_band_strength} shows the comparison between the experimental (dotted lines) and theoretical (solid and dashed lines) band strengths of the four water bands as a function of the concentration of HCOOH and NH$_3$. From Figure \ref{fig:HCOOH-NH3_band_strength}, it is evident that the experimental strength of the libration and bending modes increases by increasing the concentration of HCOOH. On the contrary, the strength of the stretching and free OH modes shows a decreasing trend. These behaviors should be compared with the B2PLYP/mug-cc-pVTZ (dashed) and B3PLYP/6-31G(d) (solid) trends. For the libration and bending band strength profiles, there is a qualitative agreement with experiments. In the case of the stretching and free OH
modes, theoretical band strength profiles deviate from experimental work. The lack of experimental data in the 3600-4000 cm$^{-1}$ range (see Figure \ref{fig:experiment}a) may have contributed to this disagreement. Concerning the comparison of the two levels of theory, it is noted that there is a rather good agreement. In case of H$_2$O-HCOOH mixture, HCOOH can act as both hydrogen bond donor and hydrogen bond acceptor. We considered both the interactions and noted that, if we consider HCOOH as H-bond acceptor, the band strength of three modes (libration, bending, and stretching) are lower with respect to case where HCOOH was treated as H-bond donor. But in the case of the free-OH mode, the band strength slope increases (See {  Figure \ref{fig:hcooh-donor-acceptor}} in the Appendix).

\begin{figure}
\centering
\begin{minipage}{0.49\textwidth}
\vskip 0.8cm
\includegraphics[width=\textwidth]{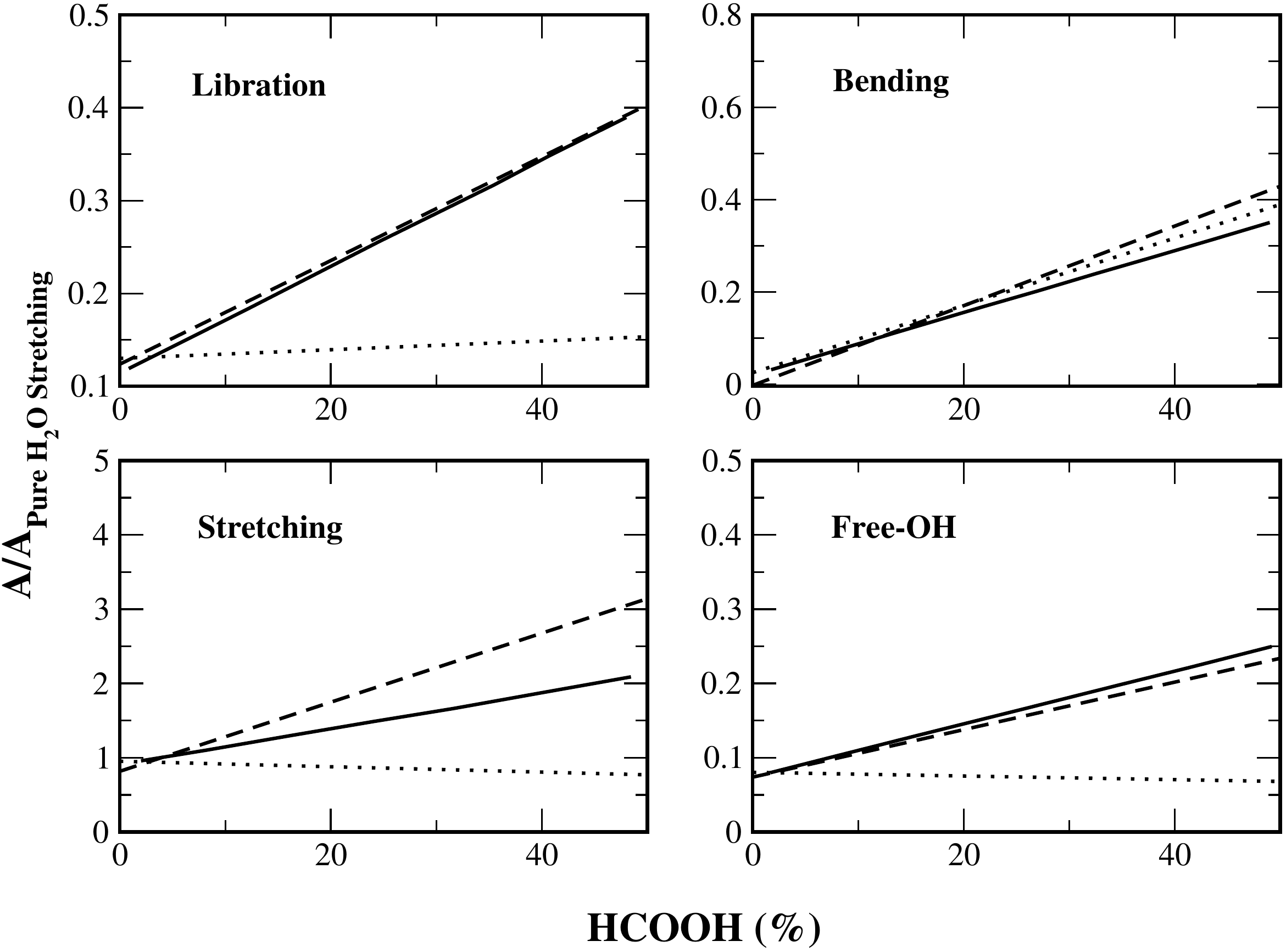}
\end{minipage}
\begin{minipage}{0.49\textwidth}
\includegraphics[width=\textwidth]{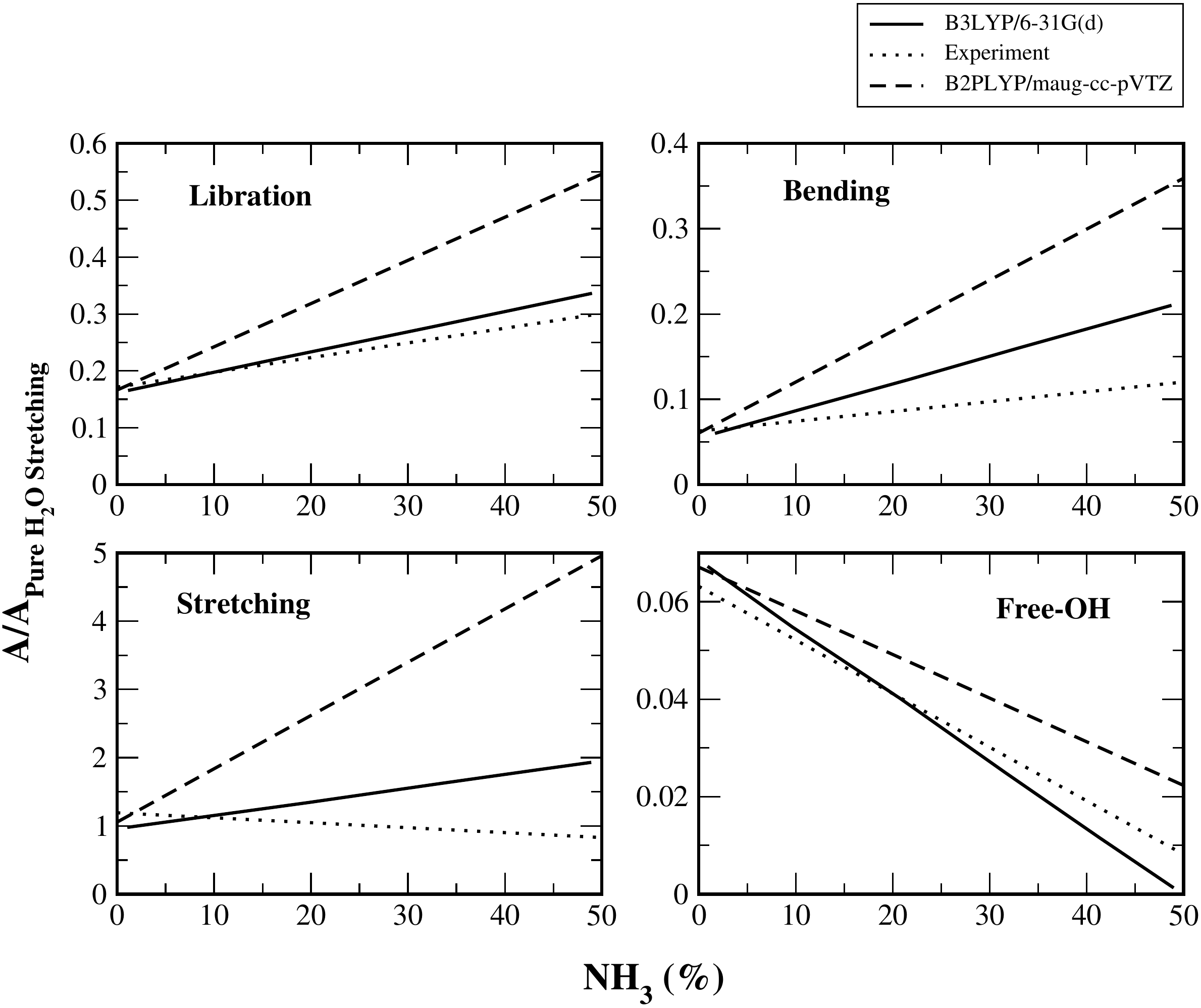}
\end{minipage}
\caption{Comparison between the calculated and experimental band strength profiles with various concentration of HCOOH and NH$_3$.}
\label{fig:HCOOH-NH3_band_strength}
\end{figure}

Moving to ammonia, the experimental data of Figure \ref{fig:HCOOH-NH3_band_strength} show that the band strength of the free-OH stretching mode nearly vanishes when a 50\% concentration of the impurity (NH$_3$) is reached. This feature interestingly supports our calculated spectra shown in the Appendix (see {  Figure \ref{fig:H2O-NH3}} last panel). Libration and bending modes have, instead, an opposite trend, with the band strength increasing by increasing the concentration of NH$_3$. The band strength of the stretching mode shows a slightly decreasing trend with the concentration of NH$_3$. From the inspection of Figure \ref{fig:HCOOH-NH3_band_strength} it is evident that both sets of theoretical results (B3LYP and B2PLYP) are in reasonably good agreement with experimental data for the libration, bending, and free-OH modes. Interestingly, the results obtained using the lower level of theory are in better agreement with experiments. In {  Figure \ref{fig:harm-anharm-compare}} (in the Appendix), the comparison of band strengths evaluated using (a) harmonic and (b) anharmonic calculations is shown. To investigate the effect of anharmonicity on the band strengths, we have only considered fundamental bands in the 0 to 3600 cm$^{-1}$ frequency range. From our experimental study on the H$_2$O-NH$_3$ system, as already mentioned, we obtained an increasing trend of the band strength for the libration, bending, and stretching modes with the increase in concentration of NH$_3$, whereas the band strength decreases for the free OH mode and tends to zero with 50\% concentration of NH$_3$. When using harmonic calculations for all four fundamental modes, trends similar to what obtained from experiment were found. But, if we consider anharmonic calculations, only the behavior of the stretching mode is well reproduced. All other modes deviate from the experimental results. While not claiming that harmonic calculations are better than the anharmonic ones, this comparison seems to suggest that the former show a better error compensation. A similar outcome has been obtained for the H$_2$O-CO system and will be briefly addressed later in the text.

Based on the comparisons discussed above, the B3LYP/6-31G(d) level of theory provides reliable results. Therefore, it has been employed in the following investigations. First of all, the comparison between computed and experimental band strengths for the H$_2$O-CH$_3$OH, CO-H$_2$O, and CO$_2$-H$_2$O mixtures will be considered to further support its suitability.}

\subsubsection{CH$_3$OH ice}
\label{CH3OH_ice}
In this work, the effect of the CH$_3$OH concentration on the band profiles of water ice has been experimentally investigated. In the case of methanol, CO$_2$ gas is still present in the system (i.e., outside the vacuum chamber) in quantities that vary in time causing negative and/or positive contributions to CO$_2$ gas-phase absorption features with respect to the background spectrum, as evident in Figure \ref{fig:experiment}c at $\sim$2340 cm$^{-1}$. Such contamination is most likely due to the dosing line, but its negligible amount should not affect the final results. Figure \ref{fig:experiment}c shows the experimental absorption spectra for various CH$_3$OH-H$_2$O ice mixtures deposited at T$=30$ K. The spectra are normalized to 1 with respect to the maximum of the O-H stretch band.

 \begin{figure}
 \centering
 \includegraphics[width=\textwidth]{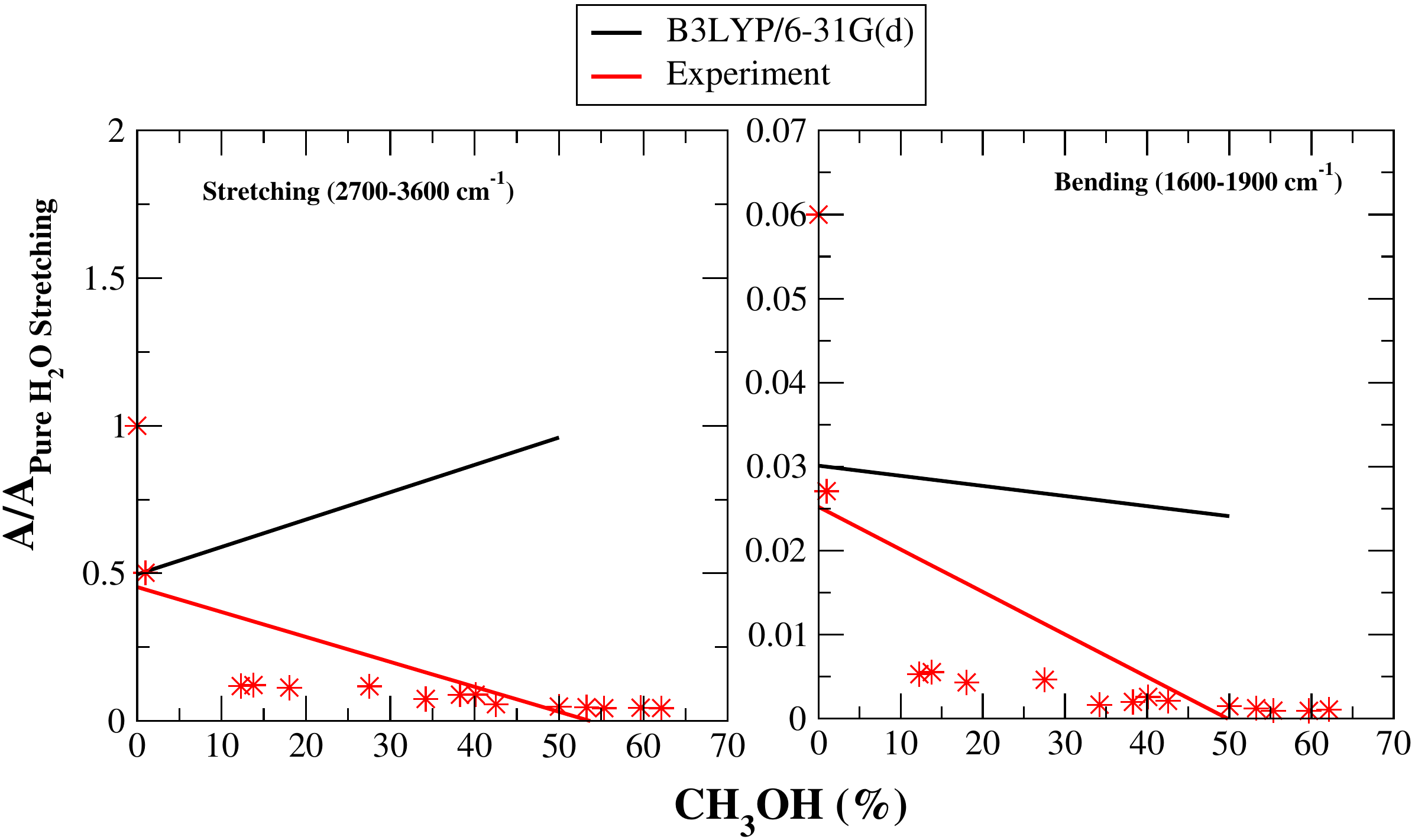}
 \caption {Comparison between calculated and experimental band strength profiles as a function of CH$_3$OH concentration. {Stars represent the experimental data points.}}
 \label{fig:CH3OH-BS-COMP}
 \end{figure}

{
Figure \ref{fig:optimized_structure}d shows the optimized structure of the $\rm{H_2O}$-$\rm{CH_3OH}$ mixture with a $4:4$ concentration ratio. It is noted that a weak hydrogen bond is expected to be formed. The simulated IR spectra for different concentrations are shown in {  Figure \ref{fig:H2O-CH3OH}} (in the Appendix). Peak positions, integral absorption coefficients, and band assignments for various H$_2$O-CH$_3$OH mixtures are collected in {  Table \ref{tab:H2O_X}} in the Appendix. The computed band strengths as a function of different concentrations are shown in Figure \ref{fig:band_strength}c.
The computed strength of the bending mode gradually increases with $\rm{CH_3OH}$ concentration (see Figure \ref{fig:CH3OH-BS-COMP}; right panel), which is in qualitative agreement with the experimental results \cite{dawe16}. In case of the stretching mode, computationally, a slight increasing trend of the band strength is noted, whereas experimental results show an opposite trend (see Figure \ref{fig:CH3OH-BS-COMP}; left panel). Because of the lack of experimental spectra, we cannot compare the band strength of the libration and free OH modes. In case of H$_2$O-CH$_3$OH mixtures, methanol can act as both hydrogen bond donor and hydrogen bond acceptor. We considered both possibilities and found that if we consider methanol as hydrogen bond donor, the band strength of all four modes show an increasing trend. On the other hand, if we consider methanol as hydrogen bond acceptor, the band strengths of three modes, namely libration, bending, and stretching, present trends similar to the previous case (where methanol acts as hydrogen bond donor), while the free-OH band shows a less pronounced behavior (see {  Figure \ref{fig:ch3oh-donor-acceptor}} in the Appendix).

\subsubsection{CO ice}
\label{CO_ice}
Figure \ref{fig:optimized_structure}e depicts the H$_2$O-CO optimized structure with a $4:4$ concentration ratio: the four CO molecules interact with the H atoms of the water molecules not involved in the hydrogen bond (interaction of the O atom of CO with the hydrogen atom of water). However, for the H$_2$O-CO system, the interaction can take place through both O and C of CO with the hydrogen atom of H$_2$O \citep{zami18}. We have considered both types of interaction and evaluated their effects on the band strengths. However we did not find any significant difference. Thus, we only discuss the band strength of the H$_2$O-CO mixture with the interaction on the O side of CO. 
{For the sake of completeness, it should be mentioned that there is also another type of interaction, which occurs between the $\pi$ bond of CO and one water-hydrogen, and it gives rise to a T shaped complex \citep{coll14}. However, according to a computational study by Collings et al. \citep{coll14}, this has a negligible effect on IR vibrational bands. As a consequence we have not investigated in detail this kind of complex.} The simulated IR absorption spectra of the four fundamental vibrational modes for various compositions are shown in the Appendix 
(see {Figure \ref{fig:H2O-CO}}). The four fundamental frequencies of water ice change significantly by increasing the concentration of CO. The most intense peak positions and the corresponding integral abundance coefficients for different H$_2$O-CO mixtures are provided in the Appendix 
(see {Table \ref{tab:H2O_X}}). In Figure \ref{fig:band_strength}d, the integrated intensities of water vibrational modes are plotted as a function of the CO concentration. It is noted that the strength of the libration, bending, and stretching modes decreases with the concentration of CO. The free OH mode shows instead a sharp increase of the band strength when increasing the CO concentration. In Table \ref{tab:linear_coeff}, the resulting linear fit coefficients are collected together with the available experimental values for H$_2$O-CO mixtures deposited at $15$ K \cite{bouw07}. It is noted that theoretical band strength slopes are in rather good agreement with experimental results \cite{bouw07}. For the H$_2$O-CO system, anharmonic calculations have also been carried out. While the band strengths of the bending and stretching modes have a similar trend as experimental data, a deviation is noted for the libration mode 
(see {Figure \ref{fig:harm-anharm-compare}} in the Appendix).

To check the effect of dispersion, B3LYP-D3/6-31G(d) calculations have been performed, with D3 denoting the correction for dispersion effects \citep{grim10}. B3LYP-D3 calculations have been carried out for H$_2$O-CO, H$_2$O-CH$_4$, H$_2$O-N$_2$, and H$_2$O-O$_2$ systems. In Figure  \ref{fig:comp-norm-dis}a, we have shown the comparison of the band strengths of different vibrational modes of water with and without the dispersion correction for the H$_2$O-CO system. The overall conclusion is that there is a good agreement with the experimental band strengths when dispersion effect is not considered. On the contrary, when the dispersion correction is included, our computed band strength profile shows a different trend. The libration and bending modes present a positive slope with the increase in impurity concentration, whereas
experimental results show a negative slope. For the free OH mode, a slight increasing trend of the band strength is obtained, whereas the experimental band strength presents a sharp increase with the concentration of CO. The band strength of the stretching mode has a similar behaviour with dispersion and without dispersion, and in agreement with the experimental result \citep{bouw07}(see Figure 3). Thus, in summary, while we are not claiming that the dispersion effects are not important for the systems investigated, we have noted that neglecting them we obtain a consistent description of the experimental behaviour (probably due to a fortuitous errors compensation).

\begin{figure}
\includegraphics[width=8.1cm]{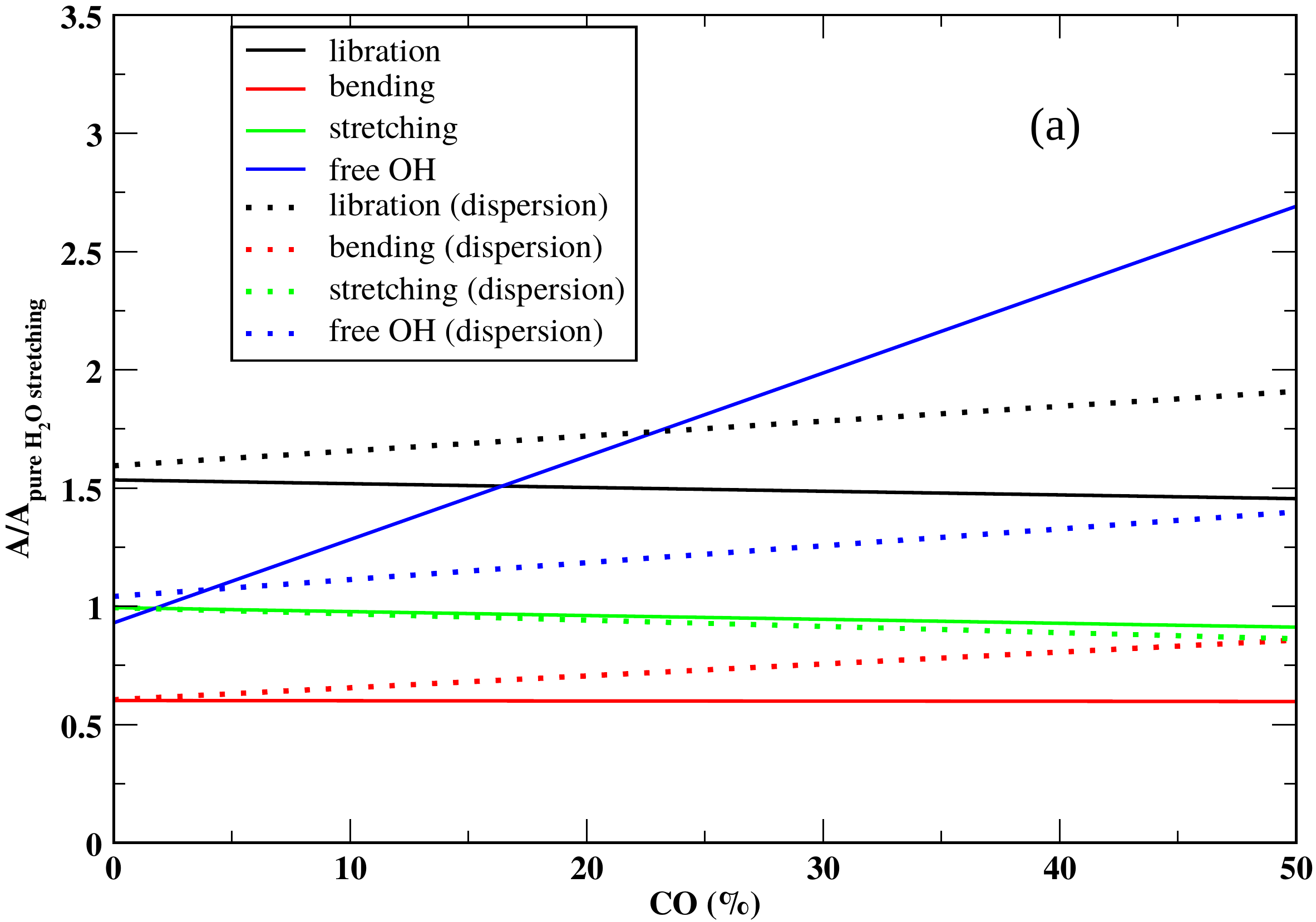}
\includegraphics[width=8.1cm]{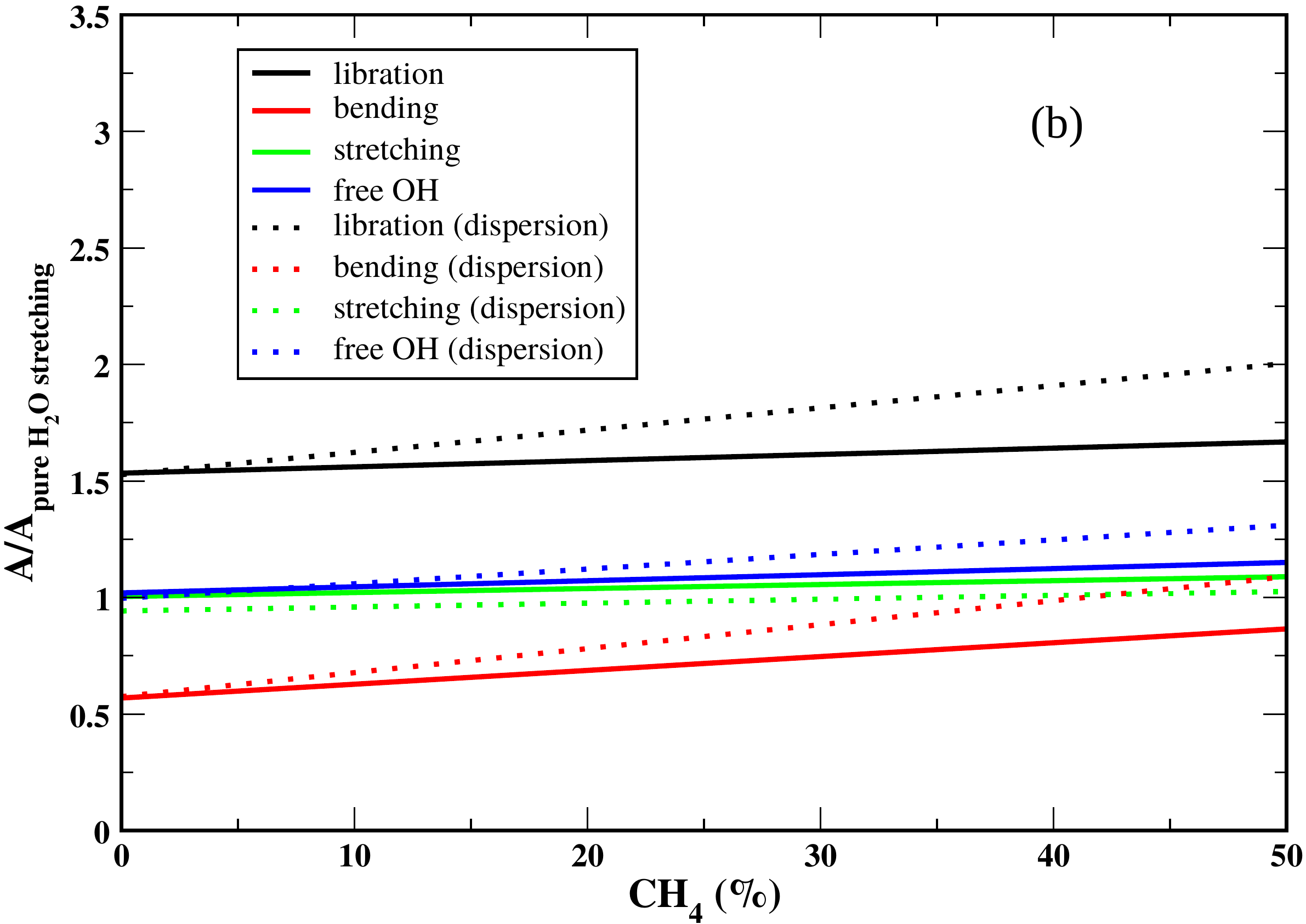}
\includegraphics[width=8.1cm]{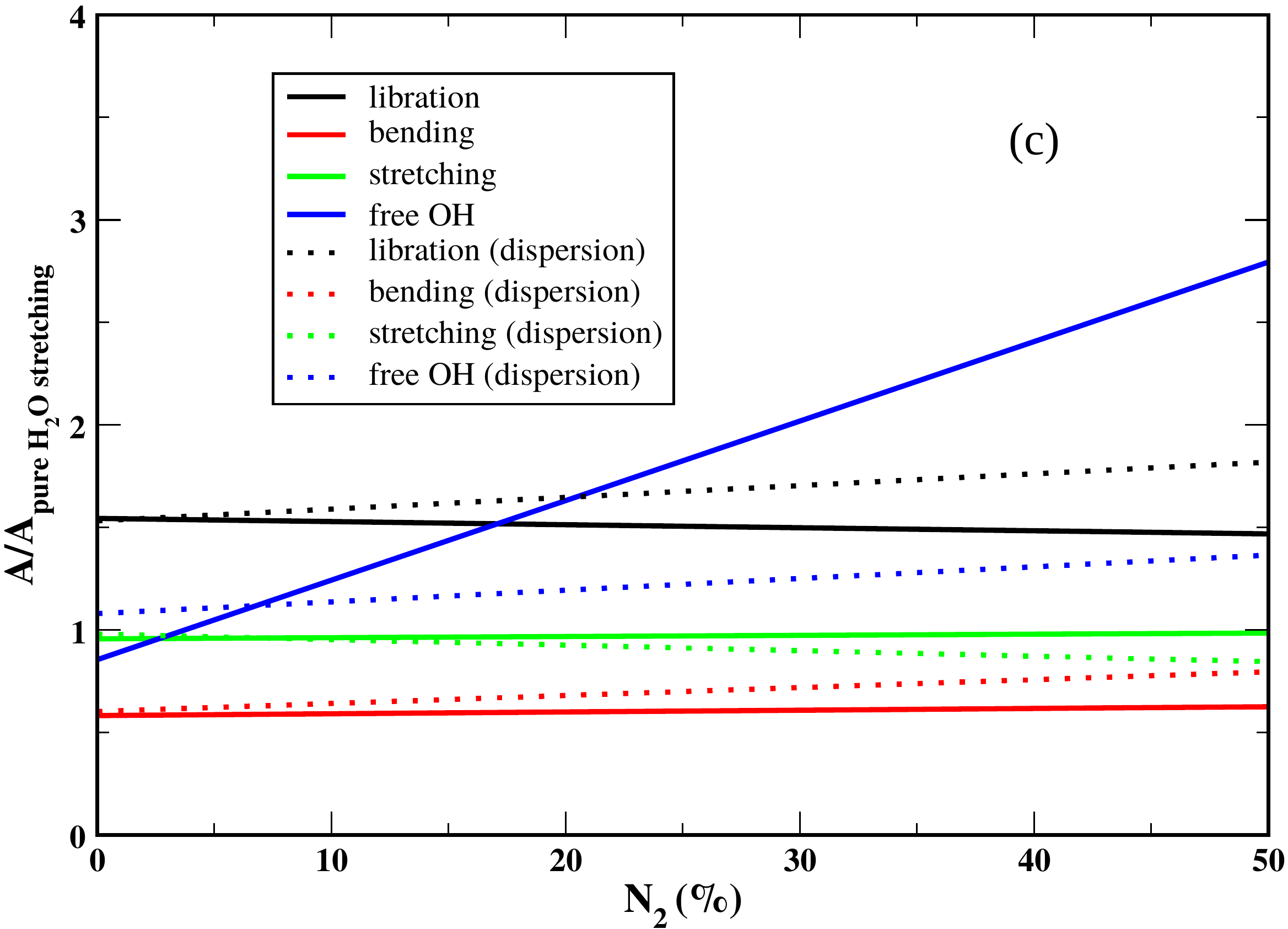}
\includegraphics[width=8.1cm]{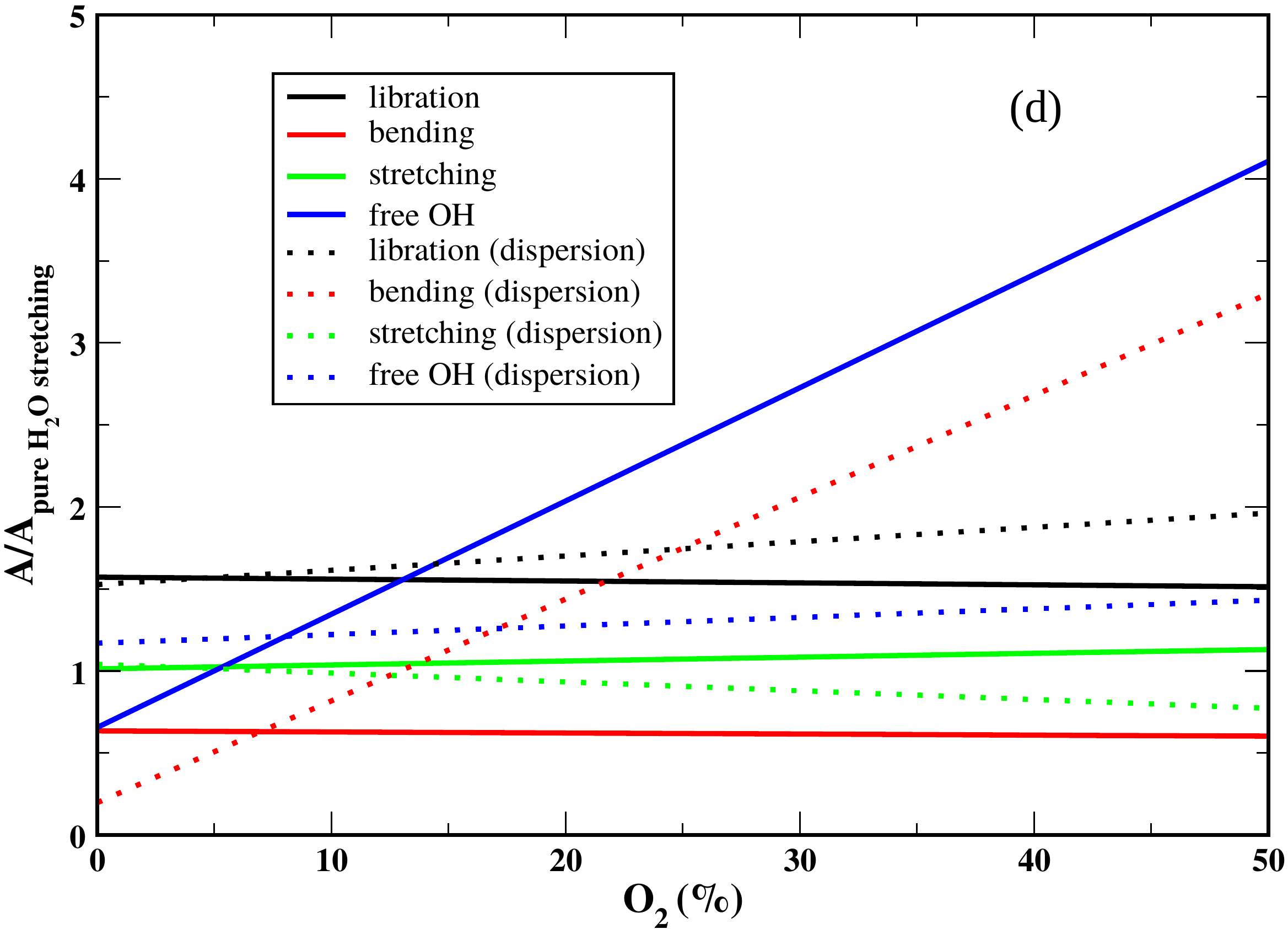}

\caption{ {Comparison of the band strengths of the four fundamental modes of water for various mixtures of (a) H$_2$O-CO, (b) H$_2$O-CH$_4$, (c) H$_2$O-N$_2$,
and (d) H$_2$O-O$_2$ by considering or not the dispersion effect.}}
\label{fig:comp-norm-dis}
\end{figure}

\begin{table}
\centering
\scriptsize{
\caption{ Linear fit coefficients for the $\rm{H_2O-X}$ (X =  HCOOH, $\rm{NH_3}$, $\rm{CH_3OH}$, CO, $\rm {CO_2}$, $\rm{H_2CO}$, $\rm{CH_4}$, OCS,
$\rm{N_2}$, and $\rm{O_2}$) mixtures.}
\label{tab:linear_coeff}
\begin{tabular}{|c|c|c|c|}
 \hline
 \hline
{\bf {Mixture}} & {\bf { Vibrational}} & \multicolumn{2}{c|}{\bf \underline{Linear coefficients}} \\
& {\bf { mode}} & {\bf Constant} & {\bf Slope} \\
 & & [$\rm 10^{-16}$ cm $\rm molecule^{-1}$] & [$\rm 10^{-19}$ cm $\rm molecule^{-1}$]\\

\hline
H$_2$O-$\rm{HCOOH}$&$\mathrm{\nu_{libration}}$&2.45 { (0.26)$^a$}&132.73 { (0.90)$^a$}\\
&$\mathrm{\nu_{bending}}$&0.58 { (0.05)$^a$}&184.25 { (14.40)$^a$}\\
&$\mathrm{\nu_{stretching}}$&1.80 { (1.90)$^a$} &48.60 { (-7.30)$^a$}\\
&$\mathrm{\nu_{free-OH}}$&0.20 { (0.16)$^a$} &9.80 { (-0.40)$^a$} \\

\hline
H$_2$O-$\rm{NH_3}$&$\mathrm{\nu_{libration}}$&0.27 { (0.34)$^a$}&6.11 { (5.00)$^a$}\\
&$\mathrm{\nu_{bending}}$&0.09 { (0.12)$^a$}&5.48 { (2.20)$^a$}\\
&$\mathrm{\nu_{stretching}}$&1.90 {(2.38)$^a$}&0.41 { (-14.4)$^a$}\\
&$\mathrm{\nu_{free-OH}}$&0.21 { (0.12)$^a$}&-4.21 { (-2.1)$^a$}\\
\hline
H$_2$O-$\rm{CH_3OH}$&$\mathrm{\nu_{libration}}$&0.25&10.0\\
&$\mathrm{\nu_{bending}}$&0.12&2.00\\
&$\mathrm{\nu_{stretching}}$&1.92&32.00\\
&$\mathrm{\nu_{free-OH}}$&0.26&2.65\\
\hline
H$_2$O-CO&$\mathrm{\nu_{libration}}$&0.30 (0.30$\pm$0.02)$ \cite{bouw07}$&-0.32 (-2.1$\pm$0.4)\cite{bouw07}\\
&$\mathrm{\nu_{bending}}$&0.12 (0.13$\pm$0.02)\cite{bouw07}&-0.016 (-1.0 $\pm$0.3)\cite{bouw07}\\
&$\mathrm{\nu_{stretching}}$&1.98 (2.0$\pm$0.1)\cite{bouw07}&-3.2 (-16$\pm$3)\cite{bouw07}\\
&$\mathrm{\nu_{free-OH}}$&0.18 (0.0)\cite{bouw07}&5.69 (1.2 $\pm$0.1)\cite{bouw07}\\
\hline
H$_2$O-$\rm{CO_2}$&$\mathrm{\nu_{libration}}$&0.3 (0.32$\pm$0.02)\cite{ober07}&2.07 (-3.2$\pm$0.4)\cite{ober07}\\
&$\mathrm{\nu_{bending}}$&0.11 (0.14$\pm$0.01)\cite{ober07}&0.12 (-0.5$\pm$0.2)\cite{ober07}\\
&$\mathrm{\nu_{stretching}}$&2.02 (2.1$\pm$0.1)\cite{ober07} &-0.22 (-22$\pm$2)\cite{ober07}\\
&$\mathrm{\nu_{free-OH}}$&0.19 (0.0)\cite{ober07}&10.02 (1.62$\pm$0.07)\cite{ober07}\\
\hline
H$_2$O-$\rm{H_2CO}$&$\mathrm{\nu_{libration}}$&0.26&5.73\\
&$\mathrm{\nu_{bending}}$&0.10&4.59\\
&$\mathrm{\nu_{stretching}}$&1.92&0.10\\
&$\mathrm{\nu_{free-OH}}$&0.13&16.53\\
\hline
H$_2$O-$\rm{CH_4}$&$\mathrm{\nu_{libration}}$&0.31&0.53\\
&$\mathrm{\nu_{bending}}$&0.11&1.18\\
&$\mathrm{\nu_{stretching}}$&2.01&3.39\\
&$\mathrm{\nu_{free-OH}}$&0.20&0.52\\
\hline
H$_2$O-$\rm{OCS}$&$\mathrm{\nu_{libration}}$&0.30&0.42\\
&$\mathrm{\nu_{bending}}$&0.11&0.23\\
&$\mathrm{\nu_{stretching}}$&1.96&2.18\\
&$\mathrm{\nu_{free-OH}}$&0.17&0.13\\
\hline
H$_2$O-$\rm{N_2}$&$\mathrm{\nu_{libration}}$&0.31&-0.30\\
&$\mathrm{\nu_{bending}}$&0.12&0.17\\
&$\mathrm{\nu_{stretching}}$&0.12&0.11\\
&$\mathrm{\nu_{free-OH}}$&0.17&7.75\\
\hline
H$_2$O-$\rm{O_2}$&$\mathrm{\nu_{libration}}$&0.31&-0.23\\
&$\mathrm{\nu_{bending}}$&0.12&-0.13\\
&$\mathrm{\nu_{stretching}}$&2.02&4.71\\
&$\mathrm{\nu_{free-OH}}$&0.13&13.80\\
\hline
\hline
\end{tabular}}
\vskip 0.2cm
{\bf Notes.} {Experimental values are provided in the parentheses. $^a$This work.}
\end{table}

\subsubsection{CO$_2$ ice}
\label{CO2_ice}
Figure \ref{fig:optimized_structure}f shows the optimized geometry of the $4:4$ mixture of H$_2$O:$\rm{CO_2}$. The absorption features of water ice for different $\rm{CO_2}$ concentrations are shown in the Appendix (see 
{Figure \ref{fig:H2O-CO2}}). The most intense frequencies for the various H$_2$O-CO$_2$ mixtures are summarized in the Appendix as well (see 
{Table \ref{tab:H2O_X}}). The trend of the band strength as a function of CO$_2$ concentrations is shown in Figure \ref{fig:band_strength}e. For the free-OH mode, a rapid increase with CO$_2$ concentration is noted, which is in good agreement with the experimental results by \citet{ober07}. Computed band strengths of the libration and bending modes also increase by increasing the $\rm{CO_2}$ concentration, which is however in contrast with the available experimental data \cite{ober07}. The band strength of the bulk stretching mode decreases instead with $\rm{CO_2}$ concentration, in reasonable good agreement with the available experiments \cite{ober07}. FTIR spectroscopy of the matrix-isolated molecular complex H$_2$O-CO$_2$ shows that CO$_2$ does not form a weak hydrogen bond with H$_2$O \cite{tso85}, but instead CO$_2$ destroys the bulk hydrogen bond network. This may cause a large decrease in the band strength of the bulk stretching mode, while the intermolecular O-H bond strength increases with the CO$_2$ concentration. Therefore, the disagreement between calculated and experimental band strengths could be thus due to the cluster size of water molecules. In Table \ref{tab:linear_coeff}, the resulting linear fit coefficients are reported, together with the available experimental values for the H$_2$O-CO$_2$ mixture deposited at $15$ K \cite{ober07}.

\subsection{Part 2. Applications}

The results discussed in previous sections suggest that the water c-tetramer structure together with harmonic B3LYP/6-31G(d) calculations 
are able to predict the experimental results presented here as well as literature data. Thus, to study the effect of other impurities 
($\rm{H_2CO}$, CH$_4$, OCS, N$_2$, and O$_2$) on pure water ice, we have further exploited this methodology. Additionally, the effect 
of impurities on the band strengths of the four fundamental bands has also been studied by considering the c-hexamer (chair) structure and the 
corresponding results are provided in the SI 
(see Figure \ref{fig:6h2o_x_band_strength}).}

\subsubsection{H$_2$CO ice}
\label{H2CO_ice}
The strongest modes of formaldehyde ($\rm{H_2CO}$) lie at $1724.14$ cm$^{-1}$ ($5.80$ $\mu$m) and $1497.01$ cm$^{-1}$ ($6.68$ $\mu$m). Figure \ref{fig:optimized_structure}g depicts the optimized structure of the $4:4$ $\rm{H_2O-H_2CO}$ mixture. The desired ratio is attained upon formation of the hydrogen bond between the O atom of $\rm{H_2CO}$ and the dangling H atoms of $\rm{H_2O}$. The effect of formaldehyde on water IR spectrum is shown in the Appendix (see 
{Figure \ref{fig:H2O-H2CO}}). Frequencies, integral absorption coefficients, and mode assignments are reported in the Appendix as well 
(see {Table \ref{tab:H2O_X}}). The band strength profiles as a function of the concentration of H$_2$O are shown in Figure \ref{fig:band_strength}f. Similar to the methanol-water mixture, all band strengths are found to increase with the concentration of formaldehyde, the free-OH stretching mode being the most affected.

\subsubsection{CH$_4$ ice}
\label{CH4_ice}
$\rm{CH_4}$ cannot be observed by means of rotational spectroscopy since it has no permanent dipole moment. The optimized structure of the $\rm{H_2O-CH_4}$ system with a $4:4$ ratio is shown in Figure \ref{fig:optimized_structure}h. The absorption IR spectra for different $\rm{H_2O-CH_4}$ mixtures are depicted in the Appendix (see 
{Figure \ref{fig:H2O-CH4}}). Peak positions, integral absorption coefficients, and band assignments are provided in the Appendix as well (see 
{Table \ref{tab:H2O_X}}). Figure \ref{fig:band_strength}g shows the band strength variations with the concentration of $\rm{CH_4}$. All band strengths marginally increase with the $\rm{CH_4}$ concentration. Figure \ref{fig:comp-norm-dis}b shows the comparison of the band strengths with and without the incorporation of corrections for accounting for dispersion effects. For all the four fundamental modes, differences are minor.

\subsubsection{OCS ice}
\label{OCS_ice}
\citet{garo10} proposed that carbonyl sulfide (OCS) is a key ingredient of the grain surface. Its abundance in ice phase may vary between 0.05 and $0.15$\% \cite{dart05}. Figure \ref{fig:optimized_structure}i shows the optimized structure of the 4:4 H$_2$O-OCS. Since oxygen is more electronegative than sulfur, the O atom of the OCS molecule is hydrogen-bonded to the water free-hydrogens. In the Appendix, 
{Figure \ref{fig:H2O-OCS}} shows the absorption IR band spectra for H$_2$O-OCS clusters with various concentrations. Figure \ref{fig:band_strength}h depicts the band strengths as a function of the concentration of OCS. Here, the free-OH mode is the most affected and its band strength increases with the concentration of OCS. All other modes roughly remain invariant by varying the amount of impurity.

\subsubsection{N$_2$ ice}
\label{N2_ice}
N$_2$ is a stable homonuclear molecule and, due to its symmetry, it is infrared inactive. However, when
embedded in an ice matrix, the crystal field breaks the symmetry, and an infrared transition is activated
around $2325.58$ cm$^{-1}$ ($4.30$ $\mu$m). Figure \ref{fig:optimized_structure}j shows the optimized geometry of the H$_2$O-$\rm{N_2}$ system with a $4:4$ ratio. The IR absorption spectra of water ice containing different amounts of $\rm{N_2}$ are shown in the Appendix 
(see {Figure \ref{fig:H2O-N2}}).
The corresponding peak frequencies and intensities are provided in the Appendix as well (see Table \ref{tab:H2O_X}). The dependence of the band strengths on the N$_2$ concentration is depicted in Figure \ref{fig:band_strength}i. It has been found that the slope of the band strength of the libration mode decreases, whereas the bending, stretching, and free OH modes show an increasing trend with the concentration of N$_2$. The linear fitting coefficients are provided in Table \ref{tab:linear_coeff}.
Figure \ref {fig:comp-norm-dis}c shows the comparison of band strengths with and without considering
the dispersion effects. It is noted that the inclusion of dispersion effect leads to small changes.

\subsubsection{O$_2$ ice}
\label{O2_ice}
Analogously to $\rm{N_2}$, $\rm{O_2}$ is a homonuclear molecule, which is infrared inactive except when it is embedded in an ice matrix \cite{ehre92,ehre98}, thus giving rise to an absorption band around $1550.39$ cm$^{-1}$ ($6.45$ $\mu$m). $\rm{O_2}$ ice is not much abundant because the largest part of the oxygen budget in the dense molecular clouds is locked in the form of $\rm{CO_2}$, CO, water ice, and silicates.
The optimized geometry of the $4:4$ H$_2$O-$\rm{O_2}$ ratio is shown in Figure \ref{fig:optimized_structure}k. IR spectra for different concentrations (Figure \ref{fig:H2O-O2}) and the corresponding peak frequencies and intensities 
(Table \ref{tab:H2O_X}) are provided in the Appendix. The dependence of band strengths upon O$_2$ concentration is shown in Figure \ref{fig:band_strength}j.
Similarly to the $\rm{N_2}$-water case, the free-OH mode is the most affected. The slope of the band strength of the libration and bending modes decreases, whereas the stretching and free-OH modes show an increasing trend with the concentration of O$_2$. The fitting coefficients for different H$_2$O-N$_2$ mixtures are provided in Table \ref{tab:linear_coeff}.

Figure \ref{fig:comp-norm-dis}d depicts the comparison of the band strengths with and without the inclusion of dispersion effects for the H$_2$O-O$_2$ system. It is evident that trend of the band strength with the impurity concentration slightly increases for the libration mode, whereas slightly decreases for the stretching mode when corrections for dispersion effects are present. In the case of the bending mode, the band strength rapidly increases, whereas the band strength rapidly decreases for the free OH mode.

\subsubsection{Comparison between various mixtures}
\label{comparison_between_various_mixtures}
\begin{figure}
\centering
\begin{minipage}{0.9\textwidth}
\vskip 0.8cm
\includegraphics[width=\textwidth]{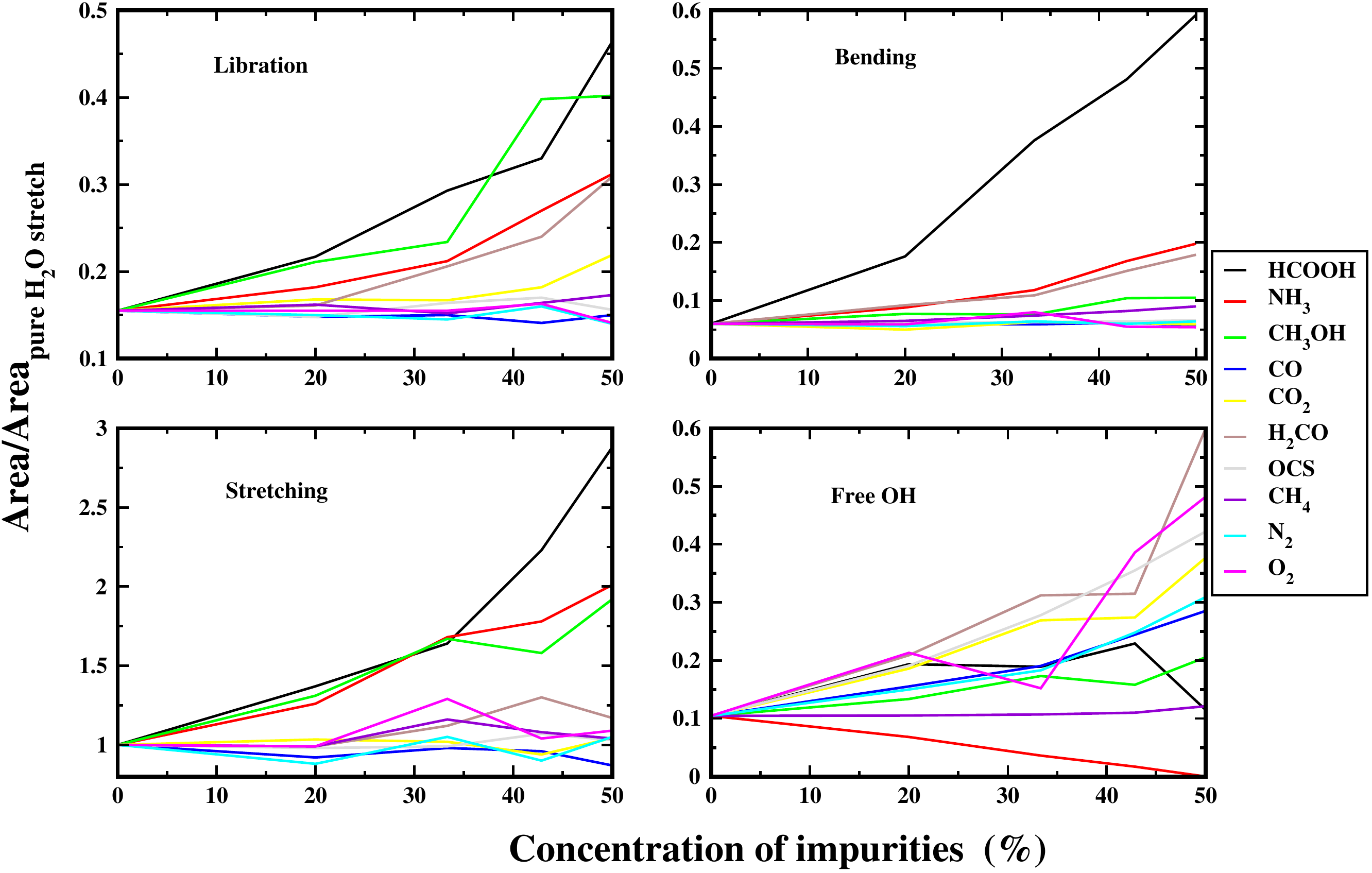}
\end{minipage}
\begin{minipage}{0.85\textwidth}
\vskip 0.8cm
\includegraphics[width=\textwidth]{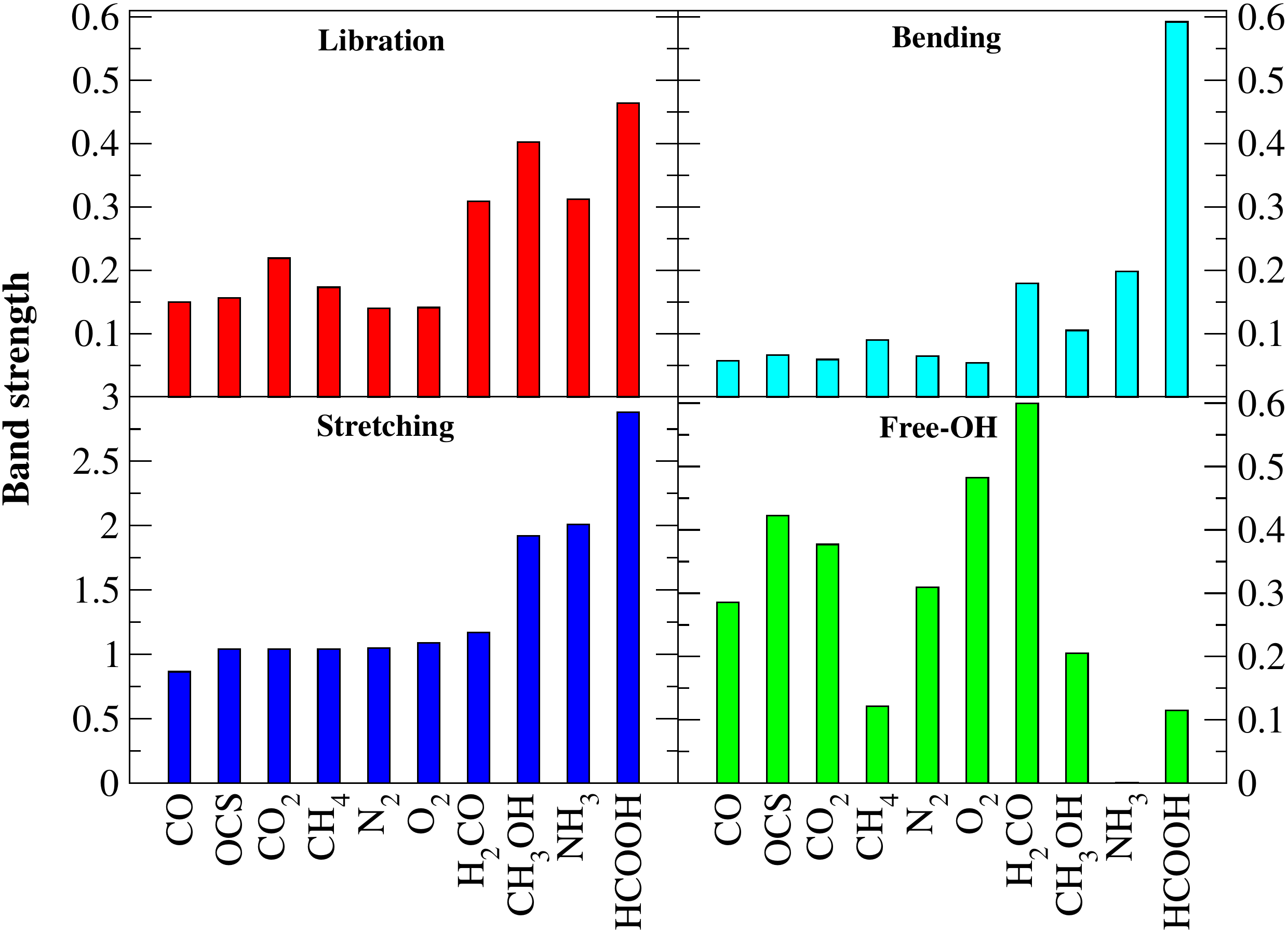}
\end{minipage}
\caption{Top panel: Effect of impurities on the four fundamental vibrational modes of water. Bottom panel: Comparison of the band strengths for
the four fundamental vibrational modes as affected by impurities.}
\label{fig:comparison_between_four_modes}
\end{figure}

To compare the effect of all impurities considered in this study on the band strength, we have plotted the band profiles of the four fundamental modes of water ice as a function
of the concentration of impurities, the results being shown in Figure \ref{fig:comparison_between_four_modes}, top panel. For all fundamental modes, band strengths increase with the concentration of $\rm{CH_3OH}$, $\rm{H_2CO}$, HCOOH, $\rm{CH_4}$. To better understand their effect, in Figure \ref{fig:comparison_between_four_modes}, bottom panel, we report the relative band strengths for the $4:4$ ratio mixtures. From this, it is clear that the libration, bending, and stretching modes are mostly affected by formic acid, while the free-OH mode is mostly affected by formaldehyde. An interesting feature is found for the free-OH mode for the $\rm{NH_3-H_2O}$ system. By increasing the $\rm{NH_3}$ concentration
with respect to pure water, the band strength of the free-OH mode decreases and disappears when the 4 : 4 concentration ratio is reached.

Figure \ref{fig:optimized_structure_6H2O} (in the SI) depicts the optimized structures of the pure water c-hexamer (chair) configuration along with those obtained for a 6:1 concentration ratio. 
{Figure \ref{fig:6h2o_x_band_strength}} (in the SI) collects the
results for the band strength variations for the c-hexamer (chair) water cluster configuration is considered, this being analogous to Figure 
\ref{fig:band_strength}.  The geometries of water clusters containing $20$ water molecules with HCOOH as an impurity in various concentrations are 
shown in Figure \ref{fig:20H2O-HCOOH} (in the SI) and the corresponding variations of the band strengths with increasing concentration of HCOOH are depicted in 
Figure \ref{fig:water_clusters-HCOOH} (in the SI). This figure also reports the comparison of band strength profiles for different water clusters. The structures of the 
20-water-molecule cluster have been taken from \citet{shim18}, and were obtained by MD-annealing calculations using classical force-fields to 
reproduce a water cluster as a model of the ASW surface. The comparison shown in Figure \ref{fig:water_clusters-HCOOH}  (in the SI) demonstrates 
that the 4H$_2$O model
provides results similar to those obtained with 6 and 20 water molecules. This furthermore confirms the validity of our approach.

\section{Conclusions}
\label{sec:conclusions}
Water ice is known to be the major constituent of interstellar icy grain mantles. Interestingly, there have been several astronomical observations \cite{boog00,kean01} of the OH stretching and HOH bending modes at $3278.69$ cm$^{-1}$ ($3.05$ $\mu$m) and $1666.67$ cm$^{-1}$ ($6.00$ $\mu$m), respectively. It is noteworthy that the intensity ratio of these two bands is very different from what obtained in laboratory experiments for pure water ice. This suggested that the presence of impurities in water ice affects the spectroscopic features of water itself. For this reason, a series of laboratory experiments were carried out in order to explain the discrepancy between observations and experiments. Furthermore, these observations prompted us to perform an extensive computational investigation aiming to evaluate the effect of different amounts of representative impurities on the band strengths and absorption band profiles of interstellar ice. We selected the most abundant impurities ($\rm{HCOOH, \ NH_3, \ CH_3OH, \ CO, \ CO_2, \ H_2CO, \ CH_4, \ OCS, \ N_2}$, and $\rm{O_2}$) and studied their  effect on four fundamental vibrational bands of pure water ice by employing different cluster models. Indeed, both the experimental and theoretical peak positions might differ from the astronomical observations. This is because the grain shape, size, and constituents, the surrounding physical conditions, and the presence of impurities play a crucial role in tuning the ice spectroscopic features. \\

Although most of the computations were performed for a cluster containing only four water molecules as
a model system (to find a trend in the absorption band strength), we demonstrated that increasing the size of the cluster would change the band strength profile only marginally. From the band strength profiles shown in 
Figure \ref{fig:CH3OH_varied_cluster_size} (in the SI), it is apparent that the stretching mode is the most affected and the bending mode is the least affected by the presence of impurities. Libration, bending, and bulk stretching modes were found to be most affected by HCOOH impurity, followed by $\rm{CH_3OH}$ and $\rm{H_2CO}$. Another interesting point to be noted is that the band strength of the free-OH stretching mode decreases with increasing concentration of $\rm{NH_3}$ and completely vanishes when the concentration of NH$_3$ becomes 50\%. Most interestingly, the experimental free-OH band profile shows a decreasing trend when water is mixed with $\rm{NH_3}$ (Figure \ref{fig:HCOOH-NH3_band_strength}, right panel), similarly to
that obtained computationally. \\

Finally, our computed and laboratory absorption spectra of water-rich ices will be part of a larger infrared ice database in support of current and 
future observations. Understanding the effect of impurities in interstellar polar ice analogs will be pivotal to support the unambiguous identification
of COMs in interstellar ice mantles by using future space missions such as JWST \cite{gibb04}.

\section{Acknowledgment}
PG acknowledges the support of CSIR (Grant No. 09/904(0013) 2K18 EMR-I).
MS gratefully acknowledges DST-INSPIRE Fellowship [IF160109] scheme.
AD acknowledges ISRO respond (Grant No. ISRO/RES/2/402/16-17).
This research was possible in part due to a Grant-In-Aid from the Higher Education Department of the Government of West Bengal.
SI acknowledges the Royal Society for financial support. ZK was supported by
VEGA -- The Slovak Agency for Science, Grant No. 2/0023/18. This work was also supported by COST Action TD1308 -- ORIGINS.

\section{Supporting Information (SI)}

Optimized structures of water clusters and impurities mixed with a 6:1 concentration ratio (Figure \ref{fig:optimized_structure_6H2O}); band strengths of the four fundamental 
vibration modes of water clusters containing impurities with various concentrations (Figure \ref{fig:6h2o_x_band_strength}); structure of water clusters
containing 20 H$_2$O molecules with HCOOH as impurity in different concentration ratio (Figure \ref{fig:20H2O-HCOOH}); comparison of the band strength of various water clusters mixed with 
HCOOH (Figure \ref{fig:water_clusters-HCOOH}); effect of the cluster size on the band strength profile (Figure \ref{fig:CH3OH_varied_cluster_size}); 
harmonic infrared frequencies and intensities of the 4H$_2$O cluster (Table \ref{tab:4H2O_B2PLYP}), 4H$_2$O/HCOOH (Table \ref{tab:4H2O_1HCOOH_B2PLYP}),
4H$_2$O/2HCOOH (Table \ref{tab:4H2O_2HCOOH_B2PLYP}), 4H$_2$O/3HCOOH (Table \ref{tab:4H2O_3HCOOH_B2PLYP}), 4H$_2$O/4HCOOH (Table \ref{tab:4H2O_4HCOOH_B2PLYP}),
4H$_2$O/NH$_3$ (Table \ref{tab:4H2O_1NH3_B2PLYP}), 4H$_2$O/2NH$_3$ (Table \ref{tab:4H2O_2NH3_B2PLYP}), 4H$_2$O/3NH$_3$ (Table \ref{tab:4H2O_3NH3_B2PLYP}),
4H$_2$O/4NH$_3$ (Table \ref{tab:4H2O_4NH3_B2PLYP}) evaluated at 
the B2PLYP/maug-cc-pVTZ level; harmonic infrared frequencies and intensities of the H$_2$O$^{QM}$ + 3 H$_2$O$^{MM}$ complex (Table \ref{tab:4H2O_MM-QM}); 
harmonic infrared frequencies and intensities of the first 4 H$_2$O$^{QM}$+16 H$_2$O$^{MM}$: complex configuration 1 (Table \ref{tab:20H2O_MM-QM_Conf_1}) and complex 
configuration 2 (Table \ref{tab:20H2O_MM-QM_Conf_2}); geometric details of optimized structures of water clusters and impurities mixed with 4:4 and 6:1 concentration 
ratios (Optimized-Structures.zip). The Supporting Information is available free of charge on the ACS Publications website.

\clearpage
\begin{center}
 \Large {{\bf Appendix}}
\end{center}

\begin{figure}
\includegraphics[width=8cm, height=6cm]{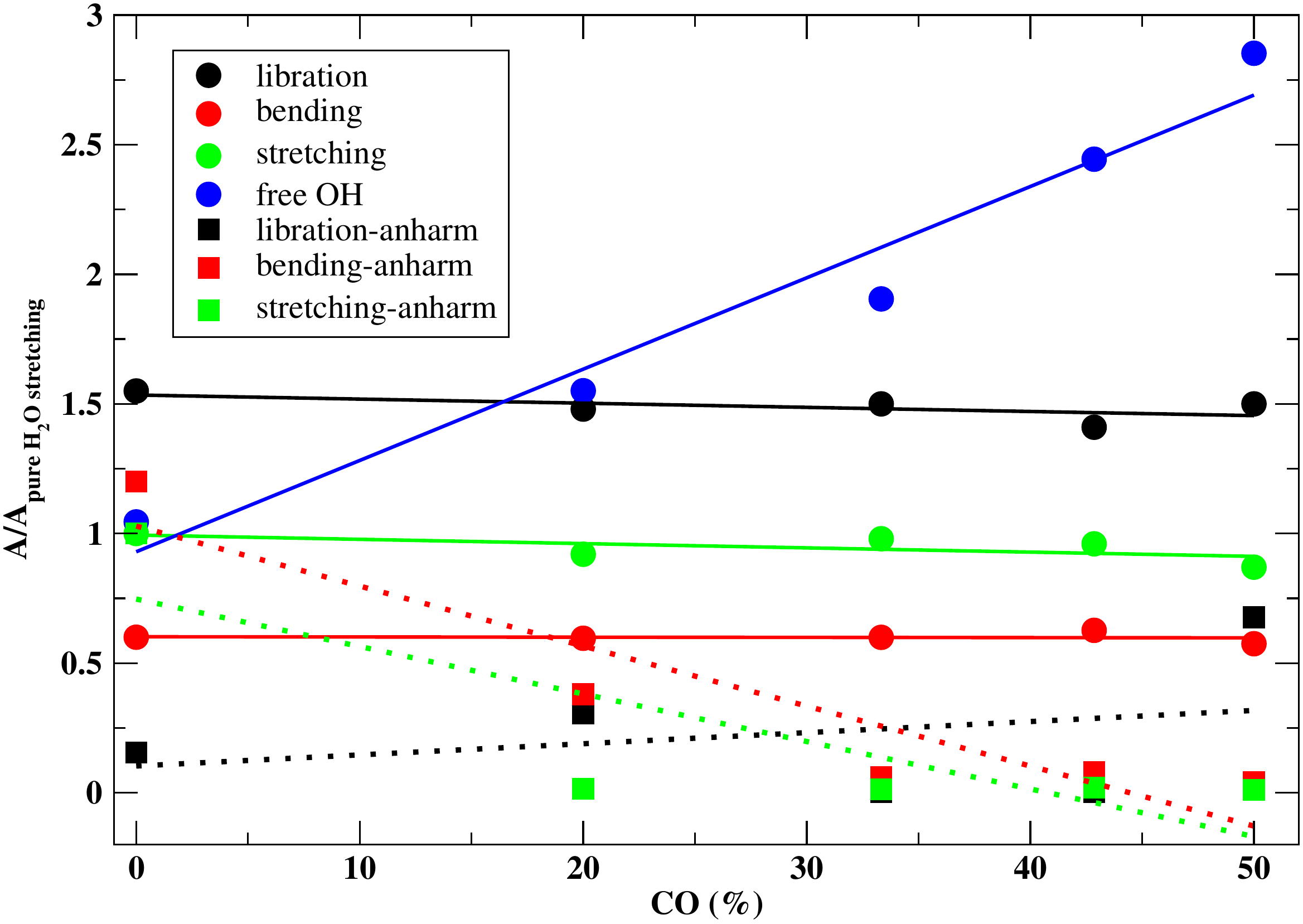}
\includegraphics[width=8cm, height=6cm]{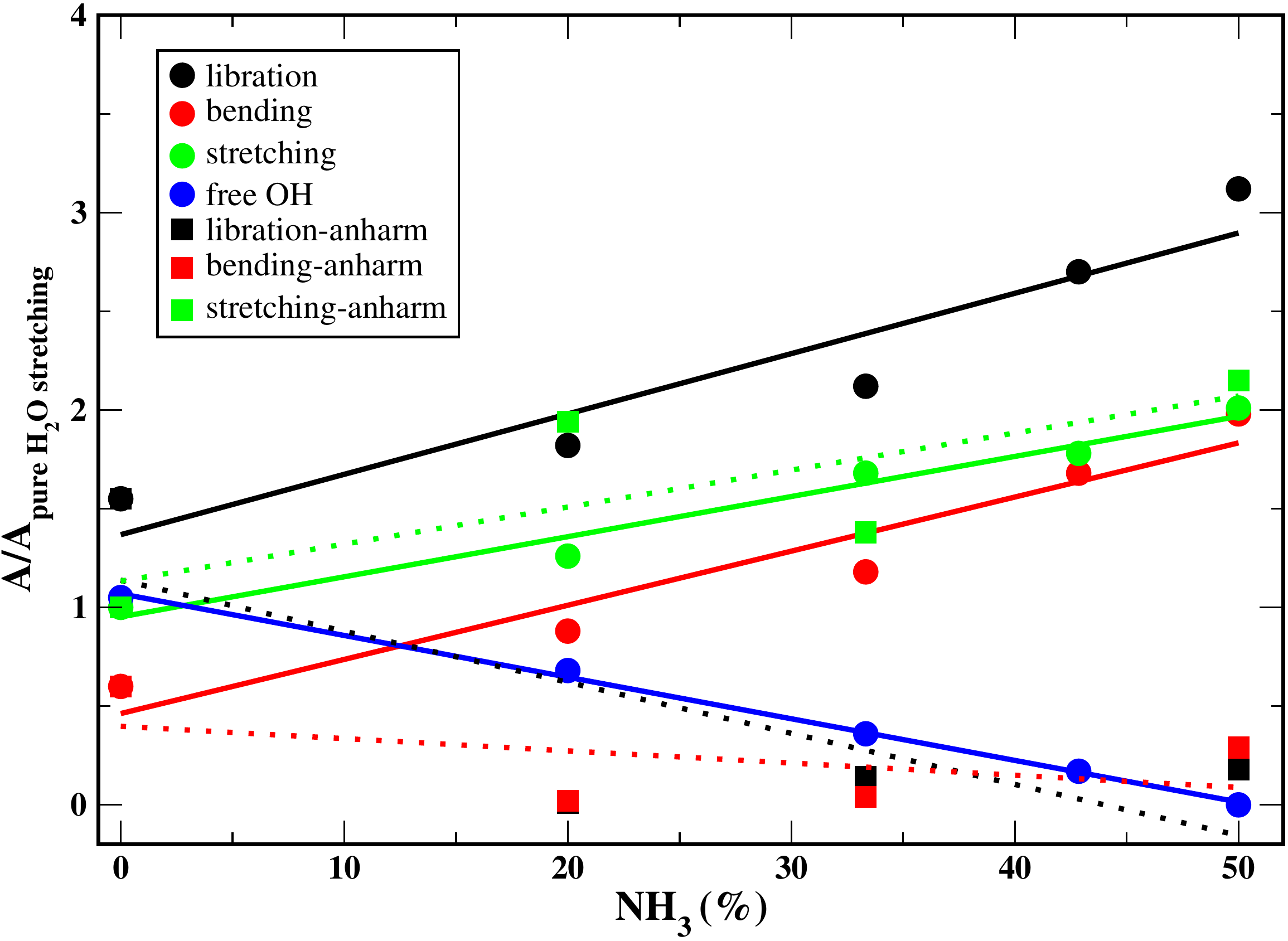}
\renewcommand{\thefigure}{A1}
\caption{The filled circles are the data points where we considered harmonic frequencies and the corresponding fitted profiles are the solid lines.
Solid filled squares represent the data sets where we considered anharmonic frequencies and the corresponding fitted results are the dotted
lines.}
\label{fig:harm-anharm-compare}
\end{figure}

\begin{figure}
\includegraphics[width=8cm, height=6cm]{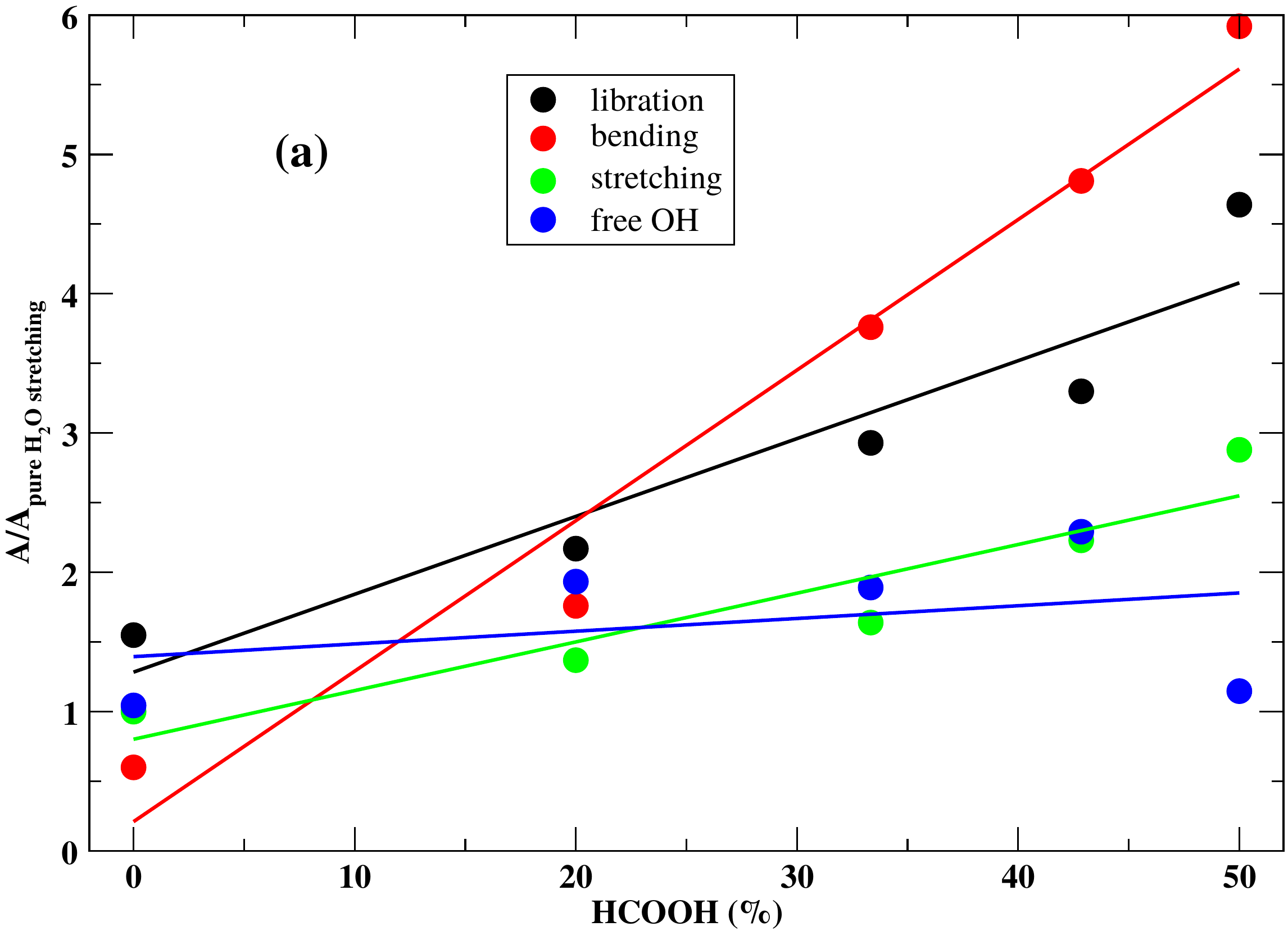}
\includegraphics[width=8cm, height=6cm]{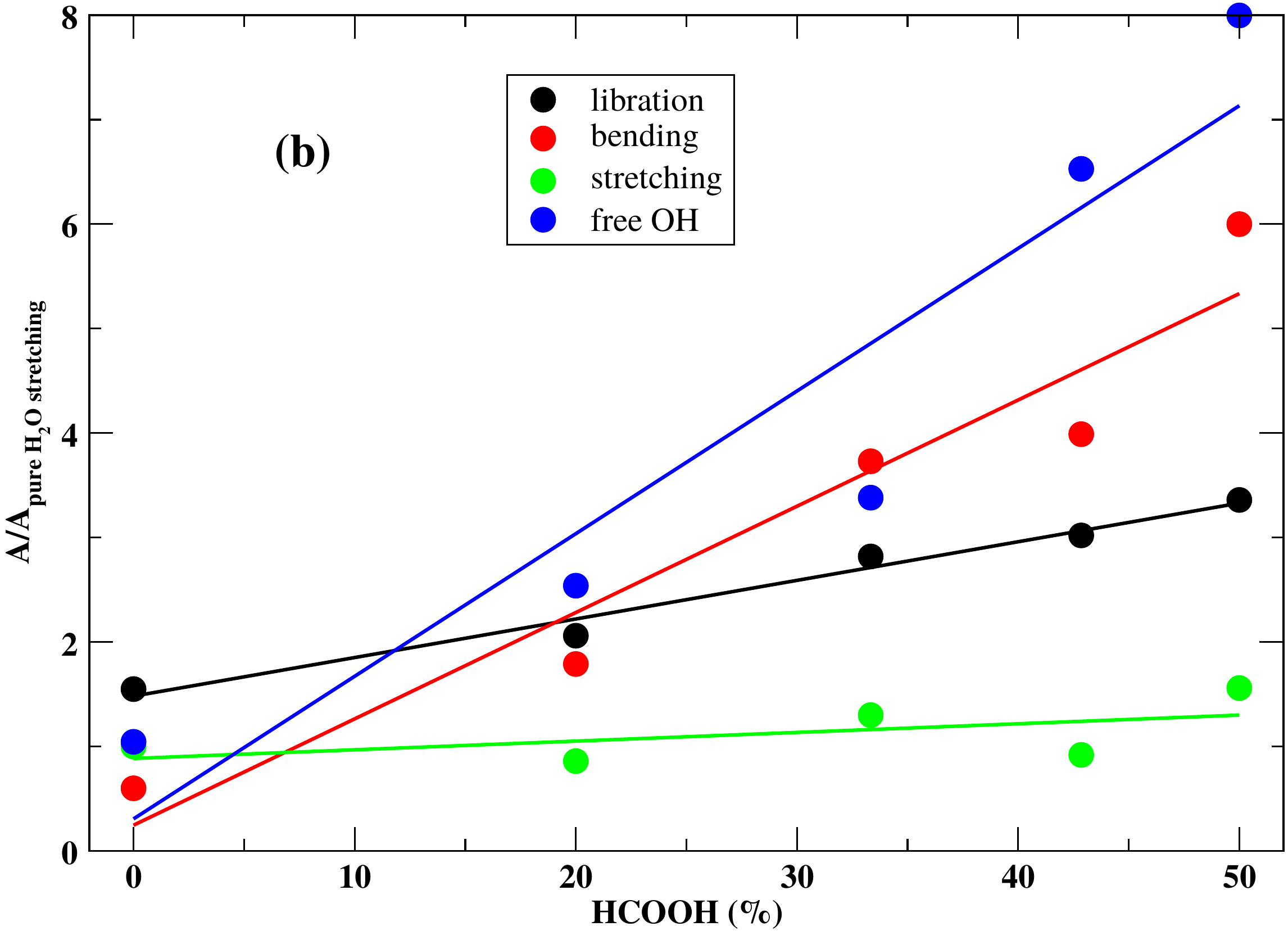}
\renewcommand{\thefigure}{A2}
\caption{ Band strength for H$_2$O-HCOOH mixtures: (a) HCOOH as hydrogen bond donor, and (b) HCOOH as hydrogen bond acceptor.}
\label{fig:hcooh-donor-acceptor}
\end{figure}

\begin{figure}
\includegraphics[width=8cm, height=6cm]{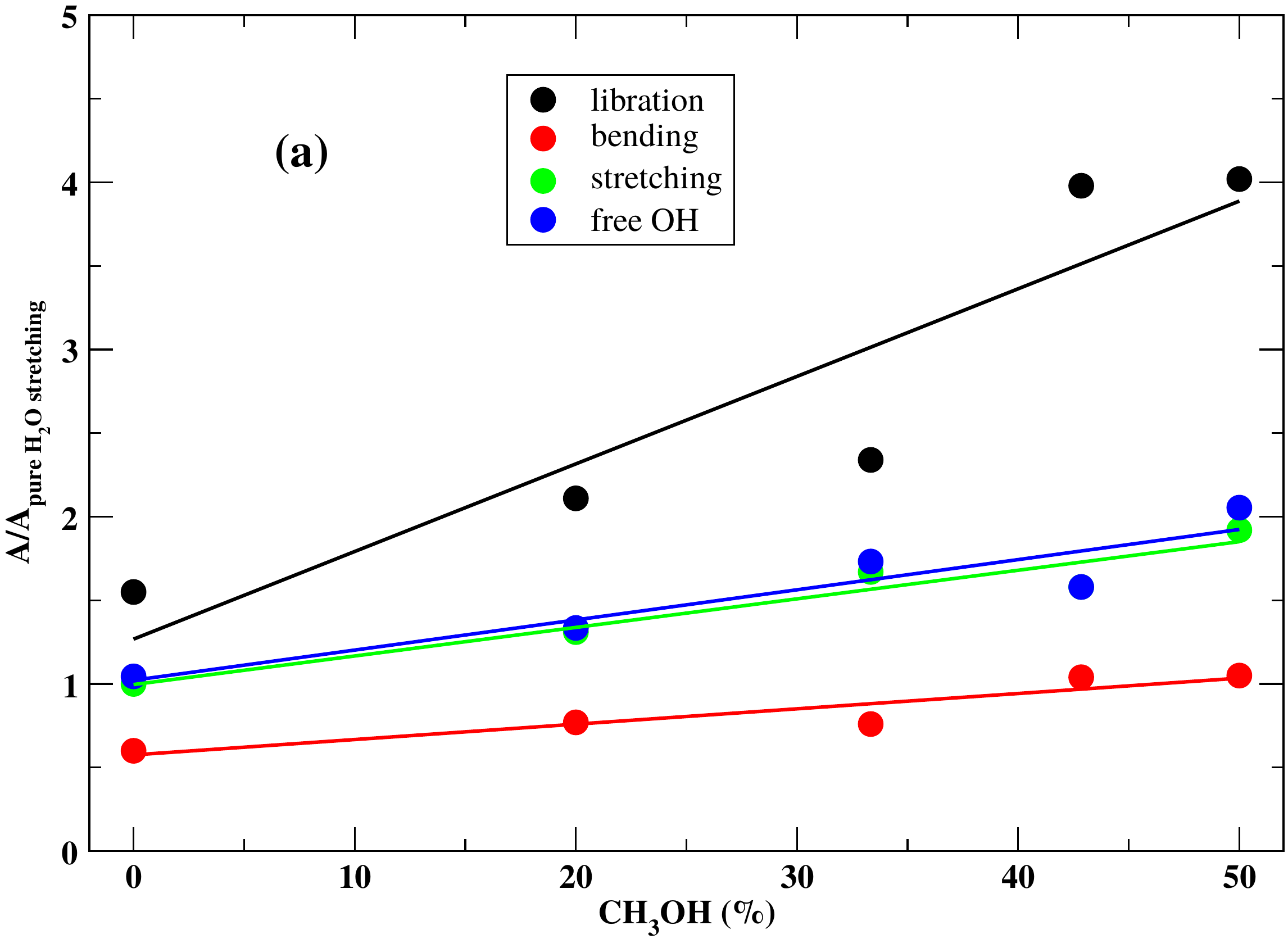}
\includegraphics[width=8cm, height=6cm]{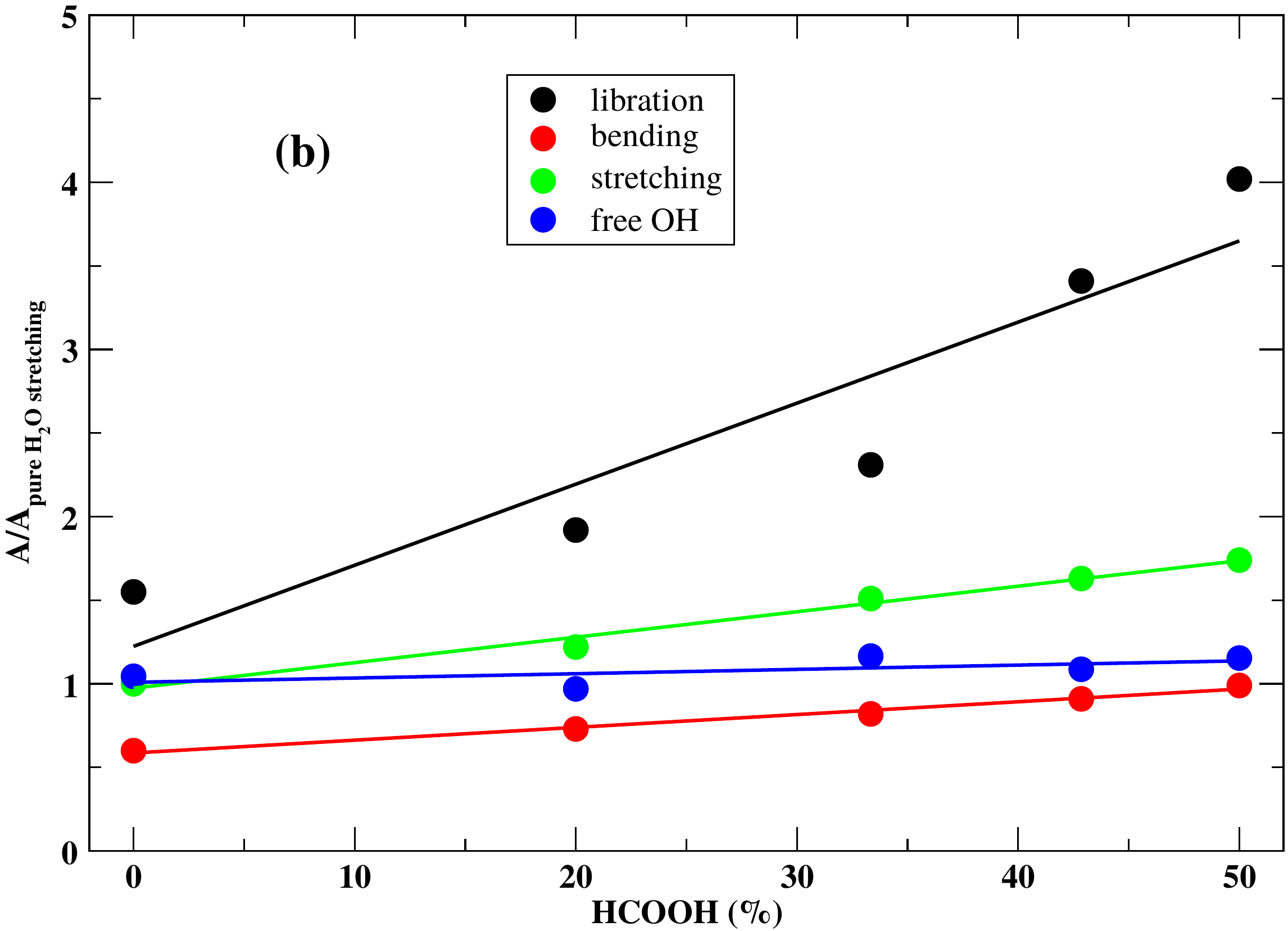}
\renewcommand{\thefigure}{A3}
\caption{Band strength for H$_2$O-CH$_3$OH mixtures: (a) CH$_3$OH as hydrogen bond donor, and (b) CH$_3$OH as hydrogen bond acceptor.}
\label{fig:ch3oh-donor-acceptor}
\end{figure}

\begin{figure}
\centering
\includegraphics[width=\textwidth]{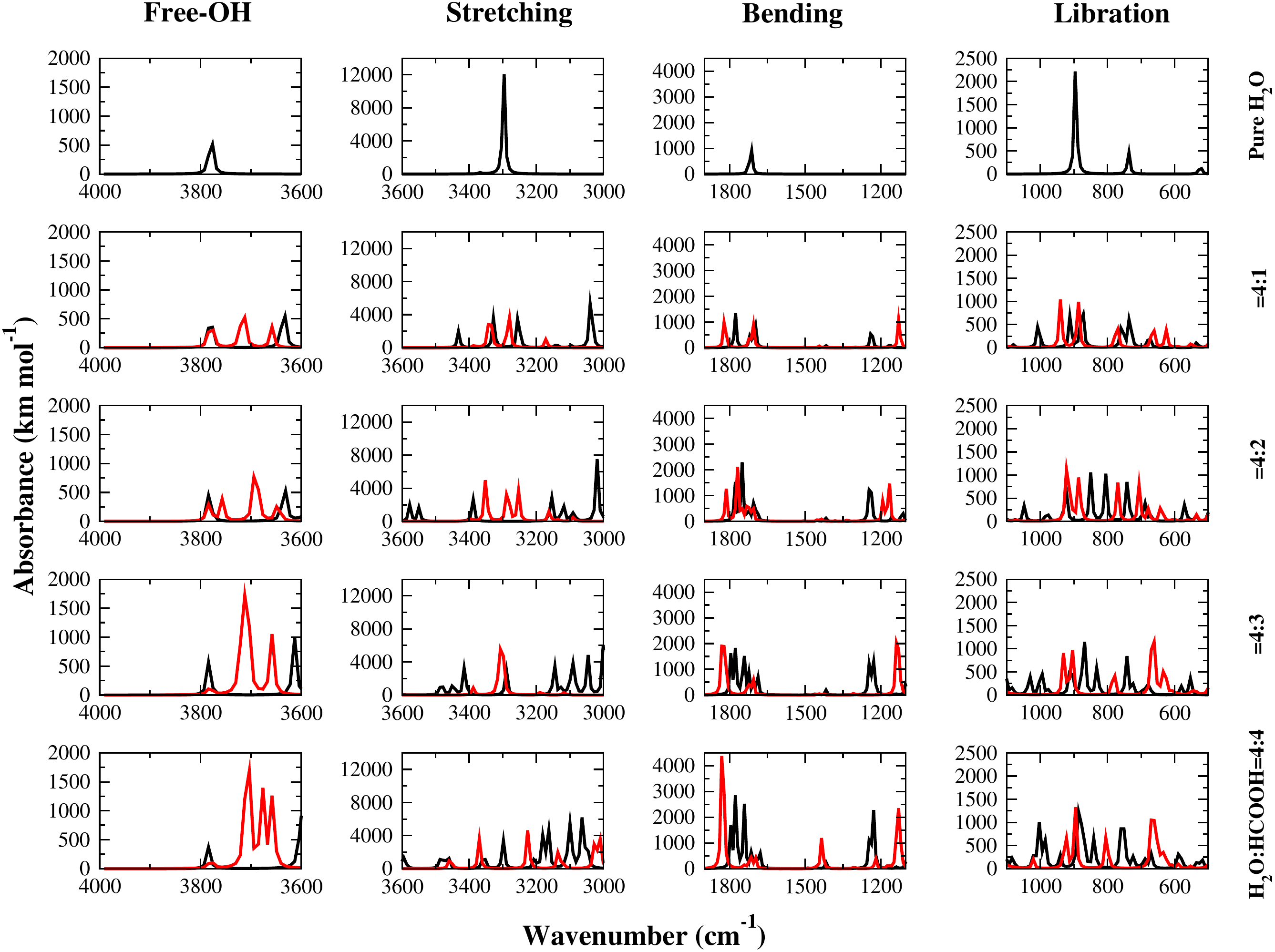}
\renewcommand{\thefigure}{A4}
\caption{Absorption spectra of the four modes for water ice for the five measured compositions, ranging
from pure water ice (top) to 4:4 H$_2$O-HCOOH mixture (bottom). Black line represent the absorbance spectra of various
concentration of H$_2$O-HCOOH, where HCOOH is used as hydrogen bond donor and for red line HCOOH is used as
hydrogen bond  acceptor.}
\label{fig:H2O-HCOOH}
\end{figure}

\begin{figure}
\centering
\includegraphics[width=\textwidth]{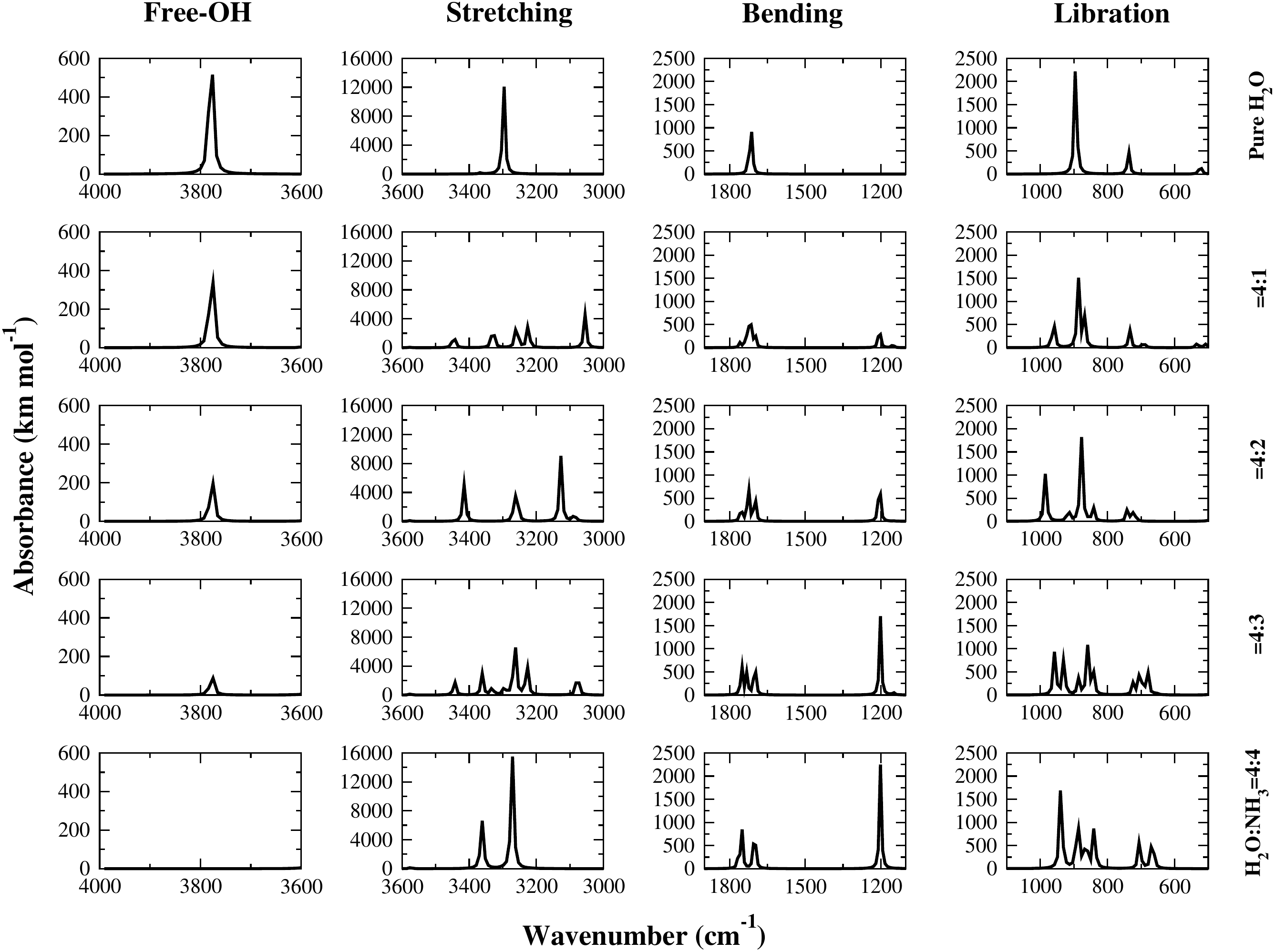}
\renewcommand{\thefigure}{A5}
\caption{Absorption spectra of the four modes for water ice for the five measured compositions, ranging
from pure water ice (top) to 4:4 H$_2$O-NH$_3$ mixture (bottom).}
\label{fig:H2O-NH3}
\end{figure}

\begin{figure}
\centering
\includegraphics[width=\textwidth]{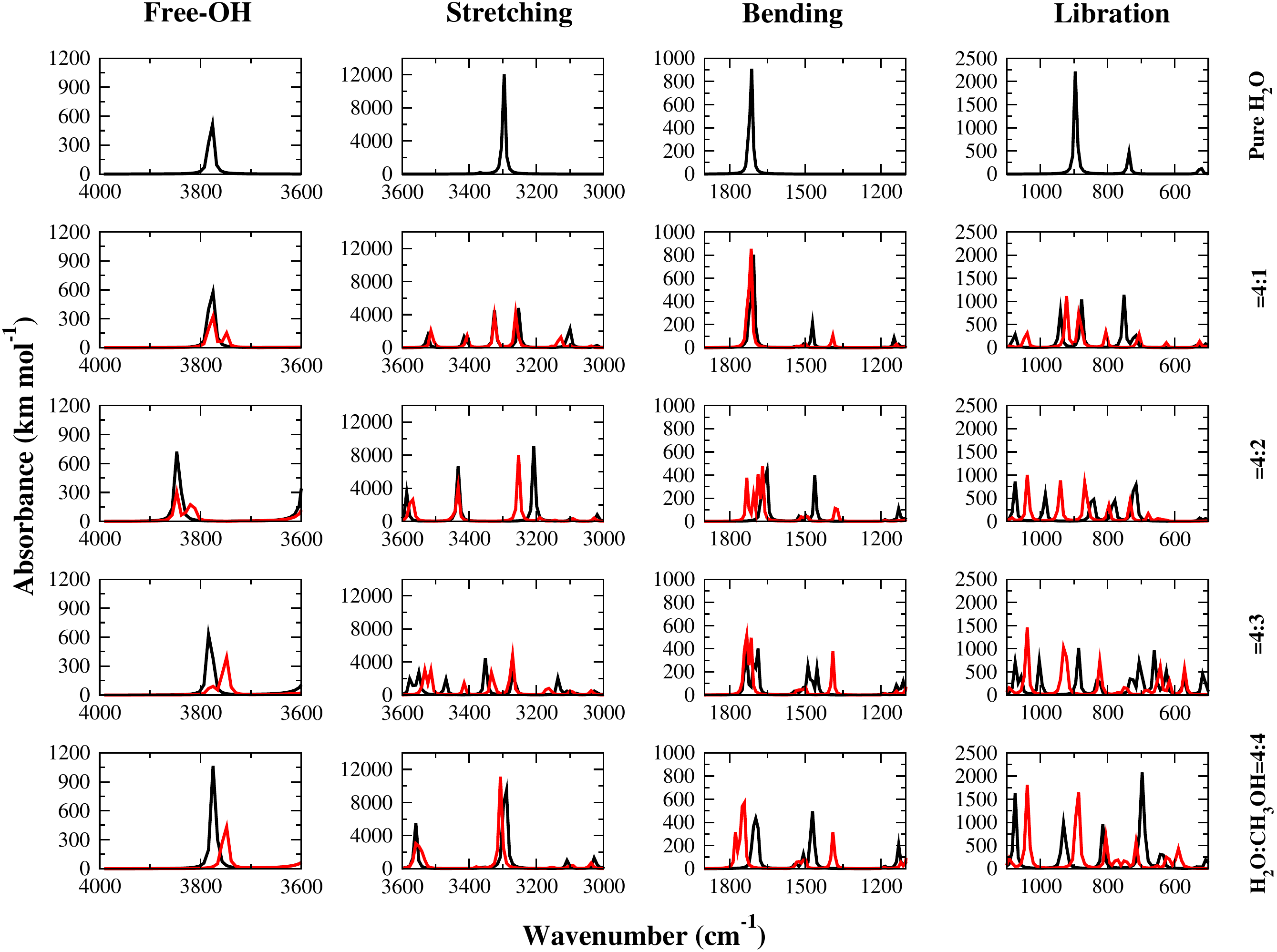}
\renewcommand{\thefigure}{A6}
\caption{Absorption spectra of the four modes for water ice for the five measured compositions, ranging from pure water ice (top) to 4:4
H$_2$O-CH$_3$OH mixture (bottom).  Black line represent the absorbance spectra of various
concentration of H$_2$O-CH$_3$OH, where CH$_3$OH is used as hydrogen bond donor and for red line CH$_3$OH is used as
hydrogen bond  acceptor.}
\label{fig:H2O-CH3OH}
\end{figure}

\begin{figure}
\centering
\includegraphics[width=\textwidth]{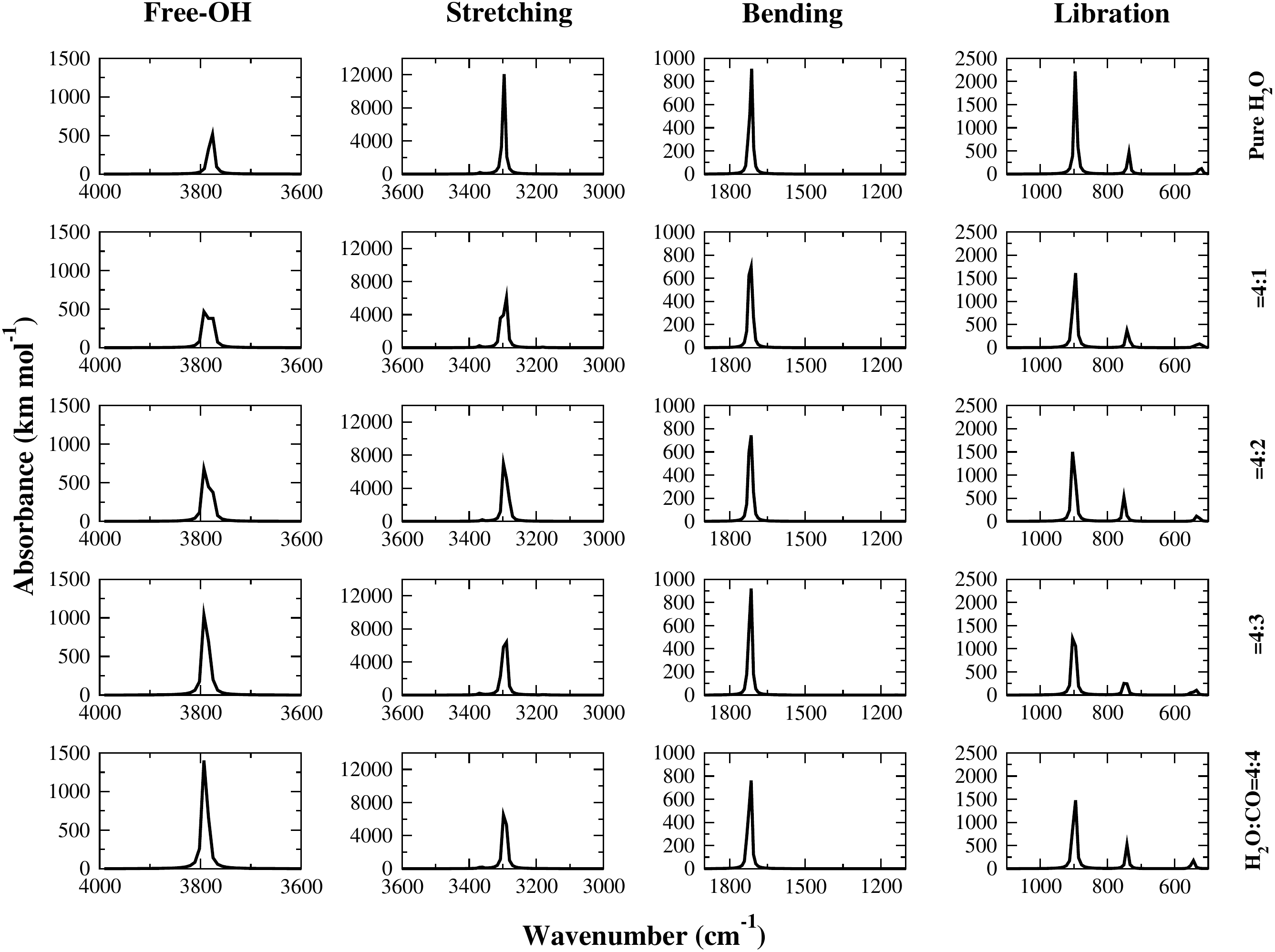}
\renewcommand{\thefigure}{A7}
\caption{Absorption spectra of the four modes for water ice for the five measured compositions, ranging from pure water ice (top) to 4:4 H$_2$O-CO mixture (bottom).}
\label{fig:H2O-CO}
\end{figure}

\begin{figure}
\centering
\includegraphics[width=\textwidth]{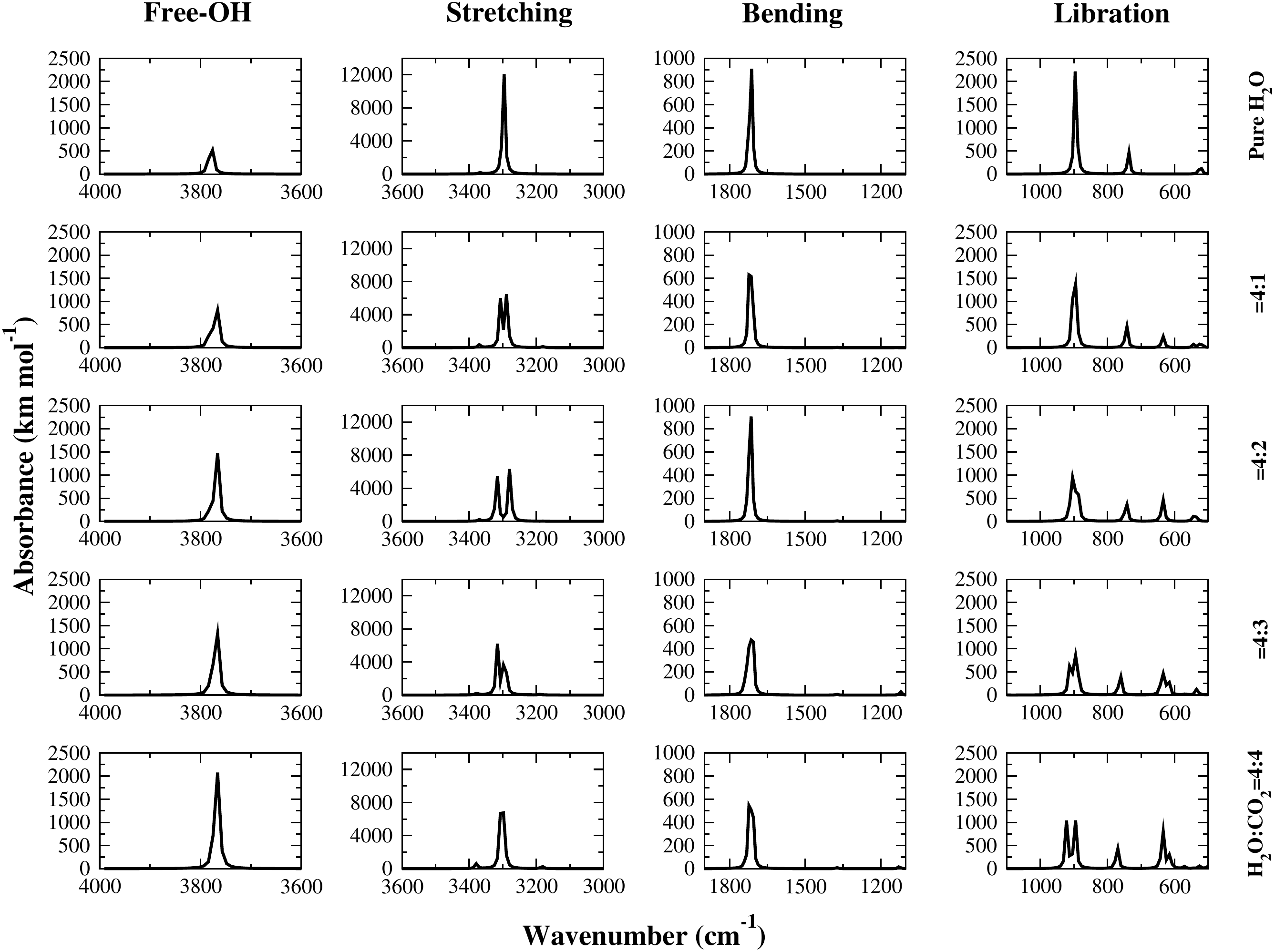}
\renewcommand{\thefigure}{A8}
\caption{Absorption spectra of the four modes for water ice for the five measured compositions, ranging from pure water ice (top) to 4:4 H$_2$O-CO$_2$ mixture (bottom).}
\label{fig:H2O-CO2}
\end{figure}

\begin{figure}
\centering
\includegraphics[width=\textwidth]{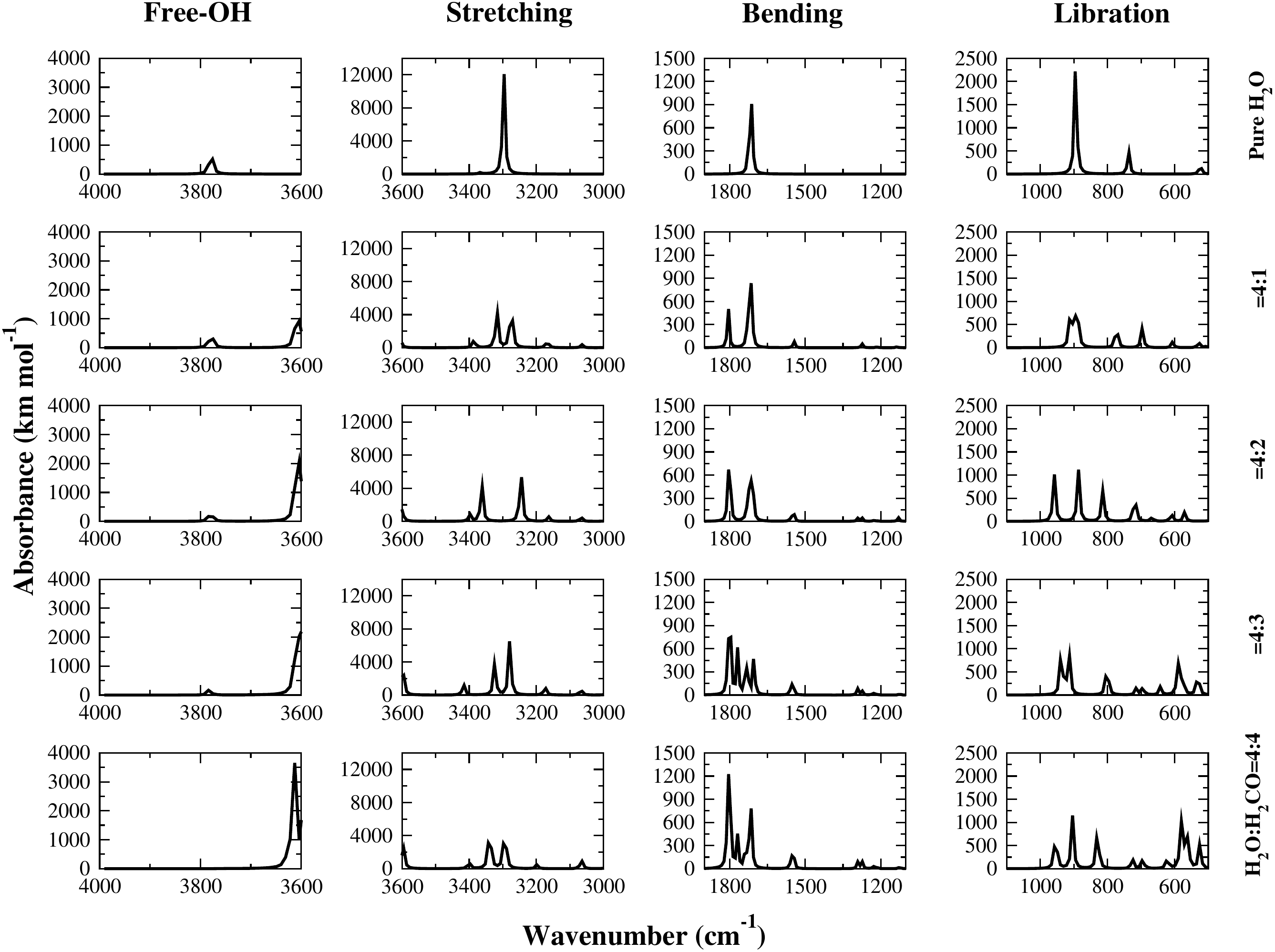}
\renewcommand{\thefigure}{A9}
\caption{Absorption spectra of the four modes for water ice for the five measured compositions, ranging from pure water ice (top) to 4:4 H$_2$O-H$_2$CO mixture (bottom).}
\label{fig:H2O-H2CO}
\end{figure}

\begin{figure}
\centering
\includegraphics[width=\textwidth]{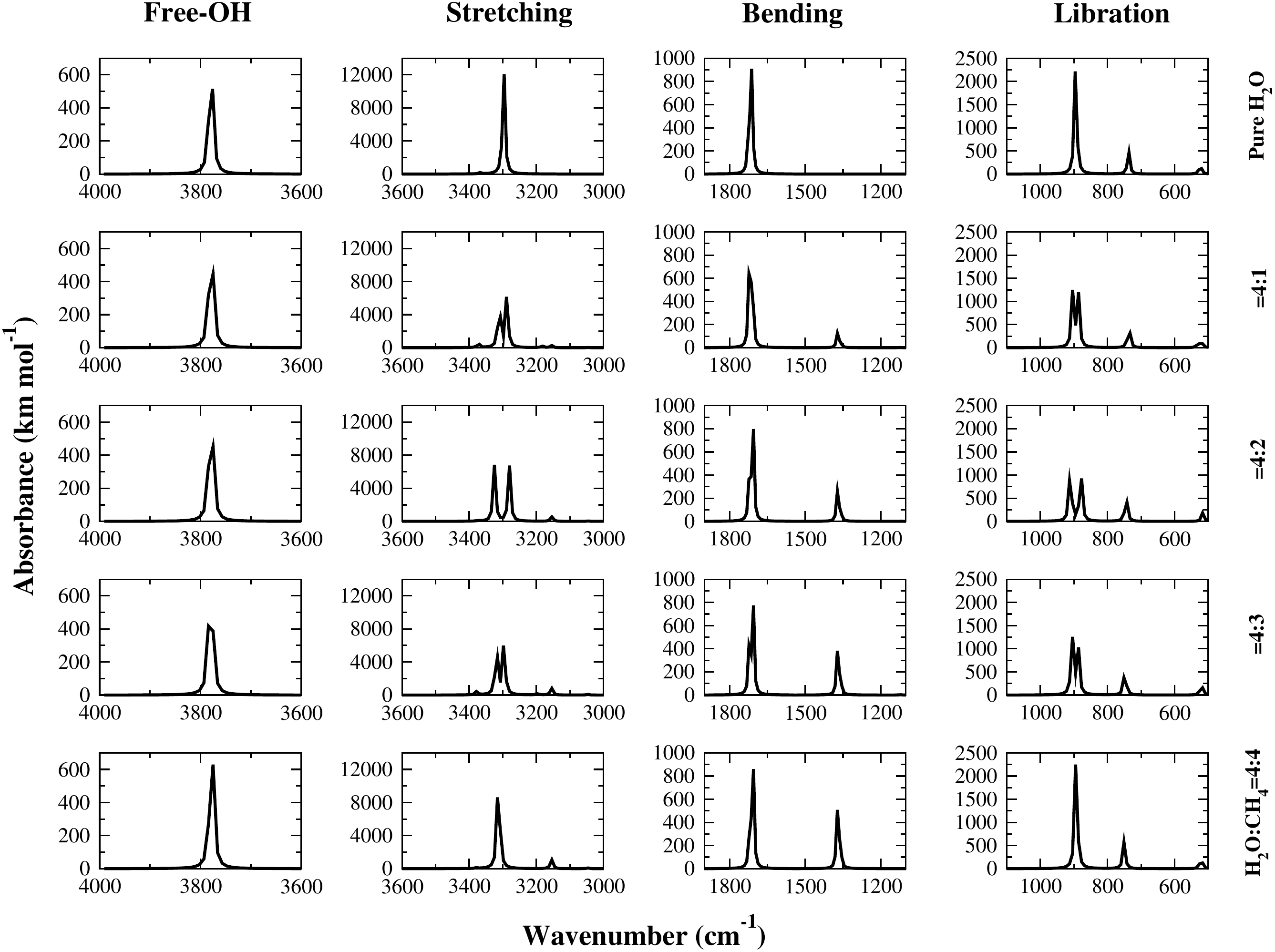}
\renewcommand{\thefigure}{A10}
\caption{Absorption spectra of the four modes for water ice for the five measured compositions, ranging from pure water ice (top) to 4:4 H$_2$O-CH$_4$ mixture (bottom).}
\label{fig:H2O-CH4}
\end{figure}

\begin{figure}
\centering
\includegraphics[width=\textwidth]{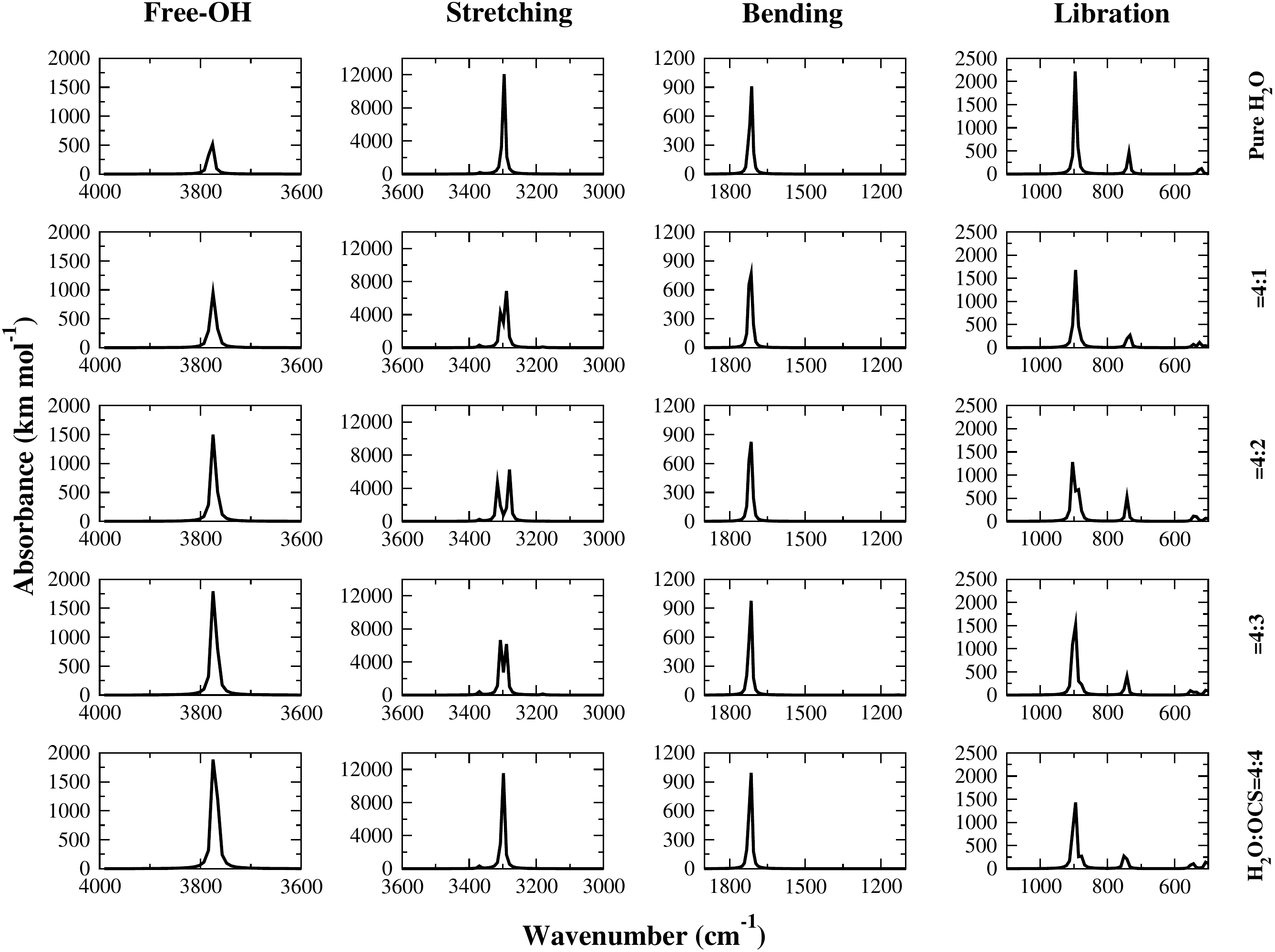}
\renewcommand{\thefigure}{A11}
\caption{Absorption spectra of the four modes for water ice for the five measured compositions, ranging from pure water ice (top) to 4:4 H$_2$O-OCS mixture (bottom).}
\label{fig:H2O-OCS}
\end{figure}

\begin{figure}
\centering
\includegraphics[width=\textwidth]{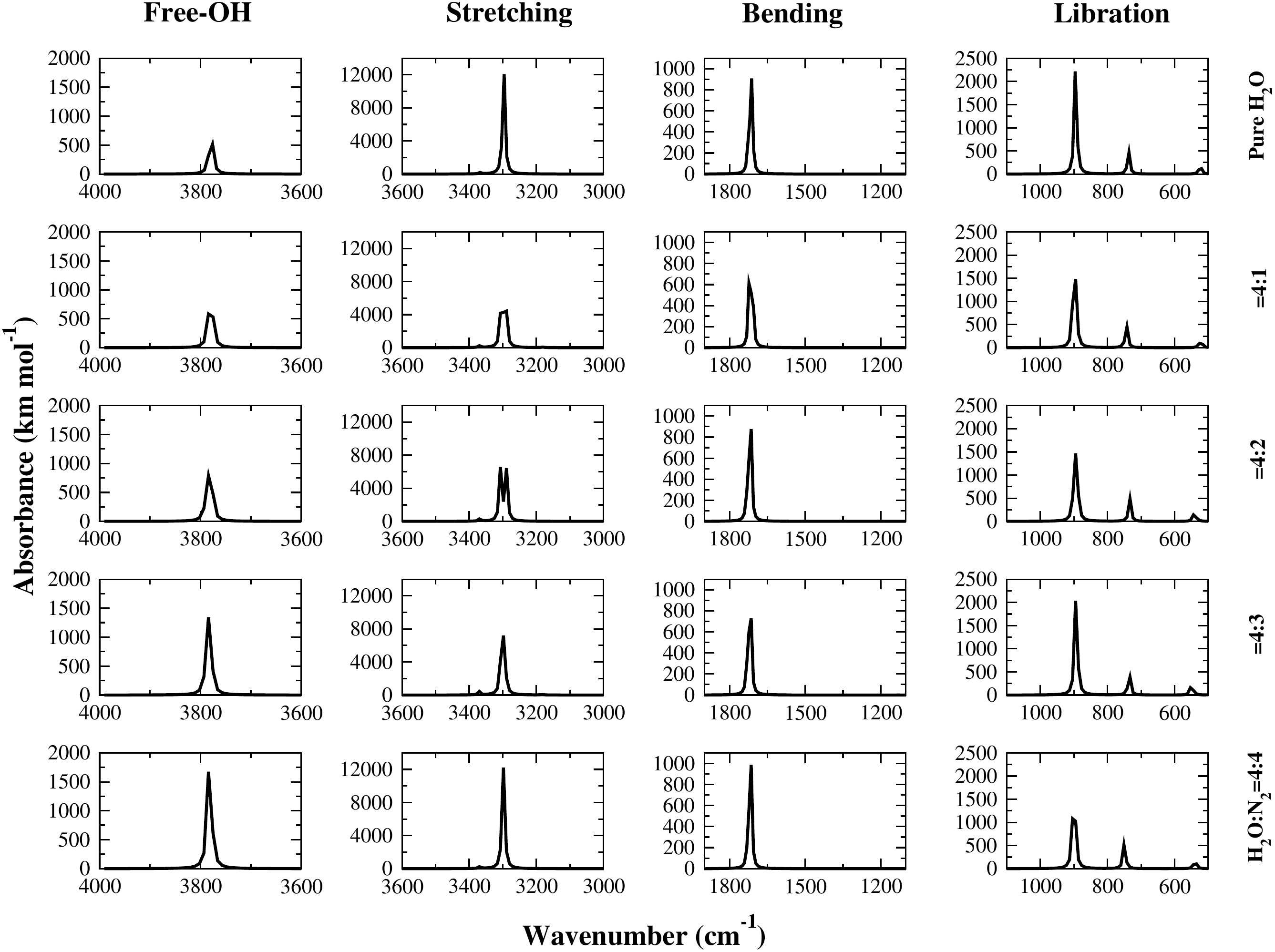}
\renewcommand{\thefigure}{A12}
\caption{Absorption spectra of the four modes for water ice for the five measured compositions, ranging from pure water ice (top) to 4:4 H$_2$O-N$_2$ mixture (bottom).}
\label{fig:H2O-N2}
\end{figure}

\begin{figure}
\centering
\includegraphics[width=\textwidth]{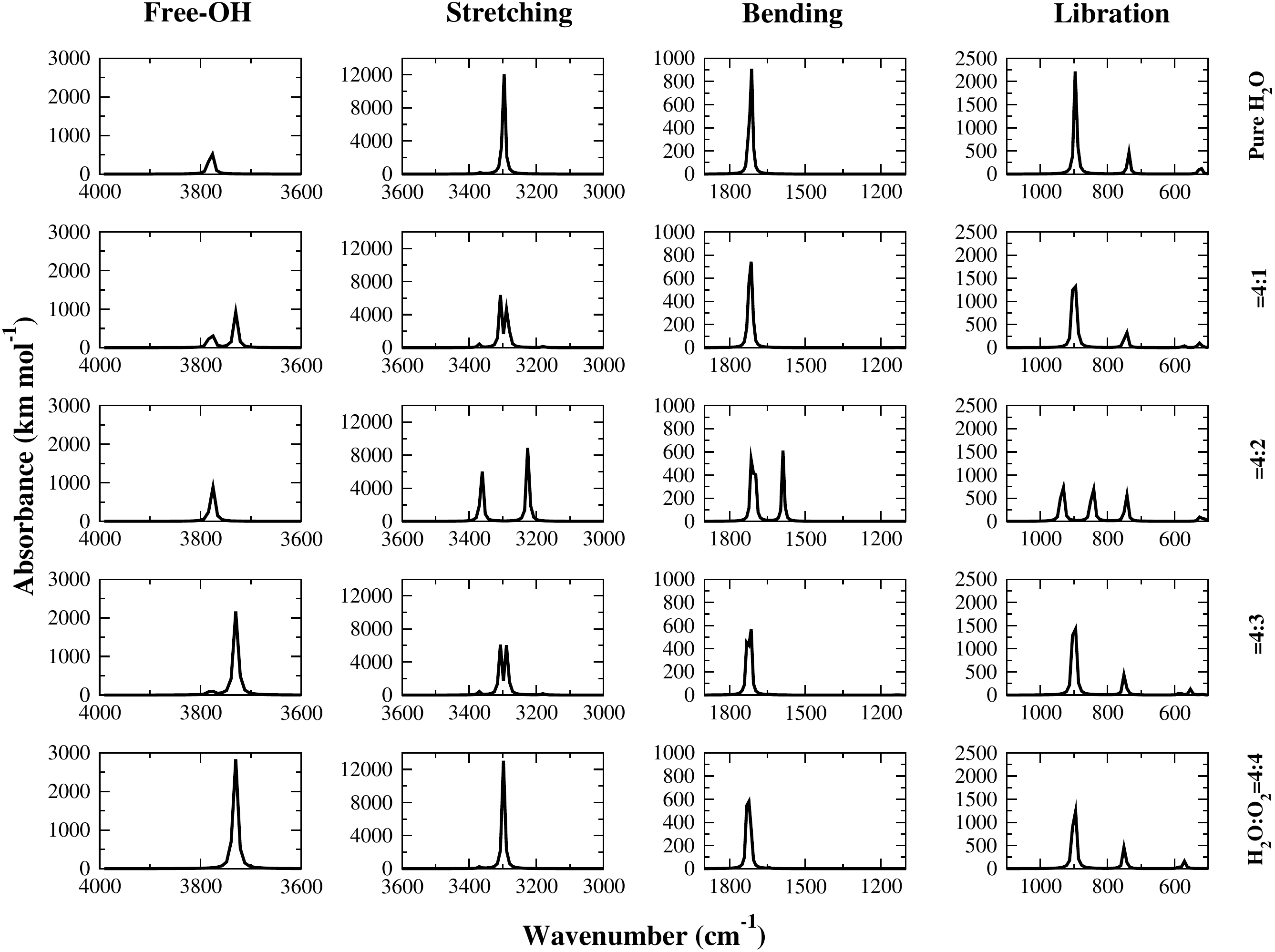}
\renewcommand{\thefigure}{A13}
\caption{Absorption spectra of the four modes for water ice for the five measured compositions, ranging from pure water ice (top) to 4:4 H$_2$O-O$_2$ mixture (bottom).}
\label{fig:H2O-O2}
\end{figure}

\clearpage
\begin{table}
{
\scriptsize{
\renewcommand{\thetable}{A1}
\caption{Comparison of band positions and band intensities as obtained by different quantum-chemical methods.}
\label{table:comparison-different-water-cluster}
\begin{tabular}{c|cc|cc|cc|cc}
\hline
{Assignment}&\multicolumn{2}{|c|}{4H$_2$O+PCM,}&\multicolumn{2}{c|}{4H$_2$O+PCM,}&\multicolumn{2}{c|}{1H$_2$O(QM)+3H$_2$O(MM)}
&\multicolumn{2}{c}{ 4H$_2$O(QM)+16H$_2$O(MM)} \\
&\multicolumn{2}{|c|}{B3PLYP/6-31G(d)}&\multicolumn{2}{c|}{B2PLYP/m-aug-cc-pVTZ}&\multicolumn{2}{c|}{+PCM, B3PLYP/6-31G(d)}&\multicolumn{2}{c}{+PCM, B3PLYP/6-31G(d)}\\
\cline{2-9}
&{Wavenumber} & {Intensity } &{Wavenumber} & {Intensity }&{ Wavenumber} & {Intensity } &{Wavenumber} & {Intensity}\\
&(cm$^{-1}$)&(km/mol)&(cm$^{-1}$) &(km/mol)&(cm$^{-1}$) &(km/mol)&(cm$^{-1}$)&(km/mol)\\
\hline
1&&&&&&&28.5053&1.6212\\
2&&&&&&&32.4112& 0.3452\\
3&&&39.3404 &0.0000&39.0097&3.5702&40.5682&1.9422\\
4&42.31&0.000&&&&&42.6937&0.9658\\
\hline
\end{tabular}}}
\end{table}

\begin{table}
{

\scriptsize{
\begin{tabular}{c|cc|cc|cc|cc}
\hline
{Assignment}&\multicolumn{2}{|c|}{4H$_2$O+PCM,}&\multicolumn{2}{c|}{4H$_2$O+PCM,}&\multicolumn{2}{c|}{1H$_2$O(QM)+3H$_2$O(MM)}
&\multicolumn{2}{c}{ 4H$_2$O(QM)+16H$_2$O(MM)} \\
&\multicolumn{2}{|c|}{B3PLYP/6-31G(d)}&\multicolumn{2}{c|}{B2PLYP/m-aug-cc-pVTZ}&\multicolumn{2}{c|}{+PCM, B3PLYP/6-31G(d)}&\multicolumn{2}{c}{+PCM, B3PLYP/6-31G(d)}\\
\cline{2-9}

&{Wavenumber} & {Intensity } &{Wavenumber} & {Intensity }&{ Wavenumber} & {Intensity } &{Wavenumber} & {Intensity}\\
&(cm$^{-1}$)&(km/mol)&(cm$^{-1}$) &(km/mol)&(cm$^{-1}$) &(km/mol)&(cm$^{-1}$)&(km/mol)\\
\hline
62&&&427.5608&44.2017&417.2017&11.4416&426.8869 &78.1914\\
63&&&&&&&432.7664&47.4011\\
64 &&&443.1204&79.6160&&&449.0251&78.8810\\
65 &&&443.1204&79.6160&&&455.8202 &9.0749\\
\hline
\end{tabular}}}
\end{table}

\begin{table}
{
\scriptsize{
\begin{tabular}{c|cc|cc|cc|cc}
\hline
{Assignment}&\multicolumn{2}{|c|}{4H$_2$O+PCM,}&\multicolumn{2}{c|}{4H$_2$O+PCM,}&\multicolumn{2}{c|}{1H$_2$O(QM)+3H$_2$O(MM)}
&\multicolumn{2}{c}{ 4H$_2$O(QM)+16H$_2$O(MM)} \\
&\multicolumn{2}{|c|}{B3PLYP/6-31G(d)}&\multicolumn{2}{c|}{B2PLYP/m-aug-cc-pVTZ}&\multicolumn{2}{c|}{+PCM, B3PLYP/6-31G(d)}&\multicolumn{2}{c}{+PCM, B3PLYP/6-31G(d)}\\
\cline{2-9}

&{Wavenumber} & {Intensity } &{Wavenumber} & {Intensity }&{ Wavenumber} & {Intensity } &{Wavenumber} & {Intensity}\\
&(cm$^{-1}$)&(km/mol)&(cm$^{-1}$) &(km/mol)&(cm$^{-1}$) &(km/mol)&(cm$^{-1}$)&(km/mol)\\
\hline
 121$^{b}$&&&&&1422.8362$^{b}$&445.8394&  1423.8136 &   320.7421\\
 122$^{b}$&&&&&&    &  1427.0557 &   239.2493\\
 123$^{b}$&&&&&&    &  1431.3527 &   280.4839\\
 124$^{b}$&&&&&&    &  1433.6066 &   107.3301\\
\hline
\end{tabular}}
\vskip 0.2cm
Notes: $^t$OH torsion; $^b$OH scissoring; $^s$OH stretching; $^f$free OH. For the conversion of km/mol to cm molecule$^{-1}$, intensity values need to
multiply by a factor $\rm{1.6603\times10^{-19}}$.}
\end{table}

\clearpage

\begin{table}
\centering
\scriptsize{
\renewcommand{\thetable}{A2}
\caption{Frequencies, integral absorbance coefficients, and normal modes of vibration of {H$_2$O-X (X = HCOOH, $\rm{NH_3}$, $\rm{CH_3OH}$, CO, $\rm {CO_2}$, $\rm{H_2CO}$, $\rm{CH_4}$, OCS, $\rm{N_2}$, and $\rm{O_2}$)}}
\label{tab:H2O_X}
\begin{tabular}{|c|>{}c|c|c|}
 \hline
 \hline
{\bf  $\rm{H_2O:X}$} & {\bf Frequency at peak}& {\bf Integral absorbance} & {\bf Modes of} \\
&{\bf in cm$^{-1}$ ($\mu$m)} & {\bf coefficient in} & {\bf vibration} \\
&& {\bf cm molecule$^{-1}$} & \\
\hline
\hline
\multicolumn{4}{|c|} {\bf HCOOH} \\
\hline
\hline
1:0.25&875.47 (11.42)&$5.98\times10^{-17}$& OH torsion \\
&1698.42 (5.89)&$5.24\times10^{-17}$& H$_2$O scissoring \\
&3326.84 (3.00)&$1.88\times10^{-16}$&OH stretching\\
&3635.02 (2.75)&$4.11\times10^{-17}$&free OH\\
\hline
\hline
\end{tabular}}
\end{table}

\begin{table}
\centering
\scriptsize{
\begin{tabular}{|c|>{}c|c|c|}
 \hline
 \hline
{\bf  $\rm{H_2O:X}$} & {\bf Frequency at peak}& {\bf Integral absorbance} & {\bf Modes of} \\
&{\bf in cm$^{-1}$ ($\mu$m)} & {\bf coefficient in} & {\bf vibration} \\
&& {\bf cm molecule$^{-1}$} & \\
\hline
\hline
\multicolumn{4}{|c|} {\bf CO} \\
\hline
\hline
1:0.25 & 895.46 (11.17) & $5.70\times10^{-17}$ & OH torsion \\
& 1711.46 (5.84) & $4.14\times10^{-17}$ & H$_2$O scissoring \\
& 3290.66 (3.04) & $3.14\times10^{-16}$ & OH stretching\\
& 3791.25 (2.64) & $2.39\times10^{-17}$ & free OH\\
\hline
\hline
\end{tabular}}
\end{table}

\begin{table}
\centering
\scriptsize{
\begin{tabular}{|c|>{}c|c|c|}
 \hline
 \hline
{\bf  $\rm{H_2O:X}$} & {\bf Frequency at peak}& {\bf Integral absorbance} & {\bf Modes of} \\
&{\bf in cm$^{-1}$ ($\mu$m)} & {\bf coefficient in} & {\bf vibration} \\
&& {\bf cm molecule$^{-1}$} & \\
\hline
\hline
\multicolumn{4}{|c|} {\bf CH$_4$} \\
\hline
\hline
1:0.25&903.11 (11.07)&$5.96\times10^{-17}$& OH torsion \\
&1710.29 (5.85)&$4.14\times10^{-17}$& H$_2$O scissoring \\
&3287.91 (3.04)&$3.06\times10^{-16}$&OH stretching\\
&3777.98 (2.65)&$2.03\times10^{-17}$&free OH\\
\hline
\hline
\end{tabular}}
\end{table}

\begin{table}
\centering
\scriptsize{
\begin{tabular}{|c|>{}c|c|c|}
 \hline
 \hline
{\bf  $\rm{H_2O:X}$} & {\bf Frequency at peak}& {\bf Integral absorbance} & {\bf Modes of} \\
&{\bf in cm$^{-1}$ ($\mu$m)} & {\bf coefficient in} & {\bf vibration} \\
&& {\bf cm molecule$^{-1}$} & \\
\hline
\hline
\multicolumn{4}{|c|} {\bf O$_2$} \\
\hline
\hline
1:0.25&893.78 (11.19)&$5.65\times10^{-17}$& OH torsion \\
&1711.86 (5.84)&$4.11\times10^{-17}$& H$_2$O scissoring \\
&3286.24 (3.04)&$3.10\times10^{-16}$ &OH stretching\\
&3729.93 (2.68)&$4.41\times10^{-17}$&free OH\\
\hline
1:0.50& 934.71 (10.70)&$6.49\times10^{-17}$& OH torsion \\
&1700.17 (5.88)&$3.78\times10^{-17}$& H$_2$O scissoring \\
&3225.04 (3.10)&$4.51\times10^{-16}$ &OH stretching\\
&3774.84 (2.65)&$2.77\times10^{-17}$&free OH\\
\hline
1:0.75&892.96 (11.20)&$5.75\times10^{-17}$& OH torsion \\
&1717.35 (5.82)&$4.04\times10^{-17}$& H$_2$O scissoring \\
&3286.59 (3.04)&$3.12\times10^{-16}$ &OH stretching\\
&3729.19 (2.68)&$8.72\times10^{-17}$&free OH\\
\hline
1:1.00&900.52 (11.10)&$5.67\times10^{-17}$& OH torsion\\
&1718.90 (5.82)&$4.18\times10^{-17}$& H$_2$O scissoring \\
&3299.52 (3.03)&$3.19\times10^{-16}$ &OH stretching\\
&3727.75 (2.68)&$1.08\times10^{-16}$&free OH\\
\hline
\hline
\end{tabular}}
\end{table}

\clearpage

\begin{suppinfo}
\begin{figure}
\centering
\includegraphics[width=\textwidth]{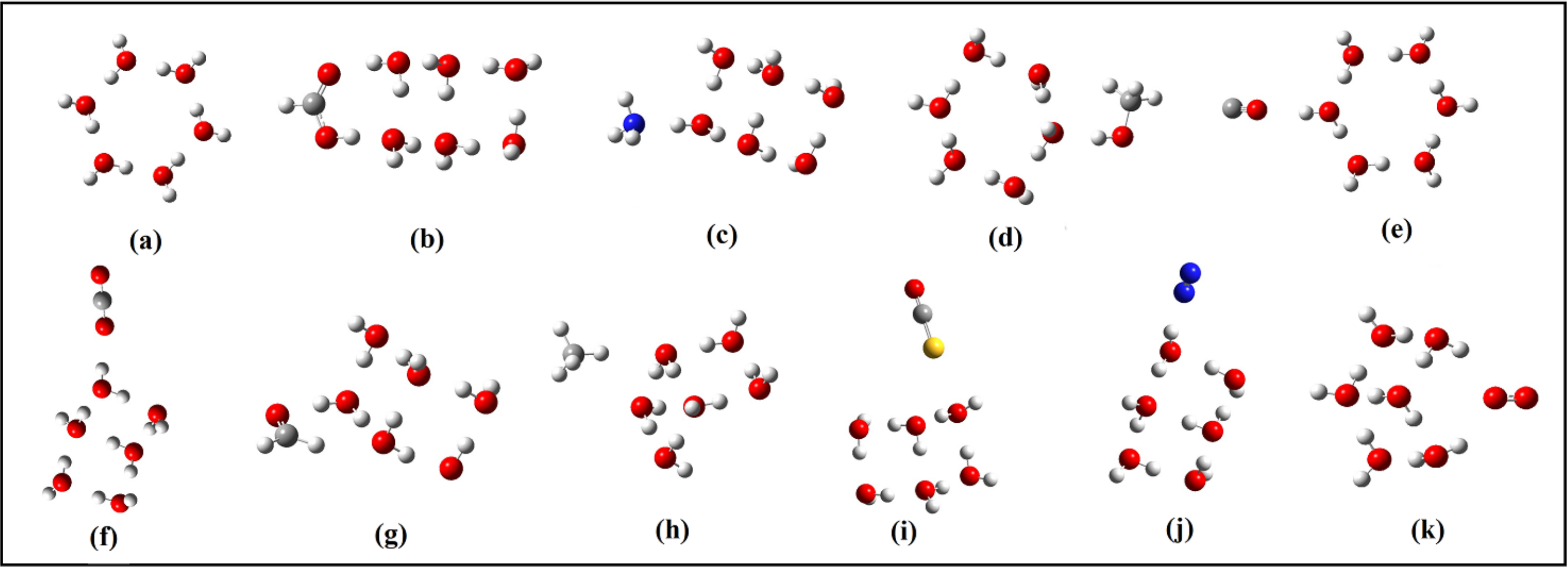}
\renewcommand{\thefigure}{S1}
\caption{Optimized structures of (a) pure water, (b) $\rm{H_2O-HCOOH}$, (c) $\rm{H_2O-NH_3}$, (d) $\rm{H_2O-CH_3OH}$, (e) $\rm{H_2O-CO}$, 
(f) $\rm{H_2O-CO_2}$, (g) $\rm{H_2O-H_2CO}$, (h) $\rm{H_2O-CH_4}$, (i) $\rm{H_2O-OCS}$, (j) $\rm{H_2O-N_2}$, and (k) $\rm{H_2O-O_2}$ clusters 
with a $6:1$ concentration ratio.}
\label{fig:optimized_structure_6H2O}
\end{figure}

\begin{figure}
\begin{minipage}{0.49\textwidth}
\includegraphics[width=\textwidth]{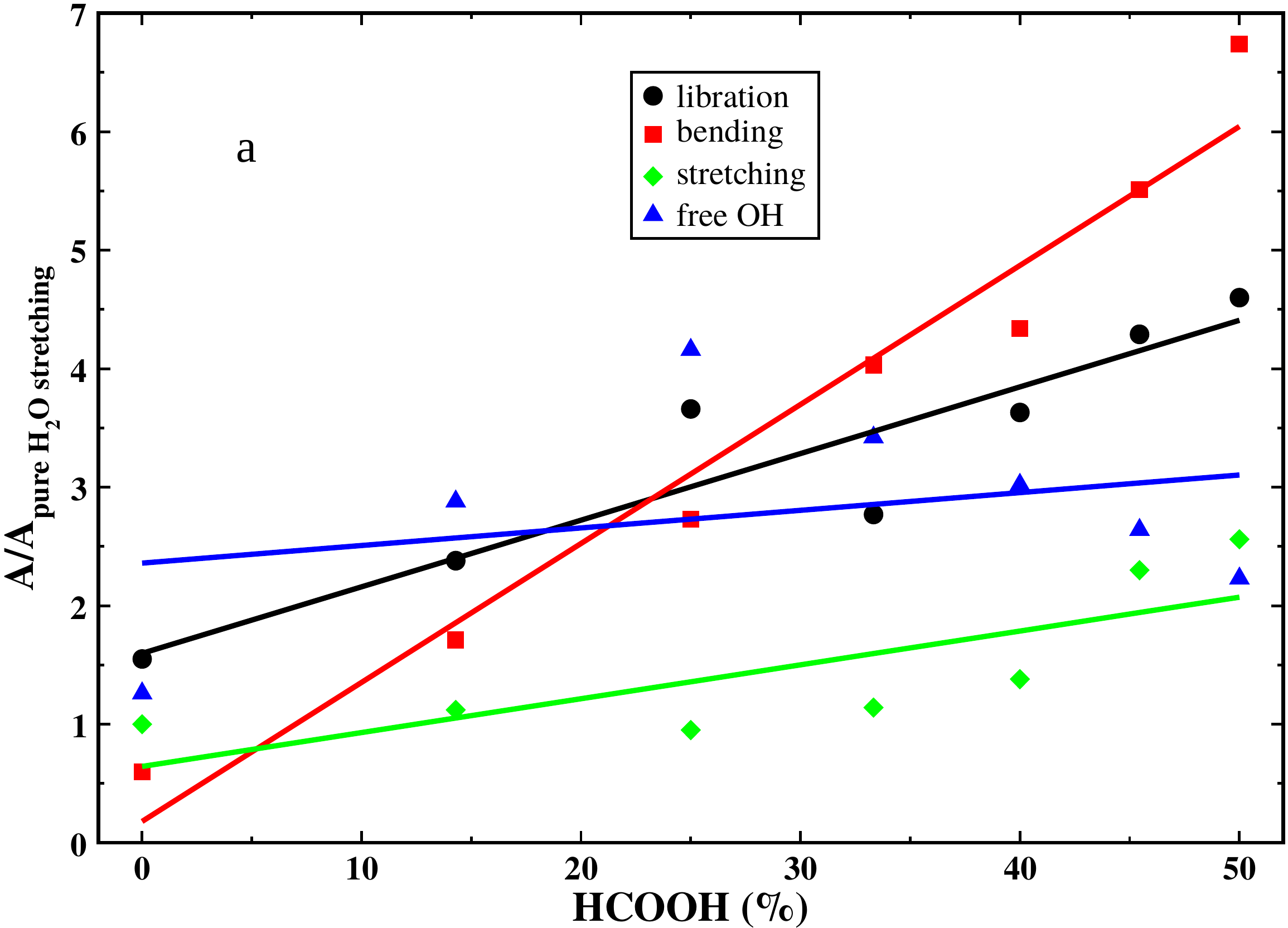}
\end{minipage}
\begin{minipage}{0.49\textwidth}
\includegraphics[width=\textwidth]{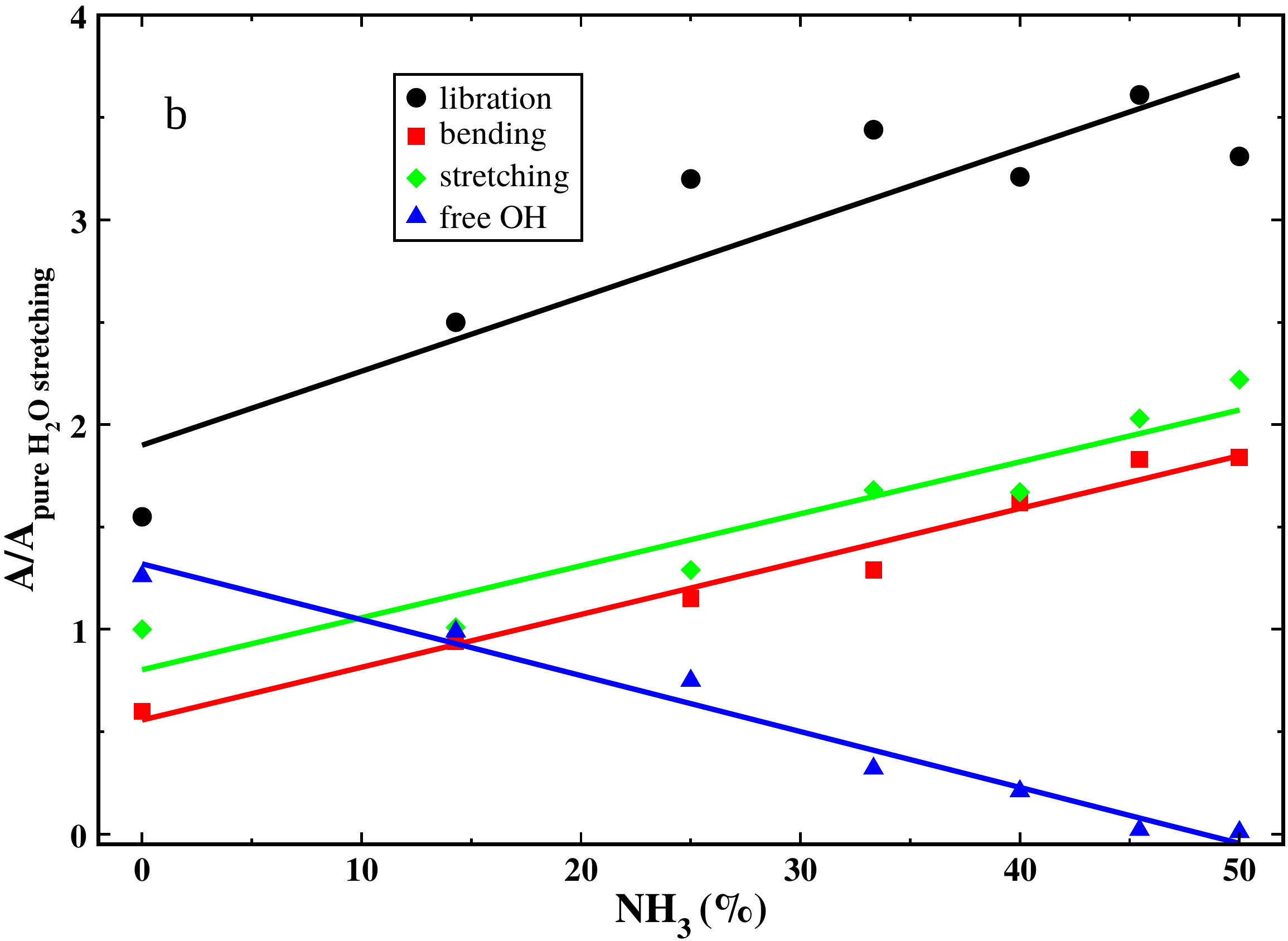}
\end{minipage}
\begin{minipage}{0.49\textwidth}
\includegraphics[width=\textwidth]{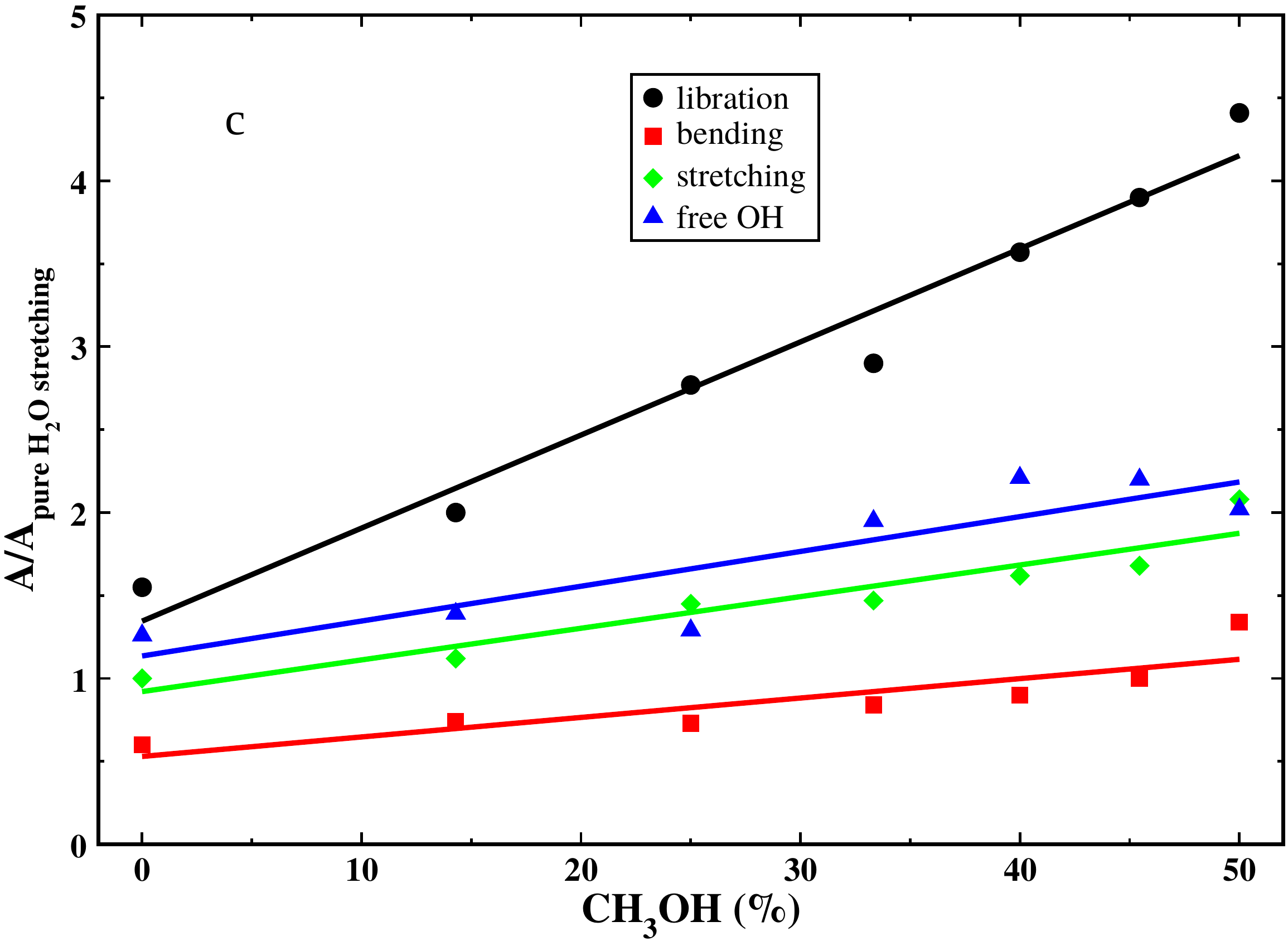}
\end{minipage}
\begin{minipage}{0.49\textwidth}
\includegraphics[width=\textwidth]{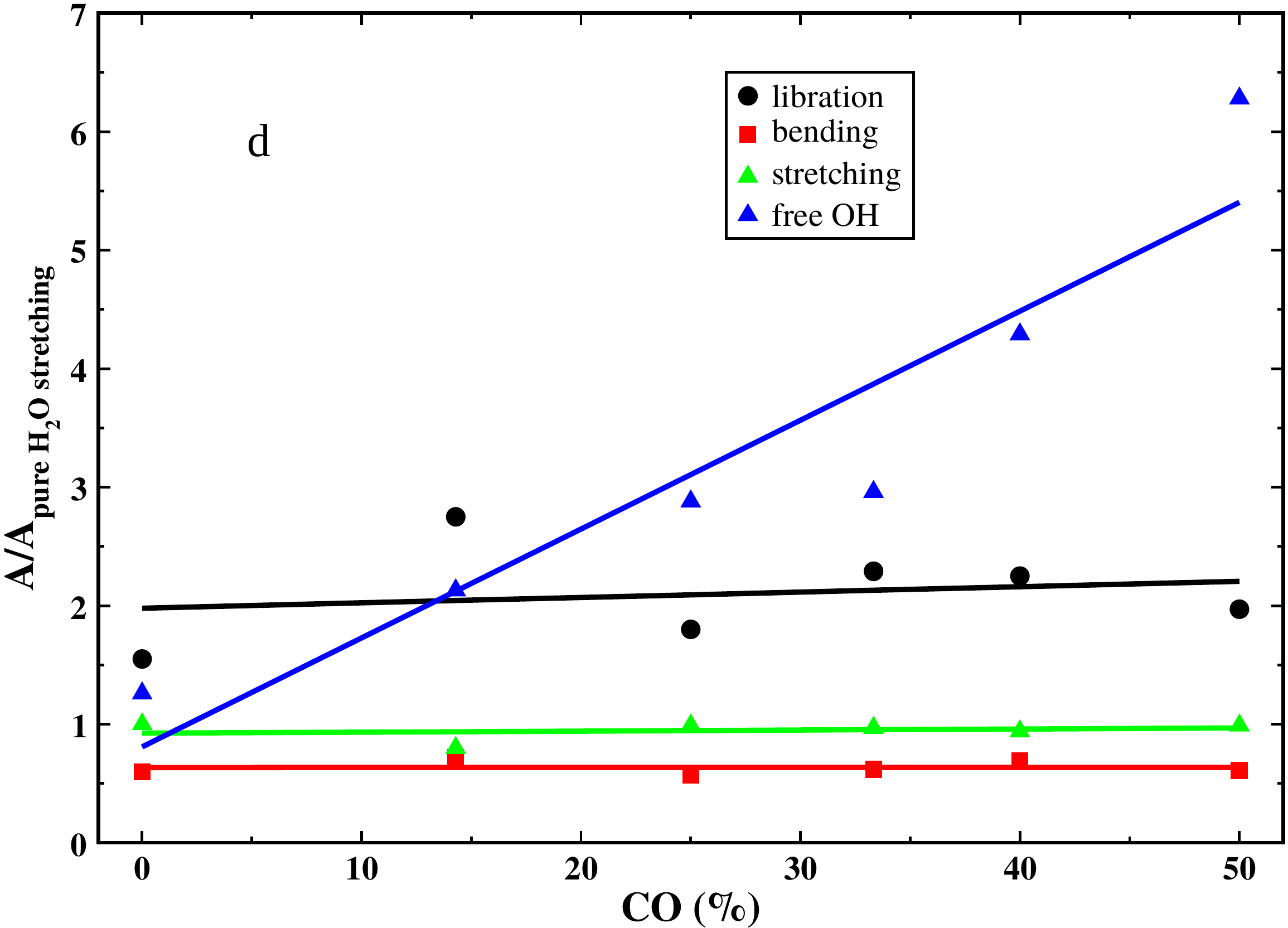}
\end{minipage}
\begin{minipage}{0.49\textwidth}
\includegraphics[width=\textwidth]{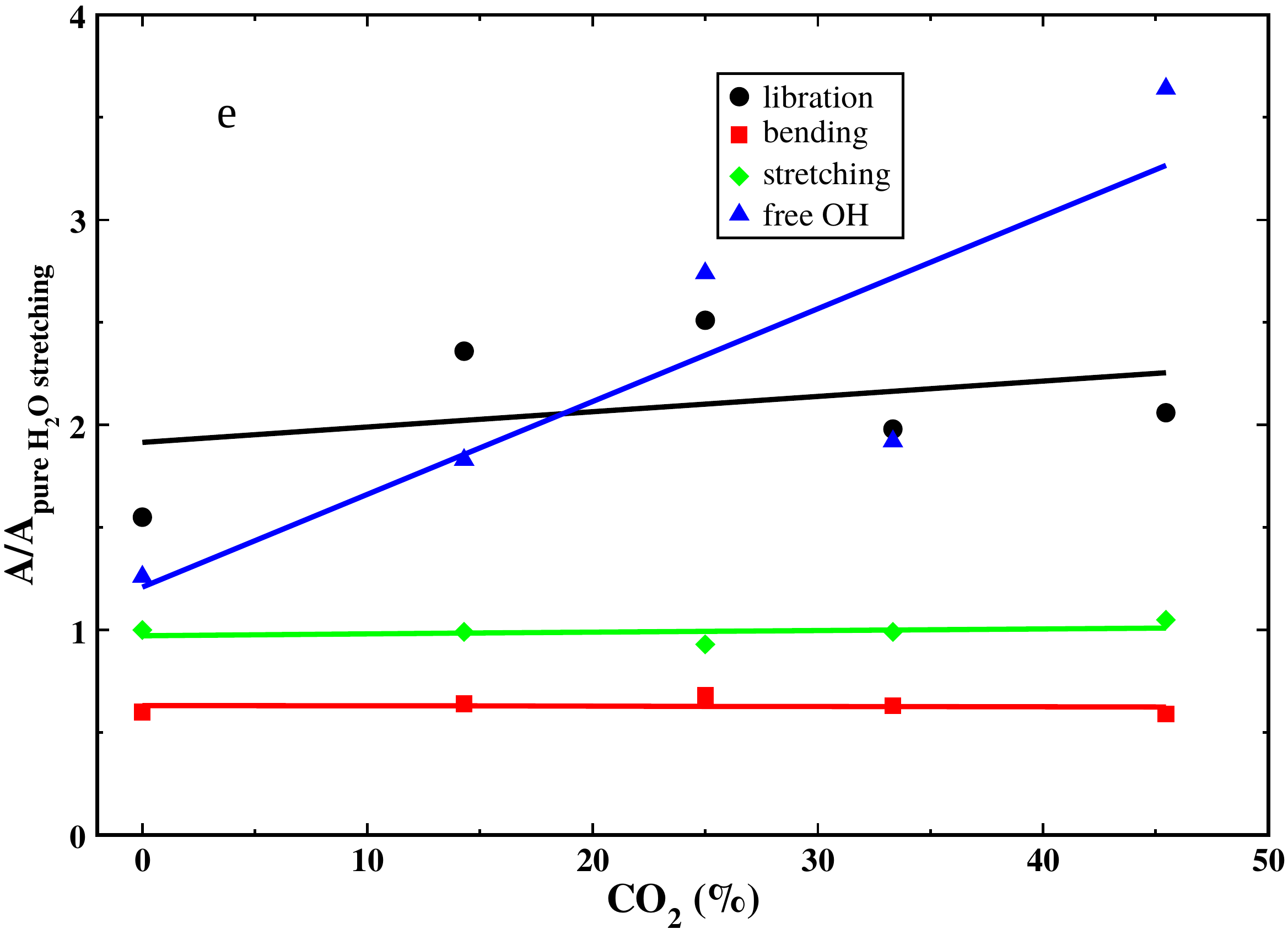}
\end{minipage}
\begin{minipage}{0.49\textwidth}
\includegraphics[width=\textwidth]{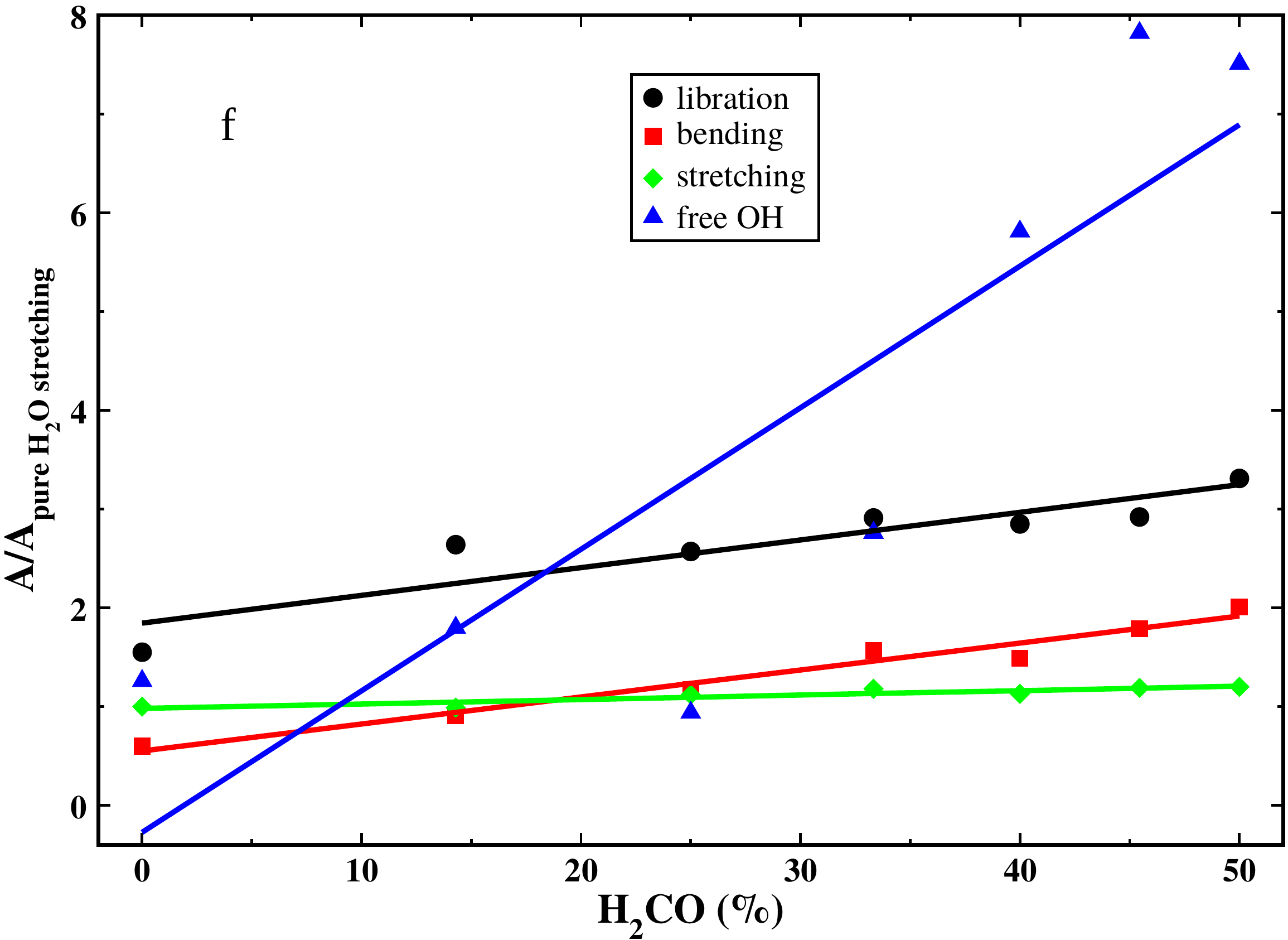}
\end{minipage}
\begin{minipage}{0.49\textwidth}
\includegraphics[width=\textwidth]{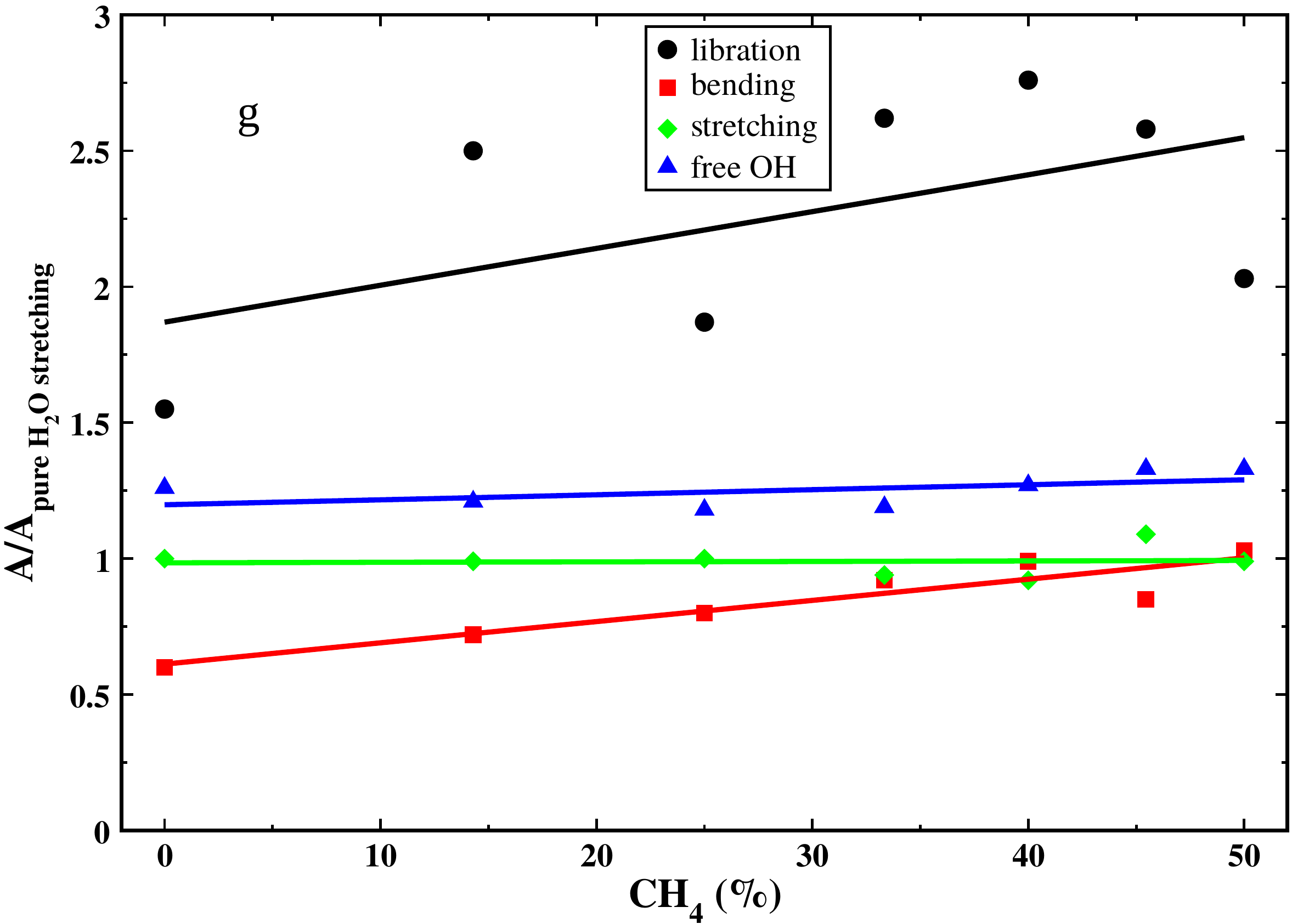}
\end{minipage}
\begin{minipage}{0.49\textwidth}
\includegraphics[width=\textwidth]{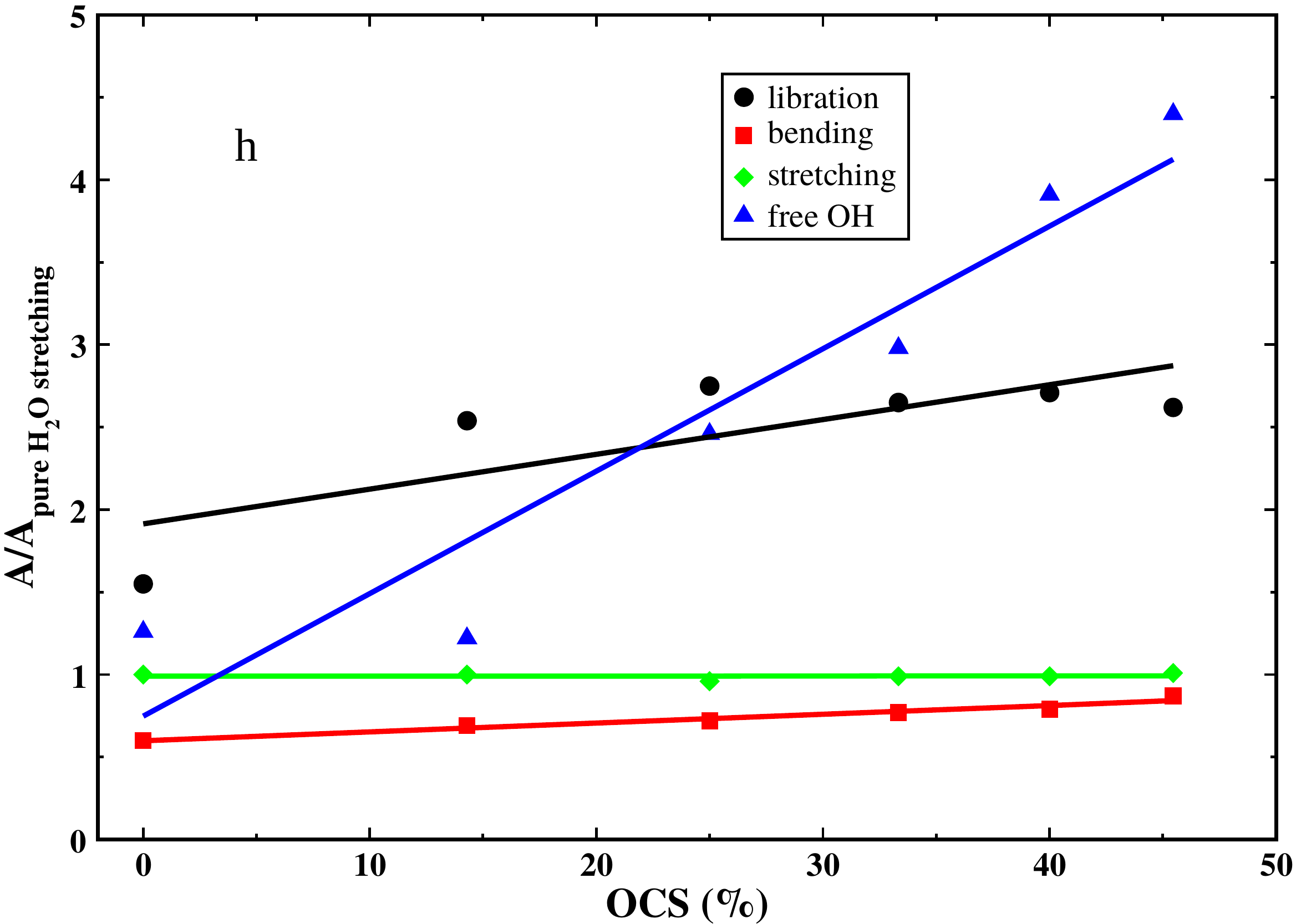}
\end{minipage}
\end{figure}
\begin{figure}
\begin{minipage}{0.49\textwidth}
\includegraphics[width=\textwidth]{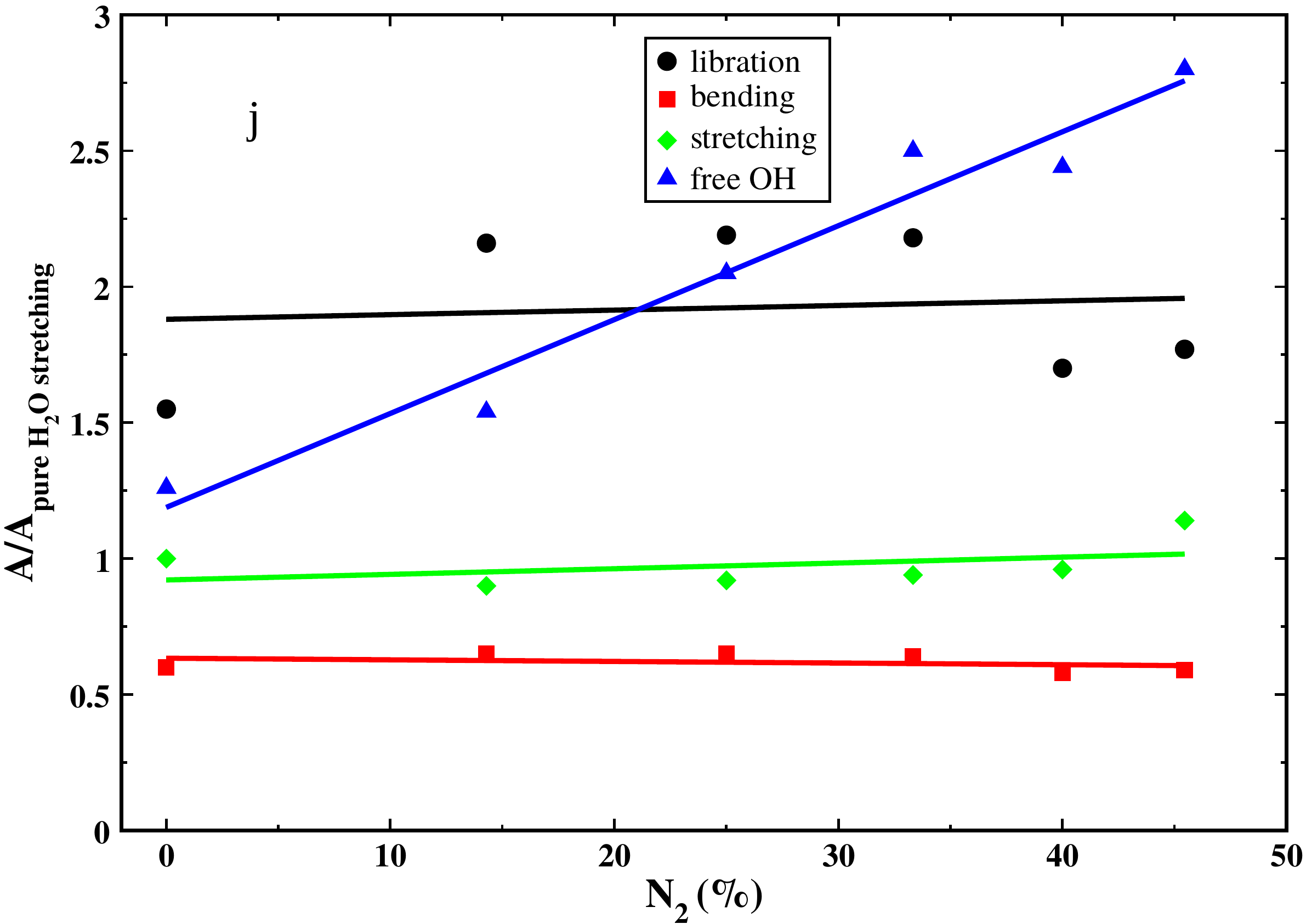}
\end{minipage}
\begin{minipage}{0.49\textwidth}
\includegraphics[width=\textwidth]{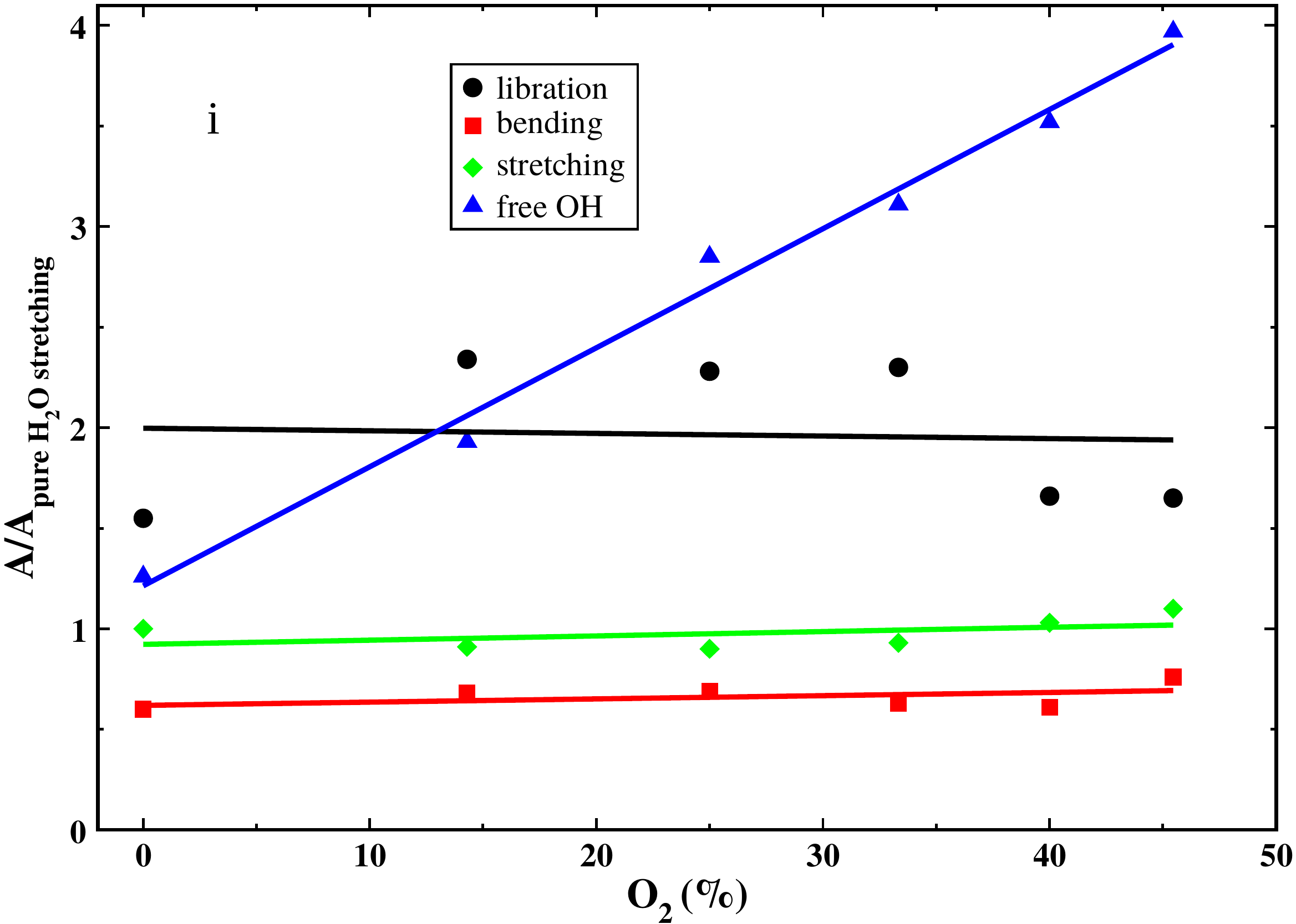}
\end{minipage}
\renewcommand{\thefigure}{S2}
\caption{Band strengths of the four fundamental vibration modes of water for 
(a) $\rm{H_2O-HCOOH}$, (b) $\rm{H_2O-NH_3}$, (c) $\rm{H_2O-CH_3OH}$, (d) $\rm{H_2O-CO}$, (e) $\rm{H_2O-CO_2}$,
(f) $\rm{H_2O-H_2CO}$, (g) $\rm{H_2O-CH_4}$, (h) $\rm{H_2O-OCS}$,
(i) $\rm{H_2O-N_2}$, and (j) $\rm{H_2O-O_2}$ clusters with various concentrations. The water c-hexamer (chair) 
configuration gas been used for pure water.}
\label{fig:6h2o_x_band_strength}
\end{figure}

\begin{figure}
\centering
\includegraphics[height=8cm,width=14cm]{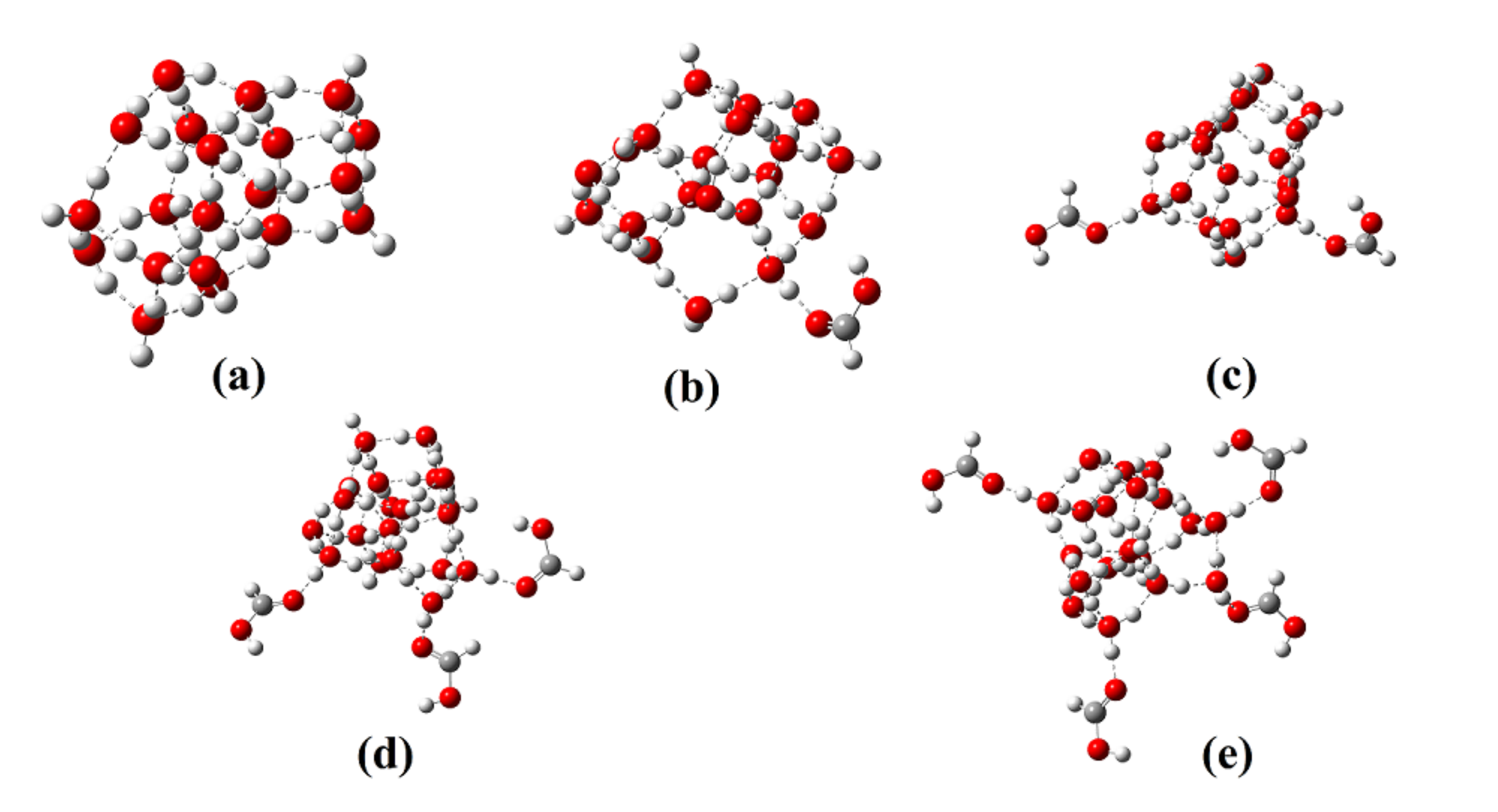}
\renewcommand{\thefigure}{S3}
\caption{Structure of water clusters containing 20 $\rm{H_2O}$ molecules with HCOOH as impurity in different concentration ratio:
(a) pure water, (b) $\rm{H_2O:HCOOH=20:1}$, (c) $\rm{H_2O:HCOOH=10:1}$, (d) $\rm{H_2O:HCOOH=6.67:1}$,
(e) $\rm{H_2O:HCOOH=5:1}$.}
\label{fig:20H2O-HCOOH}
\end{figure}

\begin{figure}
\centering
\includegraphics[height=7cm,width=10cm]{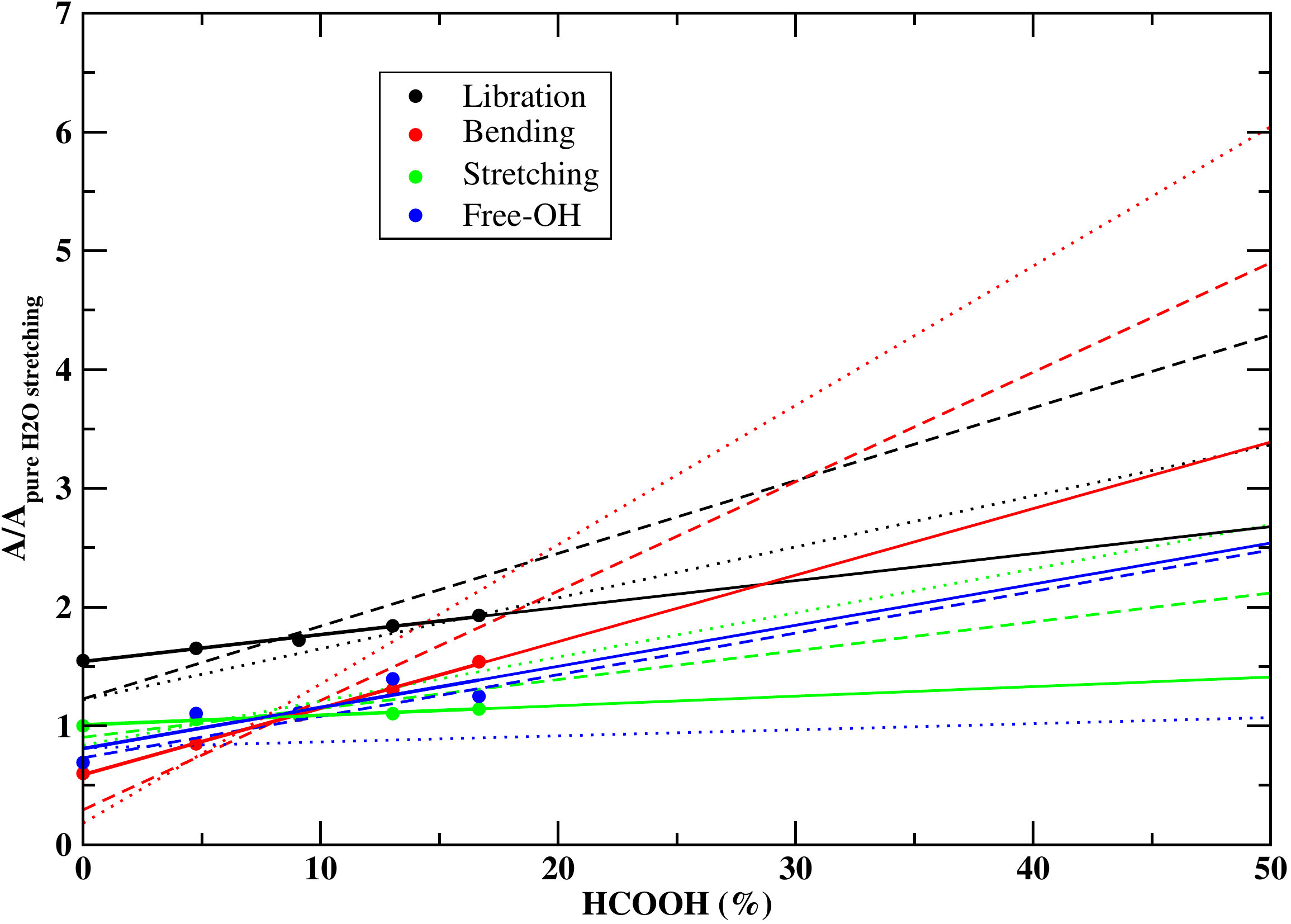}
\renewcommand{\thefigure}{S4}
\caption{Comparison of the band strength of the four fundamental vibrational modes for water clusters containing 20$\rm{H_2O}$, 6H$_2$O, and 4H$_2$O molecules with HCOOH as an impurity in different concentrations. Solid lines represent the band strength profiles for 20H$_2$O cluster, dotted 
lines for the water c-hexamer (chair) (6H$_2$O), and dashed lines for water c-tetramer (4H$_2$O).}
\label{fig:water_clusters-HCOOH}
\end{figure}

\begin{figure}
\vskip 1cm
\centering
\includegraphics[width=\textwidth]{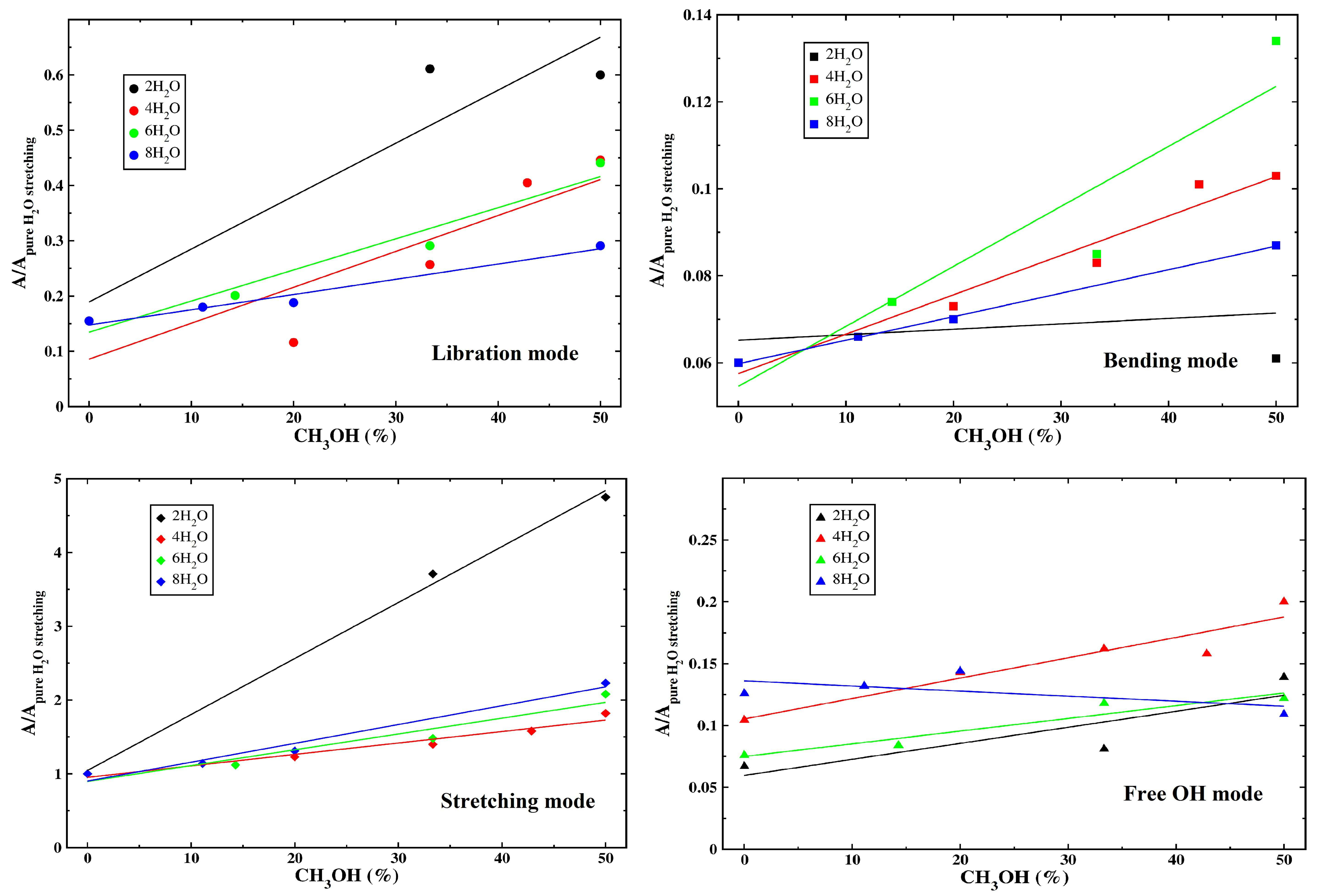}
\renewcommand{\thefigure}{S5}
\caption {Effect of the cluster size on the band strength profile.}
\label{fig:CH3OH_varied_cluster_size}
\end{figure}

\begin{twocolumn}

\begin{table}
\centering
\scriptsize{
\renewcommand{\thetable}{S1}
\caption{Harmonic infrared frequencies and intensities of the complex 4H$_2$O evaluated at the B2PLYP/maug-cc-pVTZ level.}
\label{tab:4H2O_B2PLYP}
\begin{tabular}{|c|c|c|}
\hline
{\bf NM} &$\omega$ {[\bf cm$^{-1}$]} & {\bf IR Int. [Km/mole]} \\
\hline
       1 &      39.3404 &       0.0000 \\
       2 &      85.5169 &       4.1241 \\
       3 &     202.1393 &       0.0000 \\
       4 &     222.9069 &     104.7500 \\
       5 &     222.9069 &     104.7500 \\
       6 &     231.0105 &      94.8938 \\
       7 &     240.0832 &       0.6734 \\
       8 &     246.4688 &     287.8694 \\
       9 &     246.4688 &     287.8694 \\
      10 &     274.1815 &       0.0000 \\
      11 &     398.6011 &       0.0000 \\
      12 &     427.5608 &      44.2017 \\
      13 &     443.1204 &      79.6160 \\
      14 &     443.1204 &      79.6160 \\
      15 &     683.4168 &     312.1539 \\
      16$^t$ &     778.3957 &     237.5051 \\
      17$^t$ &     778.3957 &     237.5051 \\
      18$^t$ &     934.3601 &       0.0000 \\
      19$^b$ &    1630.3695 &     176.7723 \\
      20$^b$ &    1647.3535 &      74.0296 \\
      21$^b$ &    1647.3536 &      74.0296 \\
      22$^b$ &    1672.6698 &       0.0000 \\
      23$^s$ &    3411.5900 &       0.0000 \\
      24$^s$ &    3477.3996 &    1793.2478 \\
      25$^s$ &    3477.3996 &    1793.2478 \\
      26$^s$ &    3522.3300 &       2.0149 \\
      27$^f$ &    3864.6871 &     144.7620 \\
      28$^f$ &    3865.9912 &     138.6232 \\
      29$^f$ &    3865.9912 &     138.6232 \\
      30$^f$ &    3867.1847 &       0.0000 \\
\hline
\end{tabular}} \\
{\bf Notes.} $^t$OH torsion; $^b$OH scissoring; $^s$OH stretching; $^f$free OH.
\end{table}

\begin{table}
\centering
\scriptsize{
\renewcommand{\thetable}{S2}
\caption{Harmonic infrared frequencies and intensities of the complex 4H$_2$O/HCOOH evaluated at the B2PLYP/maug-cc-pVTZ level.}
\label{tab:4H2O_1HCOOH_B2PLYP}
\begin{tabular}{|c|c|c|}
\hline
{\bf NM} &$\omega$ {[\bf cm$^{-1}$]} & {\bf IR Int. [Km/mole]} \\
\hline
       1 &       7.4652 &       2.8797 \\
       2 &      21.6554 &       0.4836 \\
       3 &      31.8296 &       0.8007 \\
       4 &      45.1056 &       0.3993 \\
       5 &      61.0883 &       3.2003 \\
       6 &      99.9662 &       7.7802 \\
       7 &     176.7899 &      26.3202 \\
\hline
\end{tabular}} \\
{\bf Notes.} $^t$OH torsion; $^b$OH scissoring; $^s$OH stretching; $^f$free OH; *vibrations contaminated by HCOOH modes.
\end{table}

\begin{table}
\centering
\scriptsize{
\renewcommand{\thetable}{S3}
\caption{Harmonic infrared frequencies and intensities of the complex 4H$_2$O/2HCOOH evaluated at the B2PLYP/maug-cc-pVTZ level.}
\label{tab:4H2O_2HCOOH_B2PLYP}
\begin{tabular}{|c|c|c|}
\hline
{\bf NM} &$\omega$ {[\bf cm$^{-1}$]} & {\bf IR Int. [Km/mole]} \\
\hline
       1 &       4.6054 &       0.4281 \\
       2 &      13.0063 &       2.9178 \\
       3 &      16.5189 &       0.8051 \\
       4 &      19.0314 &       2.3774 \\
       5 &      20.9593 &       2.6393 \\
       6 &      33.4976 &       0.4467 \\
       7 &      36.7713 &       1.8500 \\
\hline
\end{tabular}} \\
{\bf Notes.} $^t$OH torsion; $^b$OH scissoring; $^s$OH stretching; $^f$free OH; *vibrations contaminated by HCOOH modes.
\end{table}

\begin{table}
\centering
\scriptsize{
\renewcommand{\thetable}{S4}
\caption{Harmonic infrared frequencies and intensities of the complex 4H$_2$O/3HCOOH evaluated at the B2PLYP/maug-cc-pVTZ level.}
\label{tab:4H2O_3HCOOH_B2PLYP}
\begin{tabular}{|c|c|c|}
\hline
{\bf NM} &$\omega$ {[\bf cm$^{-1}$]} & {\bf IR Int. [Km/mole]} \\
\hline
       1 &       4.8748 &       0.5689 \\
       2 &       8.5054 &       0.7327 \\
       3 &      10.9091 &       1.7844 \\
       4 &      14.1204 &       2.2282 \\
       5 &      16.8372 &       1.6310 \\
       6 &      19.1432 &       3.8603 \\
       7 &      22.9673 &       0.5549 \\
\hline
\end{tabular}}
\end{table} 

\begin{table}
\centering
\scriptsize{
\begin{tabular}{|c|c|c|}
\hline
{\bf NM} &$\omega$ {[\bf cm$^{-1}$]} & {\bf IR Int. [Km/mole]} \\
\hline
   61 &    1741.9079 &     575.8421 \\
      62 &    3085.0941 &     261.1337 \\
      63 &    3085.5154 &      98.0625 \\
      64 &    3110.4187 &       7.6777 \\
      65$^{s*}$ &    3215.3096 &    2288.1686 \\
      66$^{s*}$ &    3232.6395 &    1833.2038 \\
      67$^{s*}$ &    3306.6741 &     311.8714 \\
      68$^{s}$ &    3451.6305 &     978.9670 \\
      69$^{s}$ &    3473.6093 &     894.3564 \\
      70$^{s}$ &    3609.3777 &     516.6451 \\
      71 &    3723.4615 &     111.6932 \\
      72$^{f}$ &    3838.2715 &     172.2543 \\
      73$^{f}$ &    3841.3048 &      97.0635 \\
      74$^{f}$ &    3857.9521 &     119.5655 \\
      75$^{f}$ &    3865.5487 &     121.3331 \\   
\hline
\end{tabular}} \\
{\bf Notes.} $^t$OH torsion; $^b$OH scissoring; $^s$OH stretching; $^f$free OH; *vibrations contaminated by HCOOH modes.
\end{table}      

\begin{table}
\centering
\scriptsize{
\renewcommand{\thetable}{S5}
\caption{Harmonic infrared frequencies and intensities of the complex 4H$_2$O/4HCOOH evaluated at the B2PLYP/maug-cc-pVTZ level.}
\label{tab:4H2O_4HCOOH_B2PLYP}
\begin{tabular}{|c|c|c|}
\hline
{\bf NM} &$\omega$ {[\bf cm$^{-1}$]} & {\bf IR Int. [Km/mole]} \\
\hline
       1 &       5.2521 &       0.5902 \\
       2 &       6.6193 &       1.2706 \\
       3 &       8.8369 &       0.8598 \\
       4 &       9.6756 &       1.0009 \\
       5 &      14.2363 &       3.0419 \\
       6 &      15.7986 &       1.1225 \\
       7 &      18.3559 &       3.0357 \\
       8 &      19.2707 &       2.7650 \\
       9 &      22.1555 &       0.9922 \\
      10 &      27.4779 &       3.1848 \\
      11 &      30.7377 &       0.4104 \\
      12 &      36.6940 &       0.3940 \\
      13 &      42.5989 &       1.4605 \\
      14 &      50.4443 &       2.9039 \\
      15 &      56.1885 &       2.0333 \\
      16 &      67.8597 &       3.6648 \\
      17 &      79.8514 &      11.6336 \\
      18 &      88.6848 &       0.8582 \\
      19 &     101.4052 &       1.0586 \\
      20 &     120.4063 &       2.7295 \\
      21 &     173.0431 &      53.0876 \\
      22 &     174.6165 &      79.9014 \\
      23 &     181.6174 &      16.9912 \\
      24 &     193.2247 &       8.7228 \\
      25 &     207.4059 &      98.2327 \\
      26 &     209.4436 &      56.3457 \\
      27 &     214.9922 &      15.9518 \\
      28 &     251.9141 &      48.3664 \\
      29 &     260.0519 &      27.0848 \\
      30 &     277.2414 &      21.6313 \\
      31 &     294.7764 &     138.3103 \\
      32 &     370.6813 &      53.2423 \\
      33 &     411.5523 &     157.3717 \\
      34 &     426.8010 &     160.0370 \\
      35 &     432.4841 &      46.3694 \\
      36 &     474.7455 &       4.1043 \\
      37 &     500.7327 &      25.1590 \\
      38 &     506.2341 &      58.6370 \\
      39 &     626.8822 &      79.1196 \\
      40 &     662.8116 &     197.7640 \\     
\hline
\end{tabular}}
\end{table}

\begin{table}
\centering
\scriptsize{
\begin{tabular}{|c|c|c|}
\hline
{\bf NM} &$\omega$ {[\bf cm$^{-1}$]} & {\bf IR Int. [Km/mole]} \\
\hline
      41$^{t*}$ &     680.2702 &     165.0245 \\
      42$^{t*}$ &     685.5338 &      74.2193 \\
      43$^{t*}$ &     685.8711 &     129.0114 \\
      44$^{t}$ &     703.1086 &     134.5453 \\
      45$^{t}$ &     777.4948 &     157.2018 \\
      46$^{t*}$ &     836.4350 &     145.9547 \\
      47 &     918.5370 &      51.5171 \\
\hline
\end{tabular}} \\
{\bf Notes.} $^t$OH torsion; $^b$OH scissoring; $^s$OH stretching; $^f$free OH; *vibrations contaminated by HCOOH modes. 
\end{table}

\begin{table}
\centering
\scriptsize{
\renewcommand{\thetable}{S6}
\caption{Harmonic infrared frequencies and intensities of the complex 4H$_2$O/NH$_3$ evaluated at the B2PLYP/maug-cc-pVTZ level.}
\label{tab:4H2O_1NH3_B2PLYP}
\begin{tabular}{|c|c|c|}
\hline
{\bf NM} &$\omega$ {[\bf cm$^{-1}$]} & {\bf IR Int. [Km/mole]} \\
\hline
       1     &      17.4723 &       1.6404 \\
       2     &      24.2573 &       2.7053 \\
       3     &      42.9718 &       9.2433 \\
       4     &      56.2828 &      14.2890 \\
       5     &      77.4675 &       5.3857 \\
       6     &     176.7386 &      14.5514 \\
       7     &     205.9991 &      61.5386 \\
\hline
\end{tabular}} \\
{\bf Notes.} $^t$OH torsion; $^b$OH scissoring; $^s$OH stretching; $^f$free OH; *vibrations contaminated by NH$_3$ modes.
\end{table}

\begin{table}
\centering
\scriptsize{
\renewcommand{\thetable}{S7}
\caption{Harmonic infrared frequencies and intensities of the complex 4H$_2$O/2NH$_3$ evaluated at the B2PLYP/maug-cc-pVTZ level.}
\label{tab:4H2O_2NH3_B2PLYP}
\begin{tabular}{|c|c|c|}
\hline
{\bf NM} &$\omega$ {[\bf cm$^{-1}$]} & {\bf IR Int. [Km/mole]} \\
\hline
       1 &      13.2137 &       3.5154 \\
       2 &      35.8031 &       4.5193 \\
       3 &      37.7152 &      15.3999 \\
       4 &      42.6399 &       0.6394 \\
       5 &      44.0908 &       1.0833 \\
       6 &      50.2282 &      25.2398 \\
       7 &      62.5104 &       1.7351 \\
\hline
\end{tabular}} \\
{\bf Notes.} $^t$OH torsion; $^b$OH scissoring; $^s$OH stretching; $^f$free OH; *vibrations contaminated by NH$_3$ modes.
\end{table}

\begin{table}
\centering
\scriptsize{
\renewcommand{\thetable}{S8}
\caption{Harmonic infrared frequencies and intensities of the complex 4H$_2$O/3NH$_3$ evaluated at the B2PLYP/maug-cc-pVTZ level.}
\label{tab:4H2O_3NH3_B2PLYP}
\begin{tabular}{|c|c|c|}
\hline
{\bf NM} &$\omega$ {[\bf cm$^{-1}$]} & {\bf IR Int. [Km/mole]} \\
\hline
       1 &      12.5470 &       2.6010 \\
       2 &      23.2847 &      18.0759 \\
       3 &      28.7571 &      23.4249 \\
       4 &      35.7710 &       0.3934 \\
       5 &      38.9667 &       0.0081 \\
       6 &      40.5045 &       3.8999 \\
       7 &      44.8244 &       0.2126 \\
\hline
\end{tabular}} \\
{\bf Notes.} $^t$OH torsion; $^b$OH scissoring; $^s$OH stretching; $^f$free OH; *vibrations contaminated by NH$_3$ modes. 
\end{table}

\begin{table}
\centering
\scriptsize{
\renewcommand{\thetable}{S9}
\caption{Harmonic infrared frequencies and intensities of the complex 4H$_2$O/4NH$_3$ evaluated at the B2PLYP/maug-cc-pVTZ level.}
\label{tab:4H2O_4NH3_B2PLYP}
\begin{tabular}{|c|c|c|}
\hline
{\bf NM} &$\omega$ {[\bf cm$^{-1}$]} & {\bf IR Int. [Km/mole]} \\
\hline
       1 &       4.9193 & 4.9769 \\
       2 &       9.0775 & 2.0942 \\
       3 &      17.7847 & 11.5017 \\
       4 &      19.1211 & 1.0599 \\
       5 &      22.4392 & 0.6988 \\
       6 &      25.9838 & 0.4032 \\
       7 &      27.6254 & 26.7443 \\
\hline
\end{tabular}}
\end{table}

\begin{table}
\centering
\scriptsize{
\begin{tabular}{|c|c|c|}
\hline
{\bf NM} &$\omega$ {[\bf cm$^{-1}$]} & {\bf IR Int. [Km/mole]} \\
\hline
      61$^s$ &    3387.6855 & 2801.0441 \\
      62$^s$ &    3394.0812 & 0.7615 \\
      63$^{s*}$ &    3473.3538 & 0.1108 \\
      64 &    3483.3149 & 13.0568 \\
      65 &    3483.4935 & 11.1778 \\
      66 &    3483.6384 & 2.6130 \\
      67$^{s*}$ &    3484.6015 & 0.0846 \\
      68$^{s}$ &    3507.3380 & 1567.4274 \\
      69$^{s}$ &    3510.2809 & 1820.7051 \\
      70$^{s}$ &    3536.4083 & 0.1714 \\
      71 &    3600.7487 & 17.4948 \\
      72 &    3600.7879 & 17.3727 \\
      73 &    3600.9757 & 17.5063 \\
      74 &    3601.0932 & 17.3475 \\
      75 &    3601.1747 & 17.5682 \\
      76 &    3601.2866 & 17.5694 \\
      77 &    3601.4175 & 17.2943 \\
      78 &    3601.5786 & 17.6734 \\
\hline
\end{tabular}} \\
{\bf Notes.} $^t$OH torsion; $^b$OH scissoring; $^s$OH stretching; $^f$free OH; *vibrations contaminated by NH$_3$ modes.  
\end{table}

\begin{table}
\centering
\scriptsize{
\renewcommand{\thetable}{S10}
\caption{Harmonic infrared frequencies and intensities of the H$_2$O$^{QM}$ + 3 H$_2$O$^{MM}$ complex (Figure 2, left panel).}
\label{tab:4H2O_MM-QM}
\label{tab:qmmm_1}
\begin{tabular}{|c|c|c|}
    \hline 
    {\bf NM} &$\omega$ {[\bf cm$^{-1}$]} & {\bf IR Int. [Km/mole]} \\
    \hline
    1 &      39.0097 &       3.5702 \\
    2 &      70.0478 &       0.9941 \\
    3 &     107.4593 &     248.8304 \\
    4 &     114.6669 &      81.5288 \\
    5 &     134.1712 &     121.9418 \\
    6 &     251.1220 &      19.7407 \\
    7 &     253.1088 &      18.8069 \\
    8 &     270.4875 &       7.7664 \\
    9 &     286.0773 &      10.8931 \\
    10 &     331.3394 &     156.5948 \\
    11 &     417.2017 &      11.4416 \\
    12 &     483.9590 &     107.2607 \\
    13$^t$ &     502.7496 &     211.3432 \\
    14$^t$ &     539.1224 &      59.4593 \\
    15$^t$ &     620.5928 &       2.9662 \\
    16$^t$ &     670.5374 &     209.1029 \\
    17$^t$ &     734.0061 &     476.6369 \\
    18$^t$ &     870.1429 &     201.9038 \\
    19$^b$ &    1407.2265 &     191.4554 \\
    20$^b$ &    1422.8362 &     445.8394 \\
    21$^b$ &    1456.6860 &      64.4541 \\
    22$^b$ &    1764.1323 &     143.3391 \\
    23$^s$ &    3399.8219 &     161.5281 \\
    24$^s$ &    3533.5836 &     119.6470 \\
    25$^s$ &    3561.1437 &     374.8984 \\
    26$^s$ &    3579.9423 &     110.7922 \\
    27$^f$ &    3643.1130 &      84.3280 \\
    28$^f$ &    3647.6874 &     450.2363 \\
    29$^f$ &    3651.3875 &      90.6885 \\
    30$^f$ &    3797.0095 &      37.5965 \\
\hline
\end{tabular}} \\
{\bf Notes.} $^t$OH torsion; $^b$OH scissoring; $^s$OH stretching; $^f$free OH
\end{table}

\begin{table}
\centering
\scriptsize{
\renewcommand{\thetable}{S11}
\caption{Harmonic infrared frequencies and intensities of the 
first 4 H$_2$O$^{QM}$+16 H$_2$O$^{MM}$ complex (see Figure 2a in the manuscript, right panel,  Configuration 1) 
evaluated at the B3LYP/6-31g(d) level.}
\label{tab:20H2O_MM-QM_Conf_1}
\begin{tabular}{|c|c|c|}
\hline
{\bf NM} &$\omega$ {[\bf cm$^{-1}$]} & {\bf IR Int. [Km/mole]} \\
\hline  
            1 &    28.5053 &     1.6212\\
            2 &    32.4112 &     0.3452\\
            3 &    40.5682 &     1.9422\\
            4 &    42.6937 &     0.9658\\
            5 &    45.5125 &     2.8291\\
            6 &    47.0457 &     2.4796\\
            7 &    51.1545 &     1.5206\\
\hline
\end{tabular}}
\end{table}

\begin{table}
\centering
\scriptsize{
\begin{tabular}{|c|c|c|}
\hline
{\bf NM} &$\omega$ {[\bf cm$^{-1}$]} & {\bf IR Int. [Km/mole]} \\
\hline
            61 &   387.1876 &     8.8358\\
            62 &   426.8869 &    78.1914\\
            63 &   432.7664 &    47.4011\\
            64 &   449.0251 &    78.8810\\
            65 &   455.8202 &     9.0749\\
            66 &   458.4089 &    52.9014\\
            67 &   461.3395 &     7.1542\\
\hline
\end{tabular}}
\end{table}

\begin{table}
\centering
\scriptsize{
\begin{tabular}{|c|c|c|}
\hline
{\bf NM} &$\omega$ {[\bf cm$^{-1}$]} & {\bf IR Int. [Km/mole]} \\
\hline           
    
    119$^{b}$ &  1415.0050 &   217.1485\\
    120$^{b}$ &  1416.7673 &   553.3807\\
    121$^{b}$ &  1423.8136 &   320.7421\\
    122$^{b}$ &  1427.0557 &   239.2493\\
    123$^{b}$ &  1431.3527 &   280.4839\\
    124$^{b}$ &  1433.6066 &   107.3301\\
    125$^{b}$ &  1436.1459 &   143.6834\\
\hline
\end{tabular}} \\
{\bf Notes.} $^t$OH torsion; $^b$OH scissoring; $^s$OH stretching; $^f$free OH.  
\end{table}

\begin{table}
\centering
\scriptsize{
\renewcommand{\thetable}{S12}
\caption{Harmonic infrared frequencies and intensities of the 
first 4 H$_2$O$^{QM}$+16 H$_2$O$^{MM}$ complex (Figure 2b, right panel,  Configuration 2) evaluated at the B3LYP/6-31g(d) level.}
\label{tab:20H2O_MM-QM_Conf_2}
\begin{tabular}{|c|c|c|}
\hline
{\bf NM} &$\omega$ {[\bf cm$^{-1}$]} & {\bf IR Int. [Km/mole]} \\
\hline
         1 &    29.4016 &     1.2924\\
         2 &    34.4375 &     0.6290\\
         3 &    40.5656 &     1.6279\\
         4 &    44.5388 &     1.2326\\
         5 &    47.5478 &     0.5182\\
         6 &    49.2499 &     4.4372\\
         7 &    52.7161 &     1.1206\\
\hline
\end{tabular}}
\end{table}

\begin{table}
\centering
\scriptsize{
\begin{tabular}{|c|c|c|}
\hline
{\bf NM} &$\omega$ {[\bf cm$^{-1}$]} & {\bf IR Int. [Km/mole]} \\
\hline
        
        61 &   381.8881 &     2.8750\\
        62 &   425.0003 &    72.9104\\
        63 &   433.1753 &    46.5032\\
        64 &   449.7627 &    54.6515\\
        65 &   453.1267 &    32.8363\\
        66 &   454.2652 &    31.3116\\
        67 &   457.6722 &    30.6715\\
\hline
\end{tabular}}
\end{table}
\clearpage
\begin{table}
\centering
\scriptsize{
\begin{tabular}{|c|c|c|}
\hline
{\bf NM} &$\omega$ {[\bf cm$^{-1}$]} & {\bf IR Int. [Km/mole]} \\
\hline
 126$^{b}$ &  1444.1623 &   105.5626\\
 127$^{b}$ &  1459.2322 &    55.3003\\
 128$^{b}$ &  1465.8496 &   252.0918\\
 129$^{b}$ &  1482.2478 &   140.3077\\
 130$^{b}$ &  1505.1507 &    42.7361\\
 131$^{b}$ &  1775.8533 &    82.3784\\
 132$^{b}$ &  1780.7273 &    68.7098\\
\hline
\end{tabular}} \\
{\bf Notes.} $^t$OH torsion; $^b$OH scissoring; $^s$OH stretching; $^f$free OH.  
\end{table}
\end{twocolumn}
\end{suppinfo}


\begin{thebibliography}{99}


\bibitem[\protect\citeauthoryear{Eddington}{1937}]{eddi37}
Eddington, A. S. Interstellar matter. {\it The Observatory} {\bf 1937}, {\it 60}, 99-103.

\bibitem[\protect\citeauthoryear{Tielens and Hagen}{1982}]{tiel82}
Tielens, A. G. G. M.; Hagen, W. Model calculations of the molecular composition of interstellar grain mantles. {\it Astron. Astrophys.} {\bf 1982}, {\it 114}, 245.

\bibitem[\protect\citeauthoryear{Woon}{2002}]{woon02}
Woon, D. Pathways to glycine and other amino acids in ultraviolet-irradiated astrophysical
ices determined via quantum chemical modeling. {\it Astrophys. J. Lett.} {\bf 2002}, {\it 571}, L177.

\bibitem[\protect\citeauthoryear{Nuevo et al.}{2014}]{nuev14}
Nuevo, M.; Sandford, S.A. The photochemistry of pyrimidine in realistic astrophysical ices
and the production of nucleobases. {\it Astrophys. J.} {\bf 2014}, {\it 793}, 125.

\bibitem[\protect\citeauthoryear{Das et al.}{2008}]{das08}
Das, A.; Acharyya, K.; Chakrabarti, S.; Chakrabarti, S. K. Formation of water and methanol in star forming molecular clouds. {\it Astron. Astrophys.} {\bf 2008}, {\it 486}, 209-220.

\bibitem[\protect\citeauthoryear{Das et al.}{2010}]{das10}
Das, A.; Acharyya, K; Chakrabarti, S. K. Effects of initial condition and cloud density on the composition of the grain mantle. {\it {Mon. Not. R. Astron. Soc.}} {\bf 2010}, {\it 409}, 789-800.

\bibitem[\protect\citeauthoryear{Das and Chakrabarti}{2011}]{das11}
Das, A.; Chakrabarti, S. K. Composition and evolution of interstellar grain mantle under the effects of photodissociation.
{\it {Mon. Not. R. Astron. Soc.}} {\bf 2011}, {\it 418}, 545-555.

\bibitem[\protect\citeauthoryear{Das et al.}{2016}]{das16}
Das, A.; Sahu, D.; Majumdar, L.; Chakrabarti, S. K. Deuterium enrichment of the interstellar grain mantle. {\it {Mon. Not. R. Astron. Soc.}} {\bf 2016}, {\it 455}, 540-551.

\bibitem[\protect\citeauthoryear{Gibb et al.}{2004}]{gibb04}
Gibb, E. L.; Whittet, D. C. B.; Boogert, A. C. A.; Tielens, A. G. G. M. Interstellar ice: the infrared space observatory legacy. \textit {Astrophys. J., Suppl. Ser.} {\bf 2004}, {\it 151}, 35.

{
\bibitem[\protect\citeauthoryear{Whittet}{2003}]{whit03}
Whittet, D. C. B. {\it Dust in the Galactic Environment}; 2nd ed.; Bristol: Institute of Physics (IOP) Publishing, 2003; Series in Astronomy and Astrophysics.}

\bibitem[\protect\citeauthoryear{Gillett and Forrest}{1973}]{gill73}
Gillett. F. C.; Forrest. W. J. Spectra of the Becklin-Neugebauer point source and the Kleinmann-Low nebula from 2.8 to 13.5 microns. \textit {Astrophys. J.} {\bf 1973}, {\it 179}, 483-491.

\bibitem[\protect\citeauthoryear{Irvine and Pollack}{1968}]{irvi68}
Irvine, W. M.; Pollack, J. B. Infrared optical properties of water and ice spheres. \textit {Icarus} {\bf 1968}, {\it 8}, 324-360.

\bibitem[\protect\citeauthoryear{Merrill et al.}{1976}]{merr76}
Merrill, K. M.; Russell, R. W.; Soifer, B. T.  Infrared observations of ices and silicates in molecular clouds. \textit {Astrophys. J.} {\bf 1976}, {\it 207}, 763-769.

\bibitem[\protect\citeauthoryear{Leger et al.}{1979}]{lege79}
Leger, A.; Klein, J.; de Cheveigne, S.; Guinet, C.; Defourneau, D.; Belin, M. {\it Astron. Astrophys.} {\bf 1979}, {\it 79}, 256-259.

\bibitem[\protect\citeauthoryear{Hagen et al.}{1979}]{hage79}
Hagen, W.; Allamandola, L.; Greenberg, J. M. Interstellar molecule formation in grain mantles:
The laboratory analog experiments, results and implications. {\it {Astrophys. Space Sci.}} {\bf 1979}, {\it 65}, 215-240.

{
\bibitem[\protect\citeauthoryear{Dartois}{2005}]{dart05}
Dartois, E. {\it The ice survey opportunity of ISO;} In ISO Science Legacy. Springer: Dordrecht, {\bf 2005}, {\it 119}, 293-310.}

{
\bibitem[\protect\citeauthoryear{van Dishoeck et al.}{1993}]{vand93}
van Dishoeck, E. F.; Blake, C. A.; Draine B. T.; Lunine, J. I.
The Chemical Evolution of Protostellar and Protoplanetary Matter. In {\it Protostars and Planets III}, Levy, E. H., Lunine, J. I., Eds.;
ISBN 0-8165-1334-1. LC QB806 .P77; University of Arizona Press: Tucson, Arizona, 1993; 163-241.}

\bibitem[\protect\citeauthoryear{Boogert et al.}{2015}]{boog15}
Boogert, A. C. Adwin; Gerakines, Perry A.; Whittet, Douglas C. B. Observations of the Icy Universe. {\it Annual Review of Astronomy and Astrophysics} {\bf 2015}, {\it 53}, 541-581.

\bibitem[\protect\citeauthoryear{Ohno et al.}{2005}]{ohno05}
Ohno, K.; Okimura, M.; Akaib, N.; Katsumotoa, Y. The effect of cooperative hydrogen bonding on the OH stretching-band shift for water clusters studied by matrix-isolation infrared spectroscopy and density functional theory. {\it Phys. Chem. Chem. Phys.} {\bf 2005}, {\it 7}, 3005-3014.

\bibitem[\protect\citeauthoryear{Bouwman et al.}{2007}]{bouw07}
Bouwman, J.; Ludwig, W.; Awad, Z.; \"{O}berg, K.L.; Fuchs, G. W.; van Dishoeck, E. F.; Linnartz, H. Band profiles and band strengths in mixed
H$_2$O:CO ices. {\it { Astron. Astrophys.}} {\bf 2007}, {\it 476}, 995.

\bibitem[\protect\citeauthoryear{\"{O}berg et al.}{2007}]{ober07}
\"{O}berg, K. I.; Fraser, H.J.; Boogert, A. C. A.; Bisschop, S. E.; Fuchs, G. W.; van Dishoeck, E. F.; Linnartz, H.
Effects of CO$_2$ on H$_2$O band profiles and band strengths in mixed H$_2$O:CO$_2$ ices. {\it { Astron. Astrophys.}} {\bf 2007}, {\it 462}, 1187.


\bibitem[\protect\citeauthoryear{Gerakines et al.}{1995}]{gera95}
Gerakines, P. A.; Schutte, W. A.; Greenberg, J. M.; van Dishoeck, E. F. The infrared band strengths of H$_2$O, CO and CO$_2$ in laboratory
simulations of astrophysical ice mixtures. {\it { Astron. Astrophys.}} {\bf 1995}, {\it 296}, 810.

\bibitem[\protect\citeauthoryear{Ehrenfreund et al.}{1997}]{ehre97}
Ehrenfreund, P.; Boogert, A. C. A.; Gerakines, P. A.; Tielens, A. G. G. M.; van Dishoeck, E. F.
Infrared spectroscopy of interstellar apolar ice analogs. {\it Astron. Astrophys} {\bf 1997}, {\it 328}, 649-669.

\bibitem[\protect\citeauthoryear{Schutte et al.}{1999}]{schu99}
Schutte, W. A.; Boogert, A. C. A.; Tielens, A. G. G. M.; Whittet, D. C. B.; Gerakines, P. A.; Chiar, J. E.; Ehrenfreund, P.; Greenberg, J. M.; van Dishoeck, E. F.; Graauw, T. Weak ice absorption features at 7.24 and 7.41 $\mu$m in the spectrum
of the obscured young stellar object W 33A. {\it Astron. Astrophys} {\bf 1999}, {\it 343}, 966.

\bibitem[\protect\citeauthoryear{Cooke et al.}{2016}]{cook16}
Cooke, I. R.; Fayolle, E.C.; \"{O}berg, K. I. CO$_2$ Infrared Phonon Modes in Interstellar Ice Mixtures. {\it Astrophys. J.} {\bf 2016}, {\it 832}, 5.

\bibitem[\protect\citeauthoryear{Soifer et al.}{1979}]{soif79}
Soifer, B. T.; Puetter, R. C.; Russell, R. W.; Willner, S. P.; Harvey, P. M.; Gillett, F. C.
The 4-8 micron spectrum of the infrared source W33 A. {\it Astrophys. J. Lett.} {\bf 1979} {\it 232}, L53-L57.

\bibitem[\protect\citeauthoryear{Mantz et al.}{1975}]{mant75}
Mantz, A. W.; Maillard, J. P.; Roh, W. B.; Narahari Rao, K. Ground state molecular constants of $\rm{^{12}C^{16}O}$.
{\it { J. Mol. Spectrosc.}} {\bf 1975}, {\it 57}, 155.

{
\bibitem[\protect\citeauthoryear{Chiar et al.}{1995}]{chia95}
Chiar, J. E.; Whittet, D. C. B.; Adamson, A. J.; Kerr, T. H. Ices in the Taurus dark cloud environment. In {\it From Gas to Stars to Dust}; Haas, M. R., Davidson, J. A., Erickson, E. F., Eds.; Astronomical Society of the Pacific Conference Series, 1995; Vol. 73, pp 75-78.}

\bibitem[\protect\citeauthoryear{Chiar et al.}{1998}]{chia98}
Chiar, J. E.; Gerakines, P. A.; Whittet, D. C. B.; Pendleton, Y.; Tielens, A. G. G. M.;
Adamson, A. J.; Boogert, A. C. A. Processing of icy mantles in protostellar envelopes.
{\it Astrophys. J.} {\bf 1998}, {\it 498}, 716.

\bibitem[\protect\citeauthoryear{D'Hendecourt and Jourdain de Muizon}{1989}]{dhen89}
D'Hendecourt, L.B.; Jourdain de Muizon, M. The discovery of interstellar carbon dioxide.
{\it Astron. Astrophys.} {\bf 1989}, {\it 223}, L5.

\bibitem[\protect\citeauthoryear{de Graauw et al.}{1996}]{degr96}
de Graauw, T., Whittet, D.C.B., Gerakines, P.A., et al. SWS observations of solid $\rm{CO_2}$ in molecular clouds. {\it Astron. Astrophys.} {\bf 1996}, {\it 315}, L345.

\bibitem[\protect\citeauthoryear{Guertler et al.}{1996}]{guer96}
Guertler, J.; Henning, T.; Koempe, C.; Pfau, W.; Kraetschmer, W.; Lemke, D.
Detection of solid $\rm{CO_2}$ towards young stellar objects. {\it Astronomische Gesellschaft Abstract Series} {\bf 1996}, {\it 12}, 107.

\bibitem[\protect\citeauthoryear{Gerakines et al.}{1999}]{gera99}
Gerakines, P. A.; Whittet, D. C. B.; Ehrenfreund, P.; Boogert, A. C. A.; Tielens, A. G. G. M.; Schutte, W. A.; Chiar, J.E.; van Dishoeck, E. F.; Prusti, T.; Helmich, F. P.; De Graauw, T. Observations of solid carbon dioxide in molecular clouds with the infrared space observatory. {\it Astrophys. J.} {\bf 1999}, {\it 522}, 357.

\bibitem[\protect\citeauthoryear{Gibb et al.}{2000}]{gibb00}
Gibb, E. L.; Whittet, D. C. B.; Schutte, W. 8.; Boogert, A. C. A.; Chiar, J. E.; Ehrenfreund, P.; Gerakines, P. A.; Keane, J. V.; Tielens, A. G. G. M.; van Dishoeck, E. F.; Kerkhof, O.
An inventory of interstellar ices toward the embedded protostar W33A. {\it Astrophys. J.} {\bf 2000}, {\it 539}, 347.

\bibitem[\protect\citeauthoryear[Baas et al. (1988)]{baas88}
Baas, F.; Grim, R. J. A.; Geballe, T. R.; Schutte, W.; Greenberg, J. M. {\it The detection of solid methanol in W33A}; Dust in the Universe, 1988; 55-60.

\bibitem[\protect\citeauthoryear{Grim et al.}{1991}]{grim91}
Grim, R. J. A.; Bass, F.; Geballe, T. R.; Greenberg, J. M.; Schutte, W. Detection of solid methanol toward W33A. {\it Astron. Astrophys.} {\bf 1991}, {\it 243}, 473.

\bibitem[\protect\citeauthoryear{Allamandola et al.}{1992}]{alla92}
Allamandola, L. J.; Sandford, S. A.; Tielens, A. G. G. M.; Herbst, T. M.
Infrared spectroscopy of dense clouds in the CH stretch region-Methanol and `diamonds'. {\it Astrophys. J.} {\bf 1992}, {\it 399}, 134-146.

\bibitem[\protect\citeauthoryear{Schutte et al.}{1996}]{schu96}
Schutte, W. A.; Gerakines, P. A.; Geballe, T. R.; van Dishoeck, E. F.; Greenberg, J. M.
Discovery of solid formaldehyde toward the protostar GL 2136: observations and laboratory simulation. {\it Astron. Astrophys.} {\bf 1996}, {\it 309}, 633.

\bibitem[\protect\citeauthoryear{Kessler et al.}{1996}]{kess96}
Kessler, M. F.; Steinz, J. A.; Anderegg, M. E.; Clavel, J.; Drechsel, G.;
Estaria, P.; Faelker, J.; Riedinger, J. R.; Robson, A.; Taylor, B. G.; Xim\'{e}nez de Ferr\'{a}n, S. The Infrared Space Observatory (ISO)
mission. {\it Astron. Astrophys.} {\bf 1996}, {\it 315}, L27-L31.

\bibitem[\protect\citeauthoryear{Kessler et al.}{2003}]{kess03}
Kessler, M. F.; Mueller, T. G.; Leech, K.; Arviset, C.; Garcia-Lario, P.; Metcalfe, L.; Pollock, A.; Prusti, T.; Salama, A. In {\it The ISO Handbook,
Volume I: ISO - Mission \& Satellite Overview}, Version 2.0; Mueller, T. G., Blommaert, J. A. D. L., Garcia-Lario, P., Eds.; ESA SP-1262, ISBN No. 92-9092-968-5, ISSN 0379-6566; Eur. Space Agency, 2003; 92.

\bibitem[\protect\citeauthoryear{Keane et al.}{2001}]{kean01}
Keane, J. V.; Boogert, A. C. A.; Tielens, A. G. G. M.; Ehrenfreund, P.; Schutte, W. A.
Bands of solid $\rm{CO_2}$ in the 2-3 $\mu$m spectrum of S 140: IRS1. {\it Astron. Astrophys.} {\bf 2001}, {\it 375}, L43-L46.

\bibitem[\protect\citeauthoryear{van Dishoeck et al.}{1995}]{vand95}
van Dishoeck, E. F.; Blake, G. A.; Jansen, D. J.; Groesbeck, T. D.
Molecular abundances and low mass star formation II. Organic and deuterated species towards IRAS 16293-2422. {\it Astrophys. J.} {\bf 1995}, {\it 447}, 760-782.

\bibitem[\protect\citeauthoryear{Ikeda et al.}{2001}]{iked01}
Ikeda, M.; Ohishi, M.; Nummelin, A.; Dickens, J. E.; Bergman, P.; Hjalmarson, $\mathring{A}$.; Irvine, W. M. Survey observations of $\rm{c-C_2H_4O}$ and $\rm{CH_3CHO}$ toward massive star-forming regions. {\it Astrophys. J.} {\bf 2001}, {\it 560}, 792.

\bibitem[\protect\citeauthoryear{Lacy et al.}{1991}]{lacy91}
Lacy, J. H.; Carr, J. S.; Evans, N. J.; Baas, F.; Achtermann, J. M.; Arens, J. F.
{\it Astrophys. J.} {\bf 1991}, {\it 376}, 556-560.

\bibitem[\protect\citeauthoryear{\"{O}berg et al.}{2008}]{ober08}
\"{O}berg, K. I.; Boogert, A. C. A.; Pontoppidan, K. M.; Blake, G. A.; Evans, N. J.; Lahuis, F.; van Dishoeck, E. F. The c2d spitzer spectroscopic survey of ices around low-mass young stellar objects. III. $\rm{CH_4}$. {\it Astrophys. J.} {\bf 2008}, {\it 678}, 1032.

\bibitem[\protect\citeauthoryear{Knacke et al.}{1982}]{knac82}
Knacke, R. F.; McCorkle, S.; Puetter, R. C.; Erickson, E. F.; Kraetschmer, W.
Observation of interstellar ammonia ice. {\it Astrophys. J.} {\bf 1982}, {\it 260}, 141-146.

\bibitem[\protect\citeauthoryear{Knacke and McCorkle}{1987}]{knac87}
Knacke, R. F.; McCorkle, S.M. Spectroscopy of the Kleinmann-Low nebula-Scattering in a solid absorption band. {\it Astron. J.} {\bf 1987}, {\it 94}, 972-976.

\bibitem[\protect\citeauthoryear{Lacy et al.}{1998}]{lacy98}
Lacy, J. H.; Faraji, H.; Sandford, S. A.; Allamandola, L. J.
Unraveling the 10 micron ``silicate'' feature of protostars: the detection of frozen interstellar ammonia. {\it Astrophys. J. Lett.} {\bf 1998}, {\it 501}, L105.

\bibitem[\protect\citeauthoryear{Palumbo et al.}{1995}]{palu95}
Palumbo, M. E.; Tielens, A. G. G. M.; Tokunaga, A. T. Solid carbonyl sulphide (OCS) in W33A.
{\it Astrophys. J.} {\bf 1995}, {\it 449}, 674-680.

\bibitem[\protect\citeauthoryear{Palumbo et al.}{1997}]{palu97}
Palumbo, M. E.; Geballe T. R.; Tielens, A. G. G. M.
Solid carbonyl sulfide (OCS) in dense molecular clouds.
{\it Astrophys. J.} {\bf 1997}, {\it 479}, 839.

\bibitem[\protect\citeauthoryear{Bieler et al.}{2015}]{biel15}
Bieler, A.; Altwegg, K.; Balsiger, H.; Bar-Nun, A.; Berthelier, J. J.; Bochsler, P.; Briois, C.; Calmonte, U.; Combi, M.; De Keyser, J.; van Dishoeck, E. F. Abundant molecular oxygen in the coma of comet 67P/Churyumov–Gerasimenko.
{\it Nature} {\bf 2015}, {\it 526}, 678.

\bibitem[\protect\citeauthoryear{Rubin et al.}{2015}]{rubi15}
Rubin, M.; Altwegg, K.; van Dishoeck, E. F.; Schwehm, G.
 Molecular oxygen in Oort cloud comet 1P/Halley.
{\it Astrophys. J. Lett.} {\bf 2015}, {\it 815}, L11.

\bibitem[\protect\citeauthoryear{Sandford et al.}{2001}]{sand01}
Sandford, S. A.; Bernstein, M. P.; Allamandola, L. J.; Goorvitch, D.; Teixeira, T. C. V. S.
The abundances of solid N2 and gaseous CO2 in interstellar dense molecular clouds. {\it Astrophys. J.} {\bf 2001}, {\it 548}, 836.

\bibitem[\protect\citeauthoryear{Herbst and van Dishoeck}{2009}]{herb09}
Herbst, E.; van Dishoeck, E. F. Complex organic interstellar molecules. {\it Annu. Rev. Astron. Astrophys.} {\bf 2009}, {\it 47}, 427.

\bibitem[\protect\citeauthoryear{Sandford et al.}{1998}]{sand98}
Sandford, S. A.; Bernstein, M. P.; Swindle, T. D. The Trapping of Noble Gases by the Irradiation and Warming of Interstellar Ice Analogs.
{\it { Meteorit. Planet. Sci.}} {\bf 1988}, {\it 33}, A135.

\bibitem[\protect\citeauthoryear{Knez et al.}{2005}]{knez05}
Knez, C.; Boogert, A. C. A.; Pontoppidan, K. M.; Kessler-Silacci, J.; van Dishoeck, E. F.; Evans, Neal J., II; Augereau, J-C.; Blake, G.
A.; Lahuis, F. Spitzer Mid-Infrared Spectroscopy of Ices toward Extincted Background Stars. {\it { Astrophys. J.}} {\bf 2005}, {\it 635}, L145-L148.

{ 
\bibitem[\protect\citeauthoryear{Choi and Cho}{2011}]{choi11}
Choi, J. H.; Cho, M. Vibrational solvatochromism and electrochromism of infrared probe molecules containing C$\equiv$O, C$\equiv$N, C=O, or C-F vibrational chromophore. {\it J. Chem. Phys.} {\bf 2011}, {\it 134}, 154513.

\bibitem[\protect\citeauthoryear{Cappelli et al.}{2011}]{capp11}
Cappelli, C.; Lipparini, F.; Bloino, J.; Barone, V. Towards an accurate description of anharmonic infrared spectra in solution within the polarizable continuum model: Reaction field, cavity field and nonequilibrium effects. {\it J. Chem. Phys.} {\bf 2011}, {\it 135}, 104505.

\bibitem[\protect\citeauthoryear{B{\l}asiak et al.}{2013}]{blas13}
B{\l}asiak, B.; Lee, H.; Cho, M. Vibrational solvatochromism: Towards systematic approach to modeling solvation phenomena. {\it J. Chem. Phys.} {\bf 2013}, {\it 139}, 044111.

\bibitem[\protect\citeauthoryear{Max et al.}{2003}]{max03}
Max, J-J.; Chapados, C. Infrared spectroscopy of acetone-water liquid mixtures. I. Factor analysis.
{\it J. Chem. Phys.} {\bf 2003}, {\it 119}, 5632.

\bibitem[\protect\citeauthoryear{Max et al.}{2004}]{max04}
Max, J-J.; Chapados, C. Infrared spectroscopy of acetone-water liquid mixtures II. Molecular model.
{\it J. Chem. Phys.} {\bf 2004}, {\it 120}, 6625.}

\bibitem[\protect\citeauthoryear{Blake and Jenniskens}{1994}]{blak94}
Jenniskens, P.; Blake, D. F. Structural transitions in amorphous water ice and astrophysical implications.
{\it Science} {\bf 1994}, {\it 265}, 753.

\bibitem[\protect\citeauthoryear{Pradzynski et al.}{2012}]{prad12}
Pradzynski, C. C.; Forck, R. M.; Zeuch, T.; Slav\'{i}\~{c}ek, P.; Buck, U. A Fully Size-Resolved Perspective on the Crystallization of Water
Clusters. {\it Science} {\bf 2012}, {\it 337}, 1529.

\bibitem[\protect\citeauthoryear{Odutola and Dyke}{1980}]{odut80} Odutola, J. A.; Dyke, T. R. Partially deuterated water dimers: Microwave spectra and structure.
{\it J. Chem. Phys.} {\bf 1980}, {\it 72}, 5062.

\bibitem[\protect\citeauthoryear{Viant et al.}{1997}]{vian97}
Viant, M. R.; Cruzan, J. D.; Lucas, D. D.; Brown, M. G.; Liu, K; Saykally, R. J. Pseudorotation in Water Trimer Isotopomers Using Terahertz
Laser Spectroscopy. {\it J. Phys. Chem. A} {\bf 1997}, {\it 101}, 9032.

\bibitem[\protect\citeauthoryear{Liu et al.}{1997}]{liu97}
Liu, K.; Brown, M. G.; Saykally, R. J. Terahertz Laser Vibration-Rotation Tunneling Spectroscopy and
Dipole moment of cage Form of the Water Hexame. {\it J. Phys. Chem. A}, {\bf 1997}, {\it 101}, 8995.

\bibitem[\protect\citeauthoryear{Nauta and Miller}{2000}]{naut00}
Nauta, K.;  Miller, R. E. Formation of Cyclic Water Hexamer in Liquid Helium: The Smallest Piece of Ice. {\it Science} {\bf 2000}, {\it 287}, 293.

\bibitem[\protect\citeauthoryear{Abascal et al.}{2005}]{abas05}
Abascal, J.L.F., Sanz, E., Garcia Fernandez, R., Vega, C., {\it J. Chem. Phys.} {\bf 2005}, {\it 122}, 234511.

\bibitem[\protect\citeauthoryear{Sil et al.}{2017}]{sil17}
Sil, M.; Gorai, P.; Das, A.; Sahu, D.; Chakrabarti, S. K. Adsorption Energies of H and H$_2$: A Quantum Chemical Study.
{\it Eur. Phys. J. D.} {\bf 2017}, {\it 71}, 45.

\bibitem[\protect\citeauthoryear{Das et al.}{2018}]{das18}
Das, A.; Sil, M.; Gorai, P.; Chakrabarti, S. K.; Loison, J-C. An Approach to Estimate the Binding Energy of Interstellar Species.
{\it Astrophys. J. S. S.} {\bf 2018}, {\it 237}, 9.

\bibitem[\protect\citeauthoryear{Nguyen et al.}{2019}]{nguy19}
Nguyen, T.; Talbi, D.; Congiu, E.; Baouche, ES.; Loison, J-C.; Dulieu, F. Experimental and Theoretical Study of the Chemical Network of the Hydrogenation of NO on Interstellar Dust Grains. {\it ACS Earth Space Chem.} {\bf 2019}, {\it 3}, 1196.

\bibitem[\protect\citeauthoryear{Puzzarini et al.}{2014}]{puzz14}
Puzzarini, C.; Ali, A.; Biczysko, M.; Barone, V. Accurate spectroscopic characterization of protonated oxirane:
a potential prebiotic species in titan's atmosphere. {\it { Astrophys. J.}} {\bf 2014}, {\it 792}, 118.

\bibitem[\protect\citeauthoryear{Barone et al.}{2015a}]{baro15a}
Barone, B.; Biczysko, B.; Puzzarini, C. Quantum Chemistry Meets Spectroscopy for Astrochemistry: Increasing Complexity toward Prebiotic
Molecules. {\it Acc. Chem. Res.} {\bf 2015a}, {\it 48}, 1413.

\bibitem[\protect\citeauthoryear{Becke}{1988}]{beck88}
Becke, A. D. Density-functional exchange-energy approximation with correct asymptotic behavior. {\it { Phys. Rev. A}} {\bf 1988}, {\it 38},
3098.

\bibitem[\protect\citeauthoryear{Lee et al.}{1988}]{lee88} Lee, C.; Yang, W.; Parr, R. G.; Development of the Colle-Salvetti correlation-energy
formula into a functional of the electron density, {\it { Phys. Rev. B}} {\bf 1988}, {\it 37}, 785.

{
\bibitem[\protect\citeauthoryear{Frisch et al.}{2013}]{fris13}
Frisch, M. J.; Trucks, G. W.; Schlegel, H. B.; Scuseria, G. E.; Robb,  M. A.; Cheeseman, J. R.; Scalmani, G.; Barone, V.; Mennucci, B.; Petersson, G. A.; Nakatsuji, H. ; Caricato, M.; Li, X.; Hratchian, H. P.; Izmaylov, A. F.; Bloino, J.; Zheng, G.; Sonnenberg, J. L.; Hada, M.; Ehara, M.; Toyota, K.; Fukuda, R.; Hasegawa, J.; Ishida, M.; Nakajima, T.; Honda, Y.; Kitao, O.; Nakai, H.; Vreven, T.; Montgomery, J. A., Jr.; Peralta, J. E.; Ogliaro, F.; Bearpark, M. J.; Heyd, J. J.; Brothers, E. N.; Kudin, K. N.; Staroverov, V. N.; Keith, T. A.; Kobayashi, R.; Normand, J.; Raghavachari, K.; Rendell, A. P.; Burant, J. C.; Iyengar, S. S.; Tomasi, J.; Cossi, M.; Rega, N.; Millam, J. M.; Klene, M.; Knox, J. E.; Cross. J. B.; Bakken, V.; Adamo, C.; Jaramillo, J.; Gomperts, R.; Stratmann, R. E.; Yazyev, O.; Austin, A. J.; Cammi, R.; Pomelli, C.; Ochterski, J. W.; Martin, R. L.; Morokuma, K.; Zakrzewski, V. G.;  Voth, G. A.; Salvador, P.; Dannenberg, J. J.; Dapprich, S.; Daniels, A. D.; Farkas, O.; Foresman, J. B.; Ortiz, J. V.; Cioslowski, J.; Fox, D. J. Gaussian 09, Revision D.01; Gaussian, Inc.: Wallingford CT, 2013.}

\bibitem[\protect\citeauthoryear{Grimme}{2006}]{grim06} Grimme, S. J. Semiempirical hybrid density functional with perturbative second-order
correlation. {\it J. Chem. Phys.} {\bf 2006}, {\it 124}, 034108.

\bibitem[\protect\citeauthoryear{Papajak et al.}{2009}]{papa09}
Papajak E.; Leverentz, H.R.; Zheng, J.; Truhlar, D. G. Efficient Diffuse Basis Sets: cc-pVxZ+ and maug-cc-pVxZ.
{\it J. Chem. Theory Comput.} {\bf 2009}, {\it 5}, 1197.

\bibitem[\protect\citeauthoryear{Biczysko et al.}{2010}]{bicz10}
Biczysko, M.; Panek, P.;  Scalmani, G.; Bloino, J.; Barone, V. Harmonic and Anharmonic Vibrational Frequency Calculations with the
Double-Hybrid B2PLYP Method: Analytic Second Derivatives and Benchmark Studies. {\it J. Chem. Theory Comput.} {\bf 2010}, {\it 6}, 2115.

{
\bibitem[\protect\citeauthoryear{Frisch et al.}{2016}]{fris16}
Frisch, M. J.; Trucks, G. W.; Schlegel, H. B.; Scuseria, G. E.; Robb, M. A.; Cheeseman, J. R.; Scalmani, G.; Barone, V.; Petersson, G. A.; Nakatsuji, H.; Li, X.; Caricato, M.; Marenich, A. V.; Bloino, J.; Janesko, B. G.; Gomperts, R.; Mennucci, B.; Hratchian, H. P.; Ortiz, J. V.; Izmaylov, A. F.; Sonnenberg, J. L.; Williams-Young, D.; Ding, F.; Lipparini, F.; Egidi, F.; Goings, J.; Peng, B.; Petrone, A.; Henderson, T.; Ranasinghe, D.; Zakrzewski, V. G.; Gao, J.; Rega, N.; Zheng, G.; Liang, W.; Hada, M.; Ehara, M.; Toyota, K.; Fukuda, R.; Hasegawa, J.; Ishida, M.; Nakajima, T.; Honda, Y.; Kitao, O.; Nakai, H.; Vreven, T.; Throssell, K.; Montgomery, J. A., Jr.; Peralta, J. E.; Ogliaro, F.; Bearpark, M. J.; Heyd, J. J.; Brothers, E. N.; Kudin, K. N.; Staroverov, V. N.; Keith, T. A.; Kobayashi, R.; Normand, J.; Raghavachari, K.; Rendell, A. P.; Burant, J. C.; Iyengar, S. S.; Tomasi, J.; Cossi, M.; Millam, J. M.; Klene, M.; Adamo, C.; Cammi, R.; Ochterski, J. W.; Martin, R. L.; Morokuma, K.; Farkas, O.; Foresman, J. B.; Fox, D. J. Gaussian 16, Revision A.03; Gaussian, Inc.: Wallingford CT, 2016.}

\bibitem[\protect\citeauthoryear{Barone et al.}{2015b}]{baro15b}
Barone, V.; Biczysko, M.; Bloino, J.; Cimino, P.; Penocchio, E.; Puzzarini, C. CC/DFT Route toward Accurate Structures and Spectroscopic
Features for Observed and Elusive Conformers of Flexible Molecules: Pyruvic Acid as a Case Study. {\it J. Chem. Theory Comput.} {\bf 2015b}, 11, 4342.

{
\bibitem[\protect\citeauthoryear{Sanford et al.}{2020}]{sanf20}
Sanford, S. A.; Nuevo, M.; Bera,P.; Lee, T. J. Prebiotic Astrochemistry and the Formation of Molecules of Astrobiological Interest in Interstellar Clouds and Protostellar Disks. {\it Chem. Rev.} {\bf 2020}, dx.doi.org/10.1021/acs.chemrev.9b00560}

\bibitem[\protect\citeauthoryear{Tomasi et al.}{2005}]{toma05}
Tomasi, J.; Mennucci, B.; Cammi, R. Quantum Mechanical Continuum Solvation Models. {\it Chem. Rev.} {\bf 2005}, {\it 105}, 2999-3094.

\bibitem[\protect\citeauthoryear{Pagliai et al.}{2017}]{pagl17}
Pagliai, M.; Mancini, G.; Carnimeo, I.; De Mitri, N.; Barone, V. Electronic absorption spectra of pyridine and nicotine in aqueous solution \
with a combined molecular dynamics and polarizable QM/MM approach. {\it { J. Comput. Chem.}} {\bf 2017}, {\it 38}, 319.

\bibitem[\protect\citeauthoryear{Chung et al.}{2015}]{chun15} Chung, L. W.; Sameera; W. M. C.; Ramozzi, R.; Page, A .J; Hatanaka, M.; Petrova, G. P.; Harris; T.V.; Li; Xin; K. Z.; Liu, F.;
Li, H-B.; Ding, L.; Morokuma; K., The ONIOM Method and Its Applications. {\it { Chem. Rev.}} {\bf 2015}, {\it 115}, 5678-5796.

\bibitem[\protect\citeauthoryear{Wu et al.}{2006}]{wu06}
Wu, Y.; Tepper, H. L.; Voth, G. A. Flexible simple point-charge water model with improved liquid-state properties. {\it { J. Chem. Phys.}} {\bf 2006}, {\it 124}, 024503.

\bibitem[\protect\citeauthoryear{Cances et al.}{1997}]{canc97}
Canc\'{e}s, E.; Mennucci, B.; Tomasi, J. A new integral equation formalism for the
polarizable continuum model: Theoretical background and applications to isotropic and anisotropic dielectrics.
{\it { J. Chem. Phys.}} {\bf 1997}, {\it 107}, 3032.

\bibitem[\protect\citeauthoryear{Dawes et al.}{2016}]{dawe16}
Dawes, A.; Mason, N. J.; Fraser, H. J. Using the C-O stretch to unravel the nature of hydrogen bonding in
low-temperature solid methanol-water condensates. {\it Phys. Chem. Chem. Phys.} {\bf 2016}, {\it 18}, 1245-1257.

\bibitem[\protect\citeauthoryear{Licari et al.}{2015}]{lica15}
Licari, D.; Baiardi, A.; Biczysko, M.; Egidi, F.; Latouche, C.; Barone, V. Implementation of a graphical user interface for the virtual multifrequency spectrometer: The VMS-Draw tool. {\it J. Comput. Chem.} {\bf 2015}, {\it 36}, 321-334.

\bibitem[\protect\citeauthoryear{Mar\'echal}{1987}]{mare87}
Mar\'echal, Y. IR spectra of carboxylic acids in the gas phase: A quantitative reinvestigation. {\it J. Chem. {Phys.}} {\bf 1987}, {\it 87}, 6344.

\bibitem[\protect\citeauthoryear{Bisschop et al.}{2007}]{biss07}
Bisschop, S. E.; Fuchs, G.W.; Boogert, A.C.A.; van Dishoeck, E. F.; Linnartz, H. Infrared spectroscopy of HCOOH in interstellar ice analogues.
{\it { Astron. Astrophys.}}, {\bf 2007}, {\it 470}, 749.

\bibitem[\protect\citeauthoryear{Mastrapa et al.}{2009}]{mast09}
Mastrapa, R. M.; Sandford, S. A.; Roush, T. L.; Cruikshank, D.P.;  Dalle Ore, C. M. Optical constants of amorphous and crystalline H$_2$O-ice:
2.5-22 $\mu$m (4000-455 cm$^{-1}$ optical constants of H$_2$O-ice. {\it { Astrophys. J.}} {\bf 2009}, {\it 701}, 1347.

{
\bibitem[\protect\citeauthoryear{D'Hendecourt}{1986}]{dhen86}
D'Hendecourt, L. B. Ultraviolet Photoprocessing and Infrared Spectroscopy of Laboratory Simulated Grain Mantles. In {\it Light on Dark Matter}, Israel, F. P., Eds.; Astrophysics and Space Science Library; Springer, Dordrecht: D. Reidel Publishing Co., 1986; 124, 247-252.}

{
\bibitem[\protect\citeauthoryear{Zamirri et al.}{2018}]{zami18}
Zamirri, L.; Casassa, S.; Rimola, A.; Segado-Centellas, M.; Ceccarelli, C.; Ugliengo, P.,
IR spectral fingerprint of carbon monoxide in interstellar water-ice models. {\it Mon. Not. R. Astron. Soc.}, {\bf 2018}, {\it 480}, 1427.}

{ 
\bibitem[\protect\citeauthoryear{Collings et al. 2014}{2014}]{coll14}
Collings, M. P.; Dever, J . W.; McCoustra, R. S; The interaction of carbon monoxide with model
astrophysical surfaces. {\it Phys. Chem. Chem. Phys.}, {\bf 2014}, {\it 16}, 3479-3492.}


\bibitem[\protect\citeauthoryear{Grimme}{2010}]{grim10} Grimme, S.; Antony, J.; Ehrlich, S.; Krieg, H.; A consistent and accurate ab initio
parametrization of density functional dispersion correction (DFT-D) for the 94 elements H-Pu {\it J. Chem. Phys.} {\bf 2010},
{\it 132}, 154104.

\bibitem[\protect\citeauthoryear{Tso and Lee}{1985}]{tso85}
Tso, T.L.; Lee, E. K. C. Role of hydrogen bonding studied by the FTIR spectroscopy of the matrix-isolated molecular complexes, dimer of water,
water.carbon dioxide, water.carbon monoxide and hydrogen peroxide.n carbon monoxide in solid molecular oxygen at 12-17 K. {\it J. Phys. Chem.}
{\bf 1985}, {\it 89}, 1612.

\bibitem[\protect\citeauthoryear{Garozzo et al.}{2010}]{garo10}
Garozzo, M.; Fulvio, D.; Kanuchova, Z.; Palumbo M. E.; Strazzulla, G. The fate of S-bearing species after ion irradiation of interstellar icy grain mantles.
{\it { Astron. Astrophys.}}, {\bf 2010}, {\it 509}, A67.

\bibitem[\protect\citeauthoryear{Ehrenfreund et al.}{1992}]{ehre92}
Ehrenfreund, P.; Breukers, R.; D'Hendecourt, L.; Greenberg, J. M. On the possibility of detecting solid O2 in interstellar grain mantles.
{\it {Astron. Astrophys.}} {\bf 1992}, {\it 260}, 431-436.

\bibitem[\protect\citeauthoryear{Ehrenfreund and van Dishoeck.}{1998}]{ehre98}
Ehrenfreund, P.; van Dishoeck, E. F. The search for solid O$_2$ and N$_2$ with ISO. {\it { Adv. Space Res.}} {\bf 1998}, 21, 15E.

\bibitem[\protect\citeauthoryear{Shimonishi et al.}{2018}]{shim18}
Shimonishi, T.; Nakatani, N.; Furuya, K.; Hama, T. Adsorption Energies of Carbon, Nitrogen, and Oxygen Atoms on the Low-temperature Amorphous Water Ice: A Systematic Estimation from Quantum Chemistry Calculations.
{\it { Astrophys. J.}} {\bf 2018}, {\it 855}, 27.

\bibitem[\protect\citeauthoryear{Boogert et al.}{2000}]{boog00}
Boogert, A. C. A.; Tielens, A. G. G. M.; Ceccarelli, C.;  Boonman, A. M. S.; van Dishoeck, E. F.; Keane, J. V.; Whittet, D. C. B.; de Graauw, Th. Infrared observations of hot gas and cold ice toward the low mass protostar Elias 29. {\it { Astron. Astrophys.}} {\bf 2000}, {\it 360}, 683-698.

\end{thebibliography}
\end{document}